\documentclass[9pt,lineno]{article}
\usepackage{geometry}
\usepackage{amsmath}
\usepackage{amsfonts}
\usepackage{amssymb}
\usepackage{bbold}
\usepackage{pdfpages}
\usepackage{authblk}
\usepackage{cleveref}
\usepackage{natbib}
\usepackage{caption}
\newcommand\myFigureWidth{.95}

\title{Learning as filtering: \\ implications for spike-based plasticity}
\author[,a,b]{\small Jannes Jegminat\thanks{jannes.ini@uzh.ch}}
\author[a,b]{\small Jean-Pascal Pfister}
\affil[a]{\footnotesize Department of Physiology, University of Bern, 3012 Bern, Switzerland}
\affil[b]{\footnotesize Institute of Neuroinformatics and Neuroscience Center Zurich, ETH and the University of Zurich, 8057 Zurich, Switzerland}

\date{\footnotesize This manuscript was compiled on \today}
\begin{document}

\maketitle

%\blfootnote{Corresponding author: jannes@ini.uzh.ch}

% At least three keywords are required at submission. Please provide three to five keywords, separated by the pipe symbol.

\section*{Abstract} % 250
Most normative models in computational neuroscience describe the task of learning as the optimisation of a cost function with respect to a set of parameters. However, learning as optimisation fails to account for a time varying environment during the learning process; and the resulting point estimate in parameter space does not account for uncertainty. Here, we frame learning as filtering, i.e., a principled method for including time and parameter uncertainty. We derive the filtering-based learning rule for a spiking neuronal network - the Synaptic Filter - and show its computational and biological relevance. For the computational relevance, we show that filtering in combination with Bayesian regression improves performance compared to a gradient learning rule with optimal learning rate in terms of weight estimation. Furthermore, the filtering-based rule outperforms gradient-based rules in the presence of model mismatch, indicating a better generalisation performance. The dynamics of the mean of the Synaptic Filter is consistent with the spike-timing dependent plasticity (STDP) while the dynamics of the variance makes novel predictions regarding spike-timing dependent changes of EPSP variability. Moreover, the Synaptic Filter explains experimentally observed negative correlations between homo- and heterosynaptic plasticity. % and link our learning rule to stochastic release.

\section*{Significance statement}
The task of learning is commonly framed as parameter optimisation. Here, we adopt the framework of learning as filtering where the task is to continuously define the uncertainty about the parameters to be learned. We apply it to synaptic plasticity in a spiking neuronal network. Filtering includes a time varying environment and parameter uncertainty on the level of the learning task. We show that learning as filtering can qualitatively explain two biological experiments on synaptic plasticity that cannot be explained by learning as optimisation. Moreover, we make a new prediction and improve performance with respect to a gradient learning rule. Thus, learning as filtering is promising candidate for learning models.

%\thispagestyle{firststyle}
%\ifthenelse{\boolean{shortarticle}}{\ifthenelse{\boolean{singlecolumn}}{\abscontentformatted}{\abscontent}}{}

\section{Introduction}
% If your first paragraph (i.e. with the \dropcap) contains a list environment (quote, quotation, theorem, definition, enumerate, itemize...), the line after the list may have some extra indentation. If this is the case, add \parshape=0 to the end of the list environment.
In computational neuroscience, most normative models frame learning as optimisation of a static cost function with respect to a set of parameters, such as synaptic efficacies \citep{stemmler1999voltage, seung2003learning, lengyel2005matching, booij2005gradient, gutig2006tempotron, urbanczik2014learning, xu2013supervised, urbanczik2014learning,bohte2002error, bohte2005reducing} 
or a neuron's excitability \citep{triesch2005synergies, triesch2005gradient}. Different cost functions have been used to reproduce or predict experimental findings, such as a measure of sparseness and information preservation
\citep{Olshausen1996},
% voltage adaption
%the regulation of %dendritic and somatic %
%voltage-dependent conductances %by maximising the information transfer between a sensory stimulus and a neuron's firing rate at the level of a single cell
mutual information \citep{stemmler1999voltage}, 
%STDP
the probability of timed postsynaptic spiking
%by maximisation the probability of post synaptic spiking in a predefined interval 
\citep{pfister2006optimal,brea2013matching}, 
the mutual information of input and output spike trains \citep{chechik2003spike,toyoizumi2005generalized}, the network sensitivity \citep{bell2005maximising} and free energy \citep{buckley2017free}.
 %(Dayan &H¨ausser, 2004; Hopfield & Brody, 2004; Bohte & Mozer, 2005; Aihara,&Gerstner, 2005a, 2005b).
%Optimality approaches such as ours will never be able to make strict predictions about the properties of neurons or synapses. Optimality criteria may, however, help to elucidate computational principles and provide in- sights into potential tasks of electrophysiological phenomena such as STDP.
%These are just some example of the rich literature that frames learning as an optimisation task.

%\red{\% first limitation: parameter uncertainty \\}
However learning as optimisation has the drawback of not taking parameter uncertainty into account \citep{mackay1992bayesian}. When few training data are available compared to the number of model parameters, the parameter space is not sufficiently constrained, i.e., multiple parameter instantiations yield comparable model performance. % on a given data set.
Optimisation selects the best performing parameter, thereby ignoring the inherent parameter uncertainty present in a (probabilistic) model. This contributes to the problem of overfitting, i.e., the resulting performance on the training data does not generalise to the testing data \citep{gal2016dropout,blundell2015weight} %,fortunato2017bayesian,li2017dropout}. 
Moreover, many decision making models require as input not only the most likely prediction but also prediction uncertainty \citep{bell1982regret}. To obtain accurate prediction uncertainty, the contribution of parameter uncertainty must be taken into account (e.g. \citep{henning2018approximating}). Thus parameter uncertainty is computationally relevant for avoiding overfitting and the estimation of prediction uncertainty.

Learning as static optimisation is further limited because it lacks a principled way of accounting for a dynamic environment during learning. %In many settings, the goal of learning is to find a mapping from inputs to outputs based on data. 
Often the data distribution is assumed to be static, i.e., independent of time. However, in many settings the environment and, thus, the data distribution, are dynamic. For example, the association between a location and the availability of food is not static when the source of food depletes over time. Dynamic environments pose the additional challenge of determining the speed of learning. A slow learner fails to adapt to quickly changing environmental statistics while an overly fast learner might disregard past data prematurely.
The question of how to account for a dynamic environment during learning is closely related to continual learning, i.e., the task of sequentially learning from multiple data sets while maintaining (testing) performance on all previously observed ones \citep{kirkpatrick2017overcoming,parisi2019continual}. Here, the dynamics of the environment translate into the sequential availability of data sets. %n the framework of learning as optimisation, which frequently assumes a static cost function, it is it unclear how to include the dynamics of the environment and how fast to learn. In particular, the speed at which the data change have no connection to the update steps in the optimisation algorithm; it can only be introduced exogenously, for instance as learning rate parameter \citep{surace2020choice}.
%  not clear how to include  cannot be included straightforwardly in the cost function of the prediction model. Moreover, most continual learning studies consider a simple switch between data sets but it is unclear how to represent more complicated dynamics in the learning framework of optimisation, e.g. a gradual transitions between data sets.

%\red{neuroscience + relevance of uncertainty \\}
In neuroscience, time and uncertainty play an important role. 
Many studies have shown that a dynamic environment affects how animals learn \citep{kraemer1997adaptive,zimmermann2018multiple}. For instance, flies learn odour association and adapt their forgetting rate of old associations \citep{shuai2010forgetting,brea2014normative}. Similar experiments have been conducted with rodents \citep{fassihi2014tactile,akrami2011tactile}. 
Uncertainty of rewards has been studied in prefrontal and cingulate cortex on the basis of reinforcement learning models \citep{rushworth2008choice}; and several neuro-modulators have been identified that influence choices under uncertainty \citep{doya2008modulators,cocker2012irrational}. The uncertainty related to a whisker stimulus is directly related to neuronal activity in rat barrel cortex \citep{stuttgen2008psychophysical}. 
Uncertainty has also been linked to neuronal codes \citep{ma2010signal,stuttgen2008psychophysical} and many aspects of perception and decision making \citep{knill2004bayesian}. In the context of plasticity, uncertainty of synaptic weights has been linked to spine turnover \citep{kappel2015synaptic}.

% show case highlevel contribution
Normative models of learning can benefit from going beyond the framework of static optimisation by including parameter uncertainty and time in the learning task. However, it remains unclear which framework could prove to be a fruitful alternative. Here, we propose learning as filtering. Filtering, developed by mathematicians in the 60's \citep{kushner1964differential,kushner1967dynamical}, is a principled way to include time and uncertainty.
It continuously computes the posterior distribution (also called the filtering distribution) of a latent variable from all the observations up to time $t$.
We apply learning as filtering to synaptic plasticity, a field in which the need for new learning paradigms has become apparent \citep{brea2016does}.

% specific contributions
In a continuous-time, spiking neuronal network, we derive the update rule for the synaptic weight distribution and call it the Synaptic Filter. The Synaptic Filter is computationally relevant because it outperforms a gradient rule with optimised learning rate parameter in a dynamic weight estimation task, confirming a previous result \citep{aitchison2015synaptic}. Going beyond performance measured in weight space, we study the predictive performance %of the output neuron 
and find that the Synaptic Filter outperforms the optimised gradient rule as well, and that it is robust to the presence of model mismatch. From the biological perspective, the Synaptic Filter makes three experimental predictions.
First, the mean synaptic weight change depends on the precise timing of pre- and postsynaptic spikes and is therefore reminiscent of spike-timing dependent plasticity (STDP), which yields long term potentiation of the synaptic strength (LTP) if the postsynaptic spikes follows the presynaptic spikes and long term depression (LTD) otherwise \citep{markram1997regulation,bi1998synaptic}. 
Normative models of STDP have provided a consistent view on the pre-post LTP lobe. Pre-before-post pairs induce LTP reinforcing causality. Therefore the time constant of LTP reflects the EPSP time constant. However, normative models do not provide a unifying view on the LTD window \citep{pfister2006optimal}. Here, we provide a novel computational rationale for the LTD lobe, namely to compensate for a change in bias.
Secondly, based on the hypothesis that EPSP variability encodes synaptic weight uncertainty \citep{aitchison2014bayesian,aitchison2015synaptic} we formulate the novel prediction that EPSP variability also changes as a function of the precise timing of the spikes. Finally, weight changes induced by joint pre- and postsynaptic activity at one synapse can induce weight changes at synapses that did not receive presynaptic input, reminiscent of the phenomenon of heterosynaptic plasticity. Indeed, our learning rule can explain the negative correlation between homo- and heterosynaptic plasticity observed in experiments \citep{royer2003conservation}.

\begin{figure}[h!]
%\begin{minipage}{\myFigureWidth\linewidth}
\begin{minipage}{\myFigureWidth\linewidth}
\begin{tabular}{ll}
{\bf (A)} & {\bf (B)} \\
\begin{minipage}{0.45\textwidth}
\includegraphics[width=\textwidth,trim={0cm 0cm 0cm 0cm},clip]{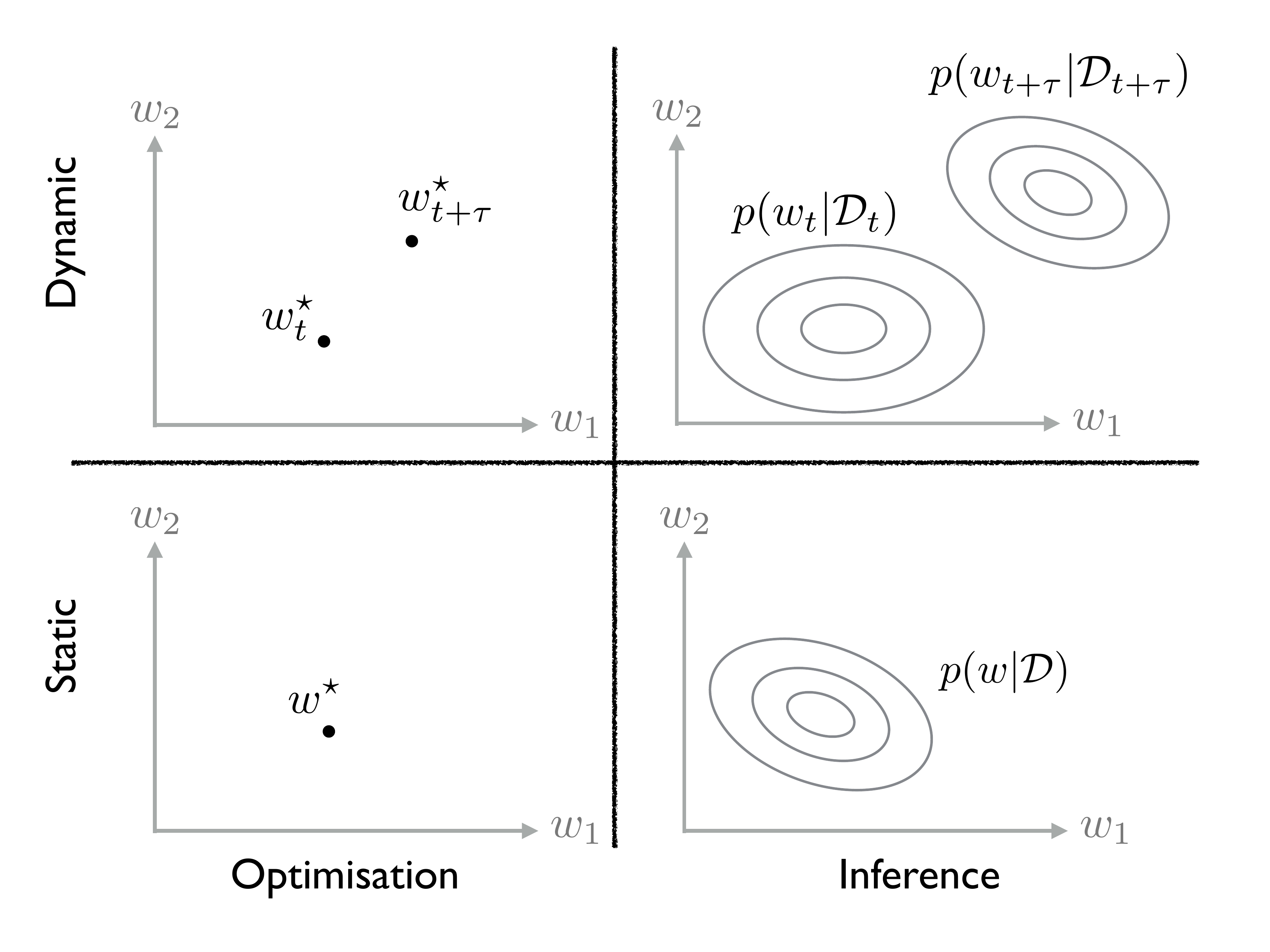}
\end{minipage} &
\begin{minipage}{0.45\textwidth}
\includegraphics[width=\textwidth]{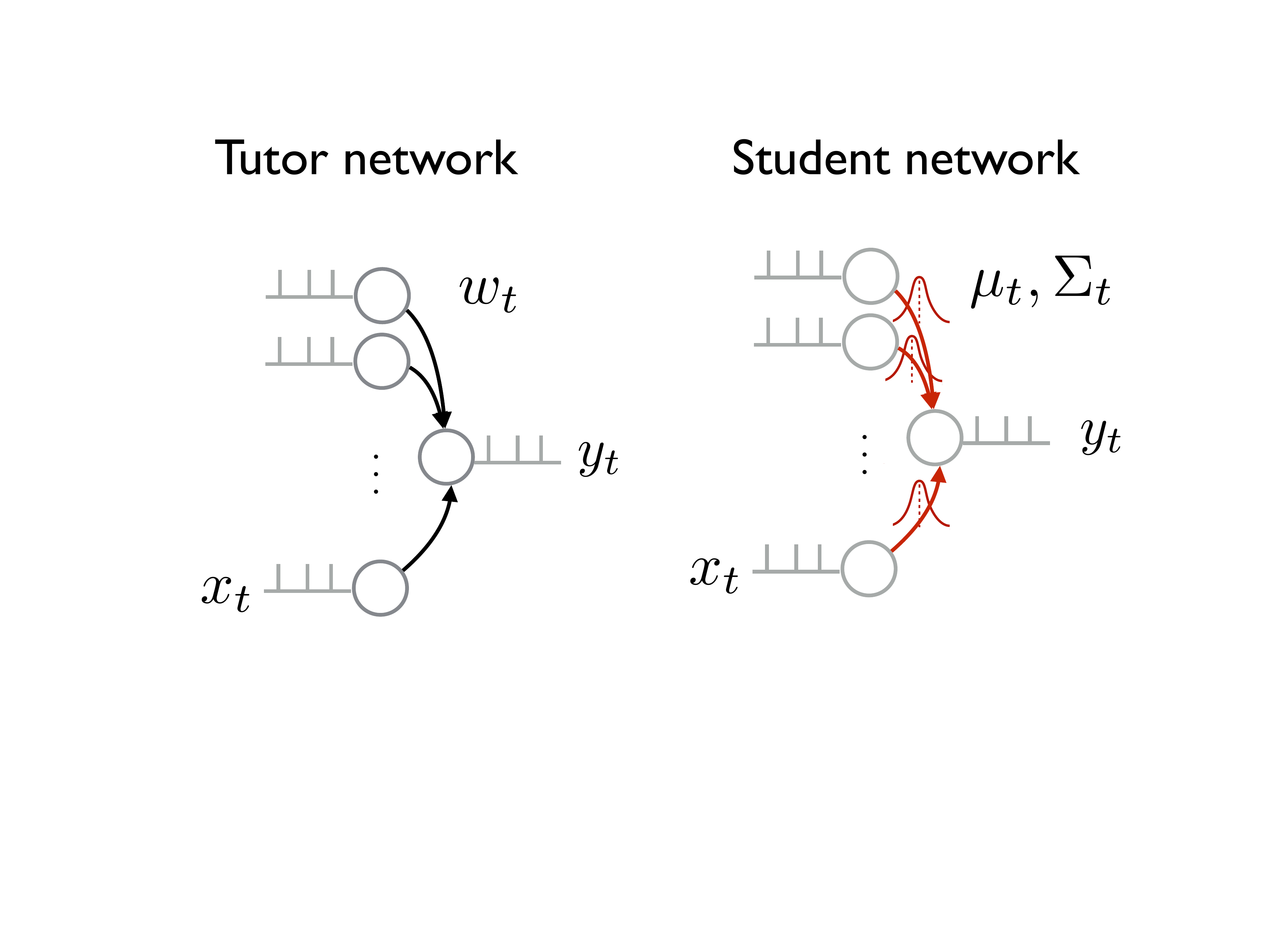}
\end{minipage}\\
{\bf (C)} & {\bf (D)} \\
\begin{minipage}{0.49\textwidth}
\includegraphics[width=\textwidth]{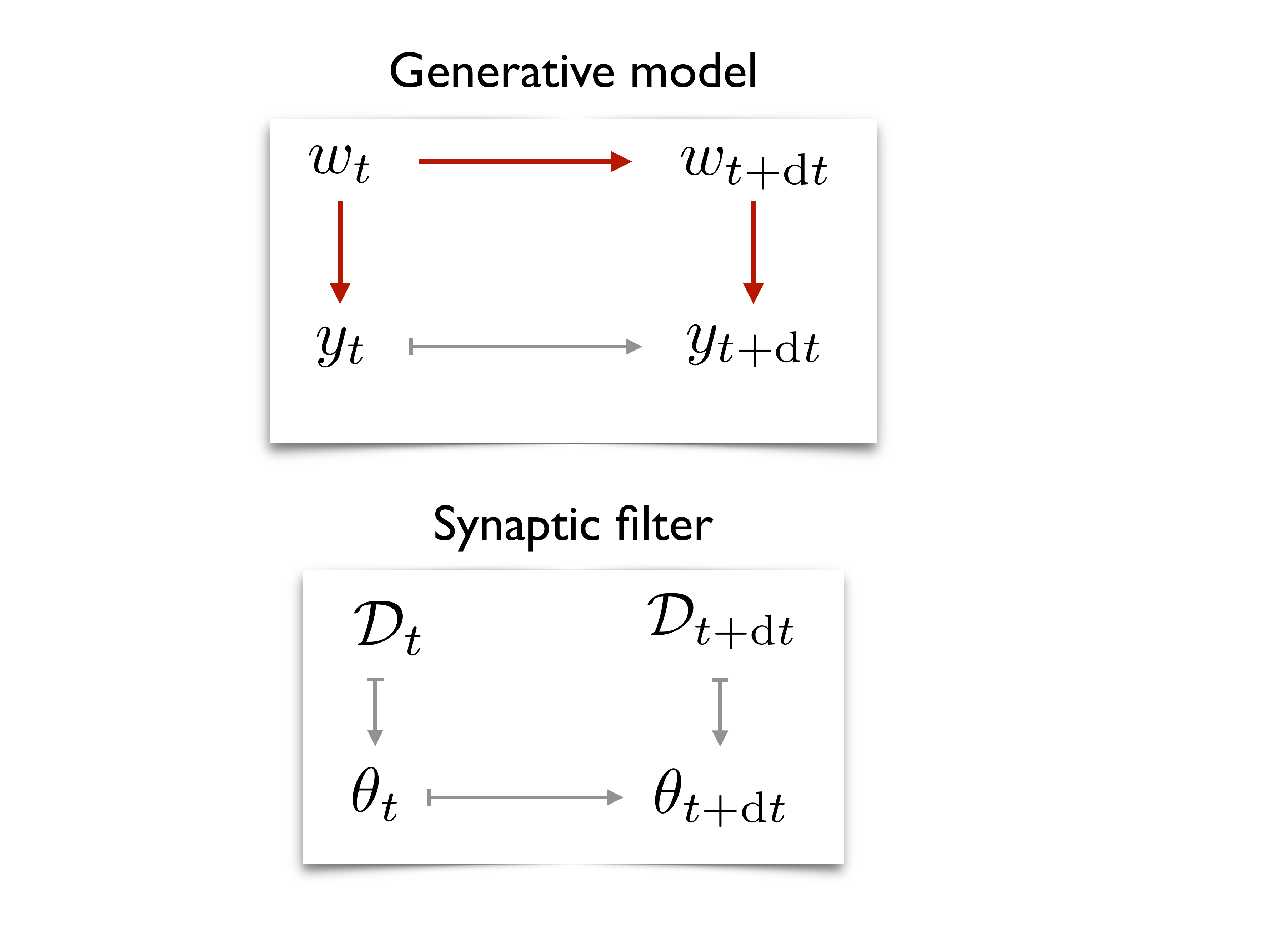}
\end{minipage} &
\begin{minipage}{0.49\textwidth}
\includegraphics[width=\textwidth]{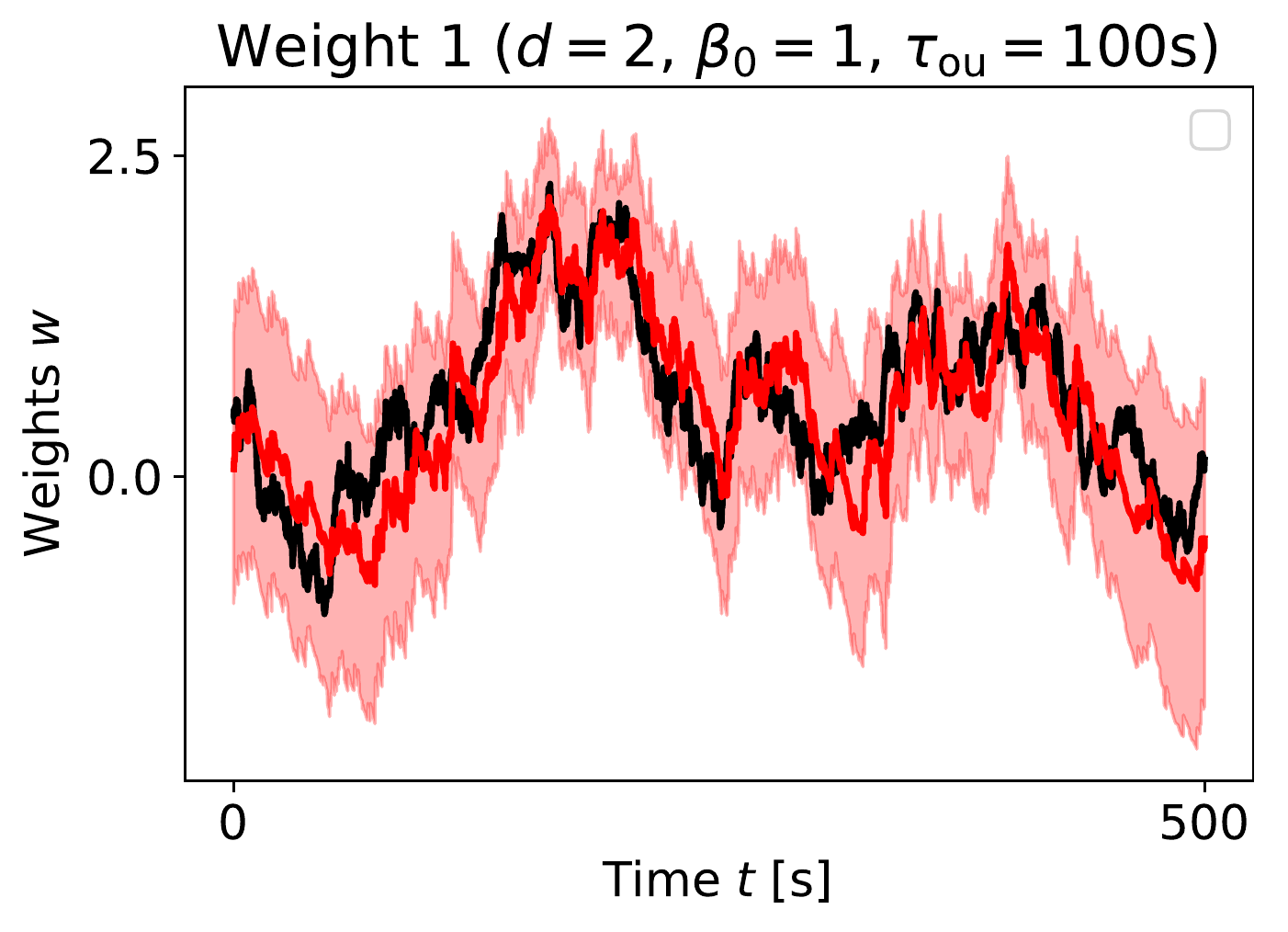}
\end{minipage}
\end{tabular}
\end{minipage}
\caption{ \label{fig:generative-model}
\textbf{(A)} Learning tasks can be static or dynamic, and deterministic or stochastic.
\textbf{(B)} The generative model %of the dynamic, stochastic synapse 
represents the assumption that the observed spike train $y_t$ was generated from a tutor network with the same input $x_t$ and hidden weights $w_t$.
\textbf{(C)} Graphical model representation of the generative model (top), the Synaptic Filter (bottom) with deterministic dependencies shown in gray and probabilistic ones in red.
\textbf{(D)} Time series of a ground truth weight (black) in the tutor network and weight distribution (red, shaded area = 2-SD) learned by the student network.
}
\end{figure}

\section{Results}
\subsection{The Synaptic Filter}

% high level intution and motivation
The goal of learning is to find predictive functions from training data $\mathcal{D}$ which map inputs $x$ to outputs $y$ typically based on a parametrised generative model. The generative model specifies how the output $y$ of the predictive function is computed from the parameters $w$ and the input $x$. Accounting for a dynamic environment on the level of the generative model, makes the data and the parameter $w$ time dependent. Accounting for the fact that many parameter instantiations are compatible with the training data corresponds to computing predictions based on parameter uncertainty. Thus, a given static generative model for learning can be generalised by including the assumption of a dynamic environment or parameter uncertainty.
%Time and uncertainty are naturally connected during learning. The link from time to uncertainty is that a dynamic data distribution renders less recent data less informative. This reduces the effective data set size, hence making parameter uncertainty important to guard against overfitting. The connection from uncertainty to time is that the speed of learning is modulated by how much information the current observation provides compared to the information already taken into account. Parameter updates with high certainty should be updated more slowly than parameter with high uncertainty.\\ 

%
%% explain fig A
Including both, time and parameter uncertainty, yields the framework of learning as filtering. \Cref{fig:generative-model} \textbf{(A)} illustrates how filtering generalises a static optimisation, which is the dominant learning framework. The learning task of static optimisation is to find a point estimate $w^\star$ in parameter space given the generative model and the data $\mathcal{D}$. By including parameter uncertainty, the learning task generalises to inferring the posterior distribution over parameters $p(w | \mathcal{D})$. 
In the limit of infinite data and for a convex parameter landscape, the parameter distribution collapses around a point, yielding similar results for parameter optimisation and inference, i.e., $p(w | \mathcal{D}) \approx \delta(w - w^\star)$. However, in many problems equivalent minima exist and the amount of data is limited such that the posterior distribution is not well approximated by a point estimate.
Another extension of static optimisation considers dynamically changing environmental statistics, i.e., the data distribution is time dependent. In this case, the learning task is to track the optimal parameter as a function of time $w^\star_t$. In a filtering framework, which includes both, parameter uncertainty and time, the task is to compute the so-called filtering distribution over the weights as a function of time $p(w_t | \mathcal{D}_t)$ given all previous observations up to time $t$.

%However, optimisation does not account for parameter uncertainty and time in a principled way. Learning as inference addresses this limitation by inferring a distribution over parameters given the data instead of point estimate. This distribution accounts for parameter uncertainty. Learning as inference is relevant when the data does not constrain the parameters sufficiently, i.e., when the data set size is small compared to the number of parameters. Another important limitation of learning as optimisation is time, i.e., it cannot account for a dynamic data generating distribution. To generalise learning as optimisation to the case of a dynamic data generating distribution, the learner has to continuously update the optimal parameter. In particular, older data has to be discounted because it is less informative about the distribution from which the data is being generated at the current moment. This reduces the effective data set size. Since a small (effective) data set size makes it important to include parameter uncertainty, time and parameter uncertainty are naturally connected. The mathematical framework that formalises the connection between time and uncertainty is filtering. The objectives of the four learning tasks, i.e., tasks with and without time, and with and without uncertainty, are illustrated in \Cref{fig:generative-model} \textbf{(A)}.
% central thing

\subsubsection*{A generative model to study spike-based plasticity}
%
%% hidden evolution
To study learning as filtering in the context of spike-based plasticity, we consider a generative model with time dependent parameters $w_t \in \mathbb{R}^{d}$, the weights. In contrast to learning as static optimisation, we do not assume that the weights are fixed. Changes in the weights reflect changes in the statistics of the environment.
Here, we assume that the weights follow an Ornstein-Uhlenbeck (OU) process with mean $\mu_{\rm{ou}} = 0$, diagonal equilibrium covariance matrix $\Sigma_{\rm ou} = \sigma^2_{\rm{ou}} \mathbb{1}$ with (non-zero elements) $\sigma_{\rm{ou}}^2=1$ and time scale $\tau_{\rm{ou}}$.
The limit of a large OU time constant, represents a static environment while the limit $\tau_{\rm ou} \rightarrow 0$ represents an environment that changes too fast for meaningful learning.

%
%% Generative probability
At each moment in time, the weights relate the input spike trains and output spike of a single neuron via the observation probability $p(y_t | w_t, x_{0:t})$. %Again, this contrasts with learning as static optimisation, because the mapping between the observations, i.e., the output spikes, $y_t$ and inputs spikes $x_t$ changes due to the SDE that governs $w_t$. 
Here, we assume that output spikes $y_t$ are generated from an inhomogeneous Poisson neuron with instantaneous firing rate given by an exponential gain function $g(u_t) = g_0 \exp(\beta u_t)$ with base rate $g_0$. The membrane potential $u_t$ is a sum of presynaptic inputs weighted by $w_t$. The determinism parameter $\beta$ controls how strongly the membrane potential affects the output spiking. For $\beta = 0$, the output spikes are independent of the membrane $u_t$ and reflect only the base rate $g_0$. With the additional assumption that the time scale of the membrane potential $\tau_{\rm m}$ is much smaller than the one of the weights, i.e., $\tau_{\rm{m}} \ll \tau_{\rm{ou}}$, we ensure the Markovianity of the generative model (see \Cref{eq:u_t tau_m << tau_ou approximation} in Materials and Methods).
The generative model is represented as graphical model in \Cref{fig:generative-model} \textbf{(C)}.
To ensure that performance metrics can be compared across various dimensions $d$, we let the determinism parameter scale with the dimension $\beta \propto \beta_0 d^{-1/2}$ such that the expected firing rate of the output neuron, and hence the rate of observations from the hidden weights, becomes independent of the dimension $d$ (see \Cref{met:sec:u-scaling} in Materials and Methods).

\subsubsection*{An assumed density filter solution: the Synaptic Filter}
%
%% Filtering task intution
The generative model can be conceptualised as the belief that a tutor network with ground truth weights $w_t$, illustrated in \Cref{fig:generative-model} \textbf{(B)}, generates the observed output spikes from given inputs. Learning as filtering corresponds to a student network that continuously computes the distribution over the ground truth weights $p(w_t | \mathcal{D}_{t})$ given the history of inputs and outputs $\mathcal{D}_t = \{ (x,y)_\tau \}_{\tau=0}^t$. %The fact that filtering includes parameter uncertainty as part of the learning task corresponds to computing a distribution over the weights rather than aiming to obtain a single point estimate of the parameter.

%
%% ADF
Generally, the filtering distribution $p(w_t | \mathcal{D}_{t})$ is intractable. Here, we obtain an approximated solution with an Assumed Density Filter (see Supplementary Information for the derivation), i.e., the exact filtering distribution $p(w_t | \mathcal{D}_t)$ is approximated with the parametric distribution $q_{\theta_t}(w_t)$. For the proposed generative model, we call the Gaussian Assumed Density Filter the \textit{Synaptic Filter}. The distribution parameters $\theta_t := (\mu_t, \Sigma_t)$ denote the mean $\mu_t$ and covariance matrix $\Sigma_t$ of the Gaussian. An Assumed Density Filtering reduces the problem to updating the distribution parameters $\theta_t$ based on observations:
\begin{align}
\label{eq:dotmu}
\dot{\mu}_t &= 
\beta \Sigma_t x_t^{\epsilon} (y_t - \gamma_t) + \tau_{\rm{ou}}^{-1} (\mu_{\rm{ou}}-\mu_t),
\\
\label{eq:dotsigma2}
\dot{\Sigma}_t &= - \beta^2 \gamma_t (\Sigma_t x_t^{\epsilon})(\Sigma_t x_t^{\epsilon})^\top + 
2 \tau_{\rm{ou}}^{-1} (\Sigma_{\rm{ou}}-\Sigma_t),
\end{align}
where $\gamma_t$ is the expected firing rate, i.e., the expectation of the firing rate $g(u_t)$ computed based on the approximated filtering distribution $q_{\theta_t}(w_t)$ and $x^\epsilon_t$ denotes the presynaptic activation (see Materials and Methods). The first term in \Cref{eq:dotmu,eq:dotsigma2} comes from the observations, which is why it scales with $\beta$. The update of the mean has the structure of 3-factor learning rule \citep{fremaux2016neuromodulated} with classical Hebbian factors, i.e., the pre-synaptic activation $x_t^\epsilon$ and the difference between observed and expected output $y_t - \gamma_t$, and the covariance $\Sigma_t$ as third factor with a modulatory function, which has been linked to the computation of surprise \citep{liakoni2019approximate}. The second term in \Cref{eq:dotmu,eq:dotsigma2} comes from the hidden dynamics of the weights, which is why it is proportional to the inverse time constant $\tau^{-1}_{\rm ou}$. For $\beta = 0$ or $\tau_{\rm ou} \rightarrow \infty$ the updates become independent of observations or the environmental dynamics respectively.
In general, the covariance update in Assumed Density Filtering depends on the observations $y_t$. However, the combination of a Gaussian filtering distribution and an exponential gain function yields an update (\Cref{eq:dotsigma2}) independent of the observations; an interesting similarity with the Kalman filter.
\Cref{fig:generative-model} \textbf{(D)} illustrate that the synaptic filter (red) successfully tracks the weights of the tutor network (black). In the Supplementary Information (Section 2), we show that the Synaptic Filter is a good approximation of the true filtering distribution.

In addition to the Synaptic Filter, we also derive and analyse the \textit{Diagnalised Synaptic Filter}, an Assumed Density Filter based on a diagonal Gaussian (see Supplementary Information). The updates \Cref{eq:dotmu,eq:dotsigma2} differ only in that the covariance matrix is diagonal and the off-diagonal updates are omitted.

%
%% Analysis
%\red{is this paragraph necessary here or distribute across B1 and B2?}
%The first term in \cref{eq:dot mu,eq:dot sigma2} comes from the likelihood and the second from the dynamics of the generative weights. For both parameters $\mu$ and $\Sigma$, learning is gated by $\beta \Sigma x^\epsilon$. The intuition is that a more deterministic neuron ($\beta$ large) conveys more information about the hidden parameters. When $\Sigma$ is diagonal, the term $\Sigma x^\epsilon$ attributes more plasticity to active and uncertain weights. The off-diagonal elements in $\Sigma$ represent correlations between the hidden weights. When non-zero, they distribute the activation $x^\epsilon$ across the updates of multiple weights, i.e., an activation in the $i^{\rm{th}}$ weight can lead to a change in the $j^{\rm{th}}$ synapse if $\Sigma_{ij} \neq 0$; even if $x^\epsilon_j = 0$. The mean update \cref{eq:dot mu} is Hebbian with the error signal $y_t - \langle g \rangle$ representing (on average) the difference between observed and predicted spike rate. The variance update is independent of the output spikes \cref{eq:dot sigma2}. The variance depends only on the pre-synaptic firing rate (similar to \citep{aitchison2015synaptic}).

%% C0 moved to SI

\subsection{MSE performance of the Synaptic Filter}
\label{res:sec:c1}
%
% definition
The natural performance metric in filtering is the normalised Mean Square Error (MSE). It quantifies how closely the mean of the filtering distribution follows the weights of the tutor in \Cref{fig:generative-model} \textbf{(B)}. %Note that a filter can have poor MSE performance while still solving the filtering problem correctly. 
The MSE is defined by $\text{MSE} := d^{-1} \langle (w_t - \mu_t)^\top (w_t - \mu_t) \rangle_t$.
% relevance
%Compared to the ability of a filter to correctly estimate the moments of the exact filtering distributions (as analysed in \Cref{res:sec:c0}), the MSE addresses the question whether a filter is useful for locating the ground truth weights. For example, when the observations $y_t$ contain no information about the hidden weights $w_t$, the exact filtering distribution is identical to the prior but it will not be useful in terms of MSE.

% outlook
In the following, the MSE performance of the Synaptic Filter and the Diagonal Synaptic Filter are evaluated for a range of values for the determinism parameter $\beta_0$ with fixed dimension $d=5$, and a range of dimensions $d$ for a fixed value of the determinism $\beta_0 = 1$. Additionally, we compare the MSE of a gradient rule with optimised learning rate against the Synaptic Filter. The (expected) observation rate across dimensions is comparable because we chose the following scaling of $\beta \propto \beta_0 d^{-1}$. This cancels the linear scaling of the membrane potential variance with dimensionality of the model (\Cref{met:sec:u-scaling} in Materials and Methods). %In all simulation, the variance of the membrane potential is effectively scaled with $d^{-1}$ to ensure that the statistics of the output spikes are comparable across dimensions (\Cref{met:sec:u-scaling} in Materials and Methods).

\begin{figure}[h!]
\begin{minipage}{\myFigureWidth\linewidth}
\begin{tabular}{ll}
% C1
{\bf (A)} & {\bf (B)} \\
\begin{minipage}{0.45\textwidth}
\includegraphics[width=\textwidth,trim={0cm 0cm 0cm 0cm},clip]{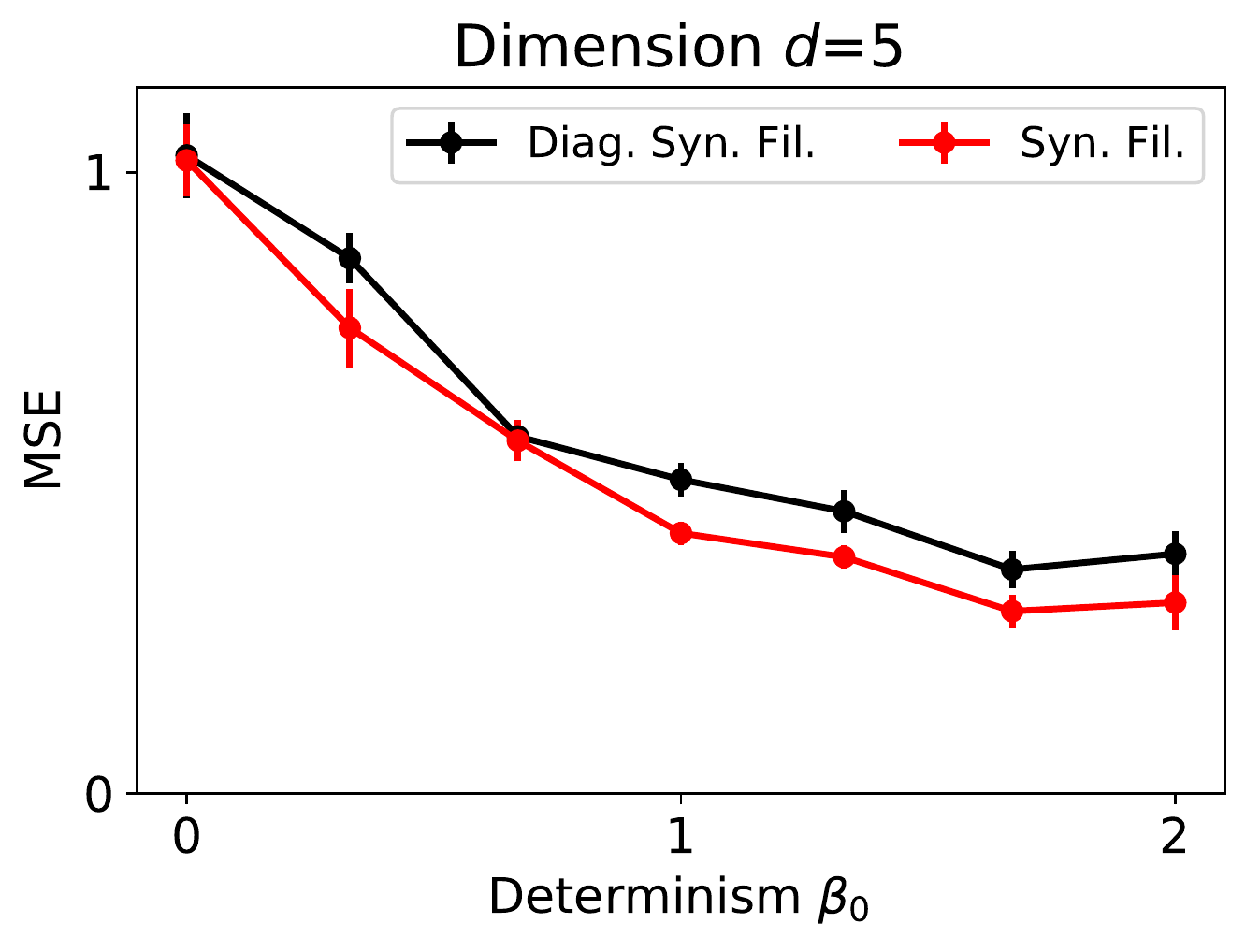}
\end{minipage} &
\begin{minipage}{0.45\textwidth}
\includegraphics[width=\textwidth]{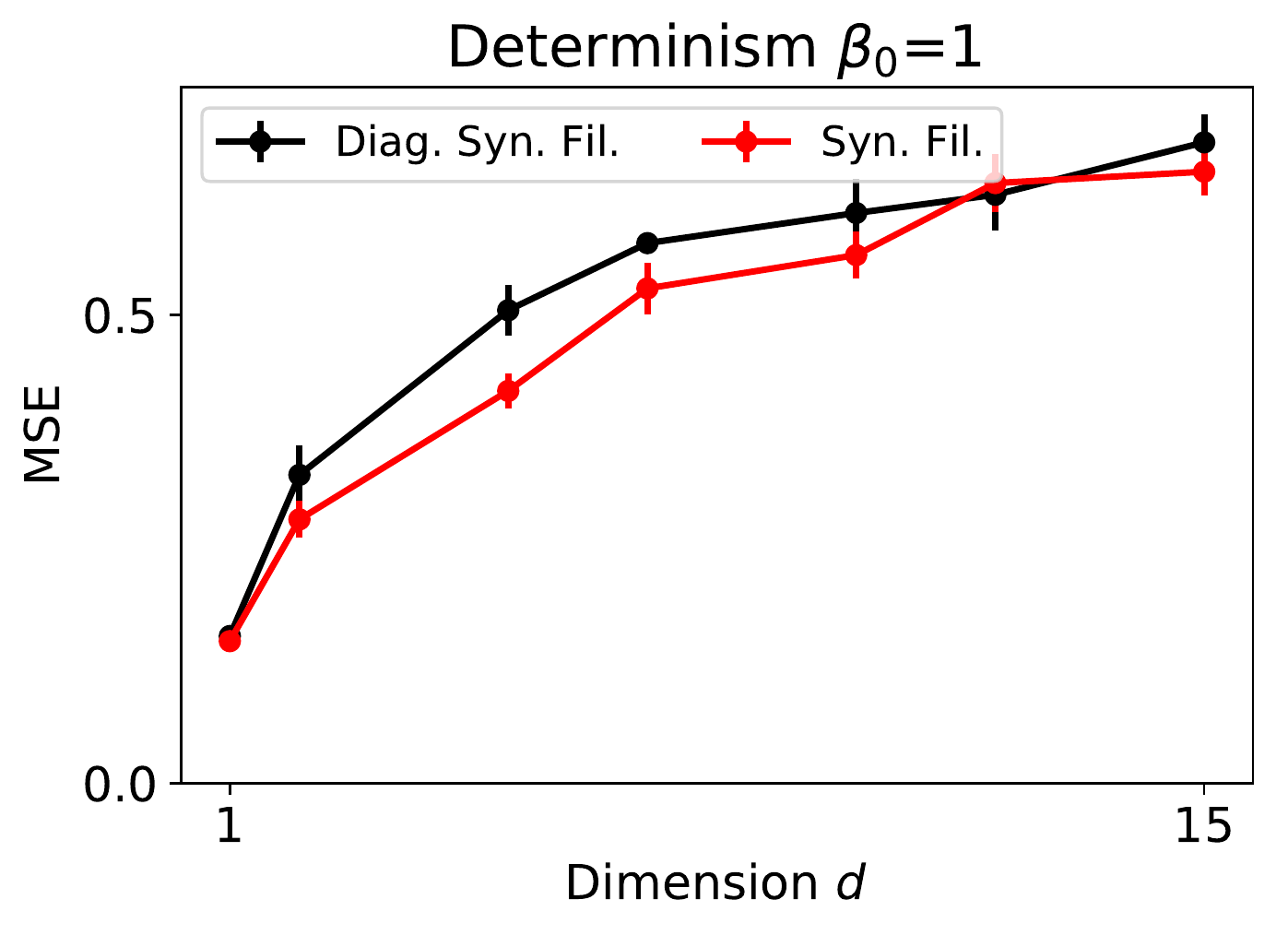}
\end{minipage}\\
{\bf (C)} & %{\bf (D)} 
\\
\multicolumn{2}{l}{
\begin{minipage}{0.9\textwidth}
\includegraphics[width=\textwidth,trim={0cm 0cm 0cm 0cm},clip]{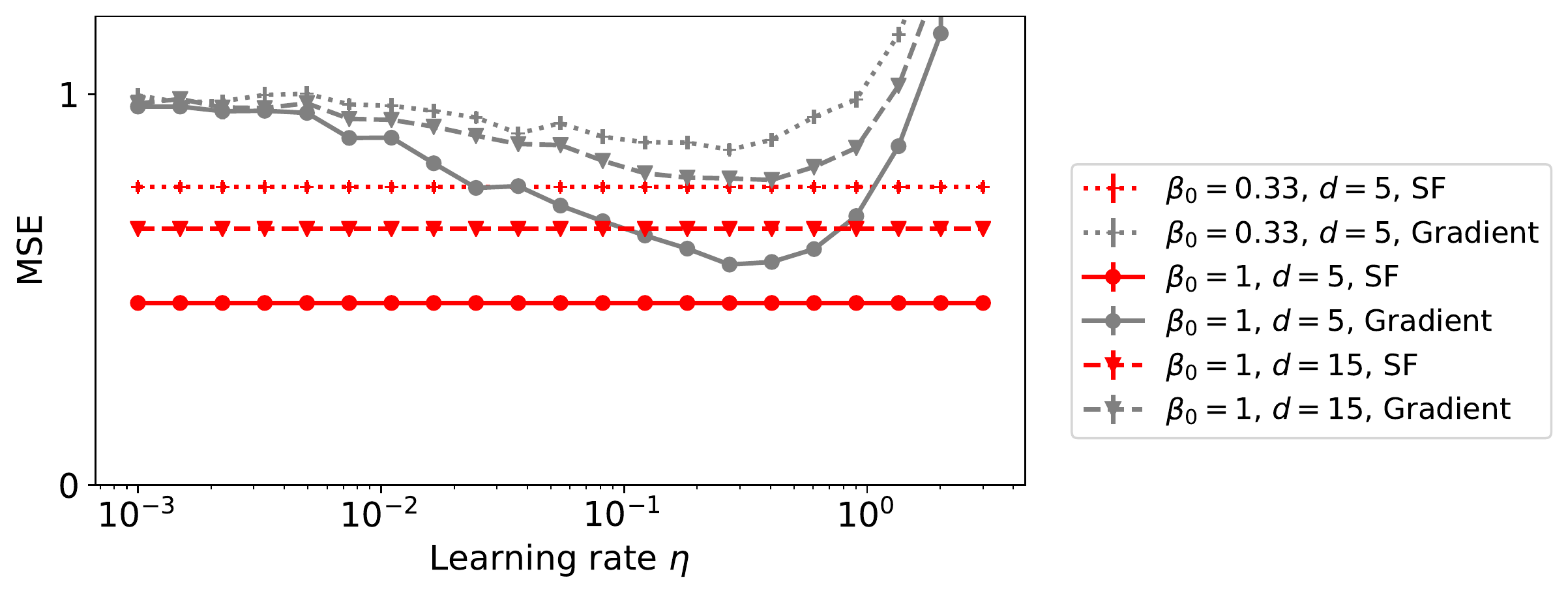}
\end{minipage}} \\
% C2
{\bf (D)} & {\bf (E)} \\
\begin{minipage}{0.45\textwidth}
\includegraphics[width=\textwidth,trim={0cm 0cm 0cm 0cm},clip]{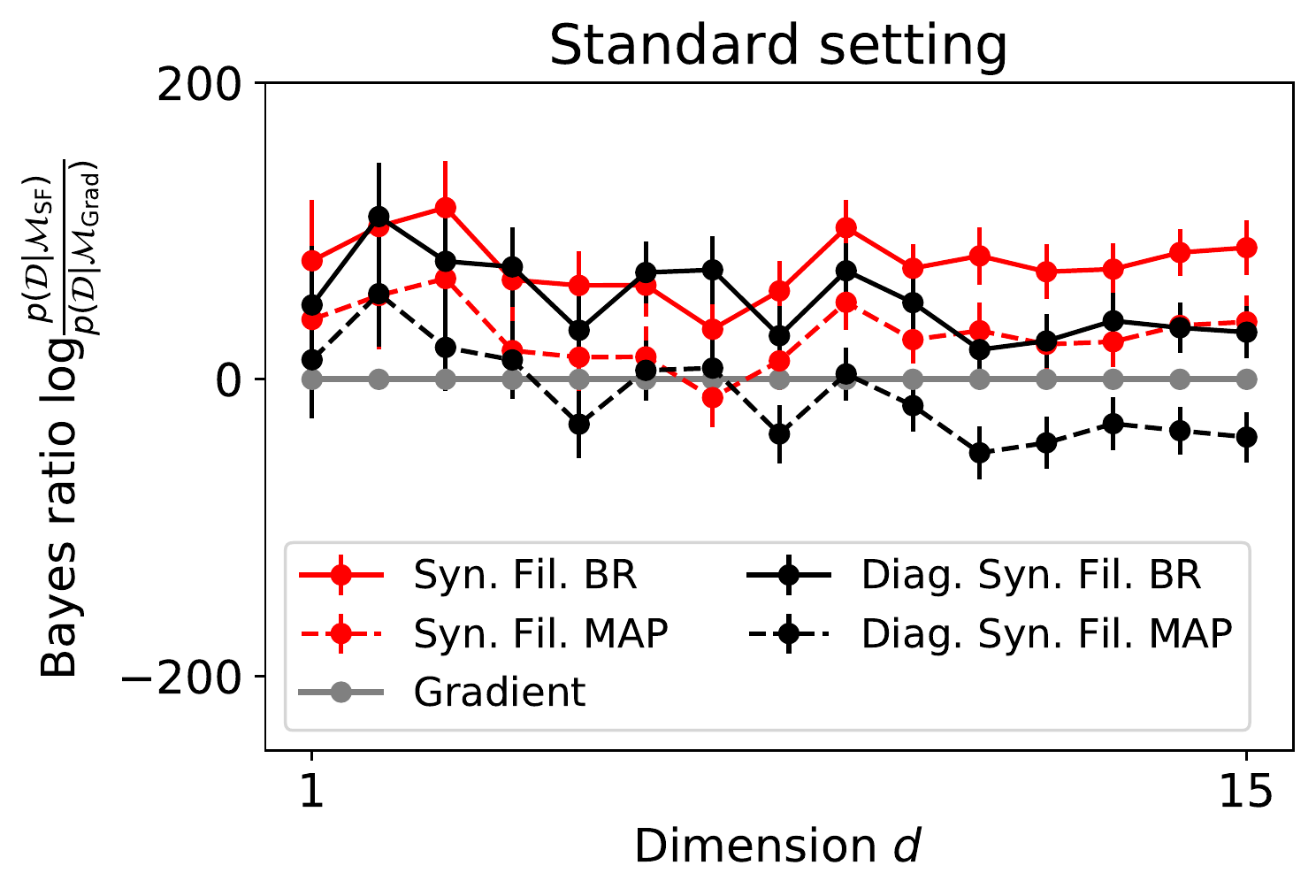}
\end{minipage} &
\begin{minipage}{0.45\textwidth}
\includegraphics[width=\textwidth]{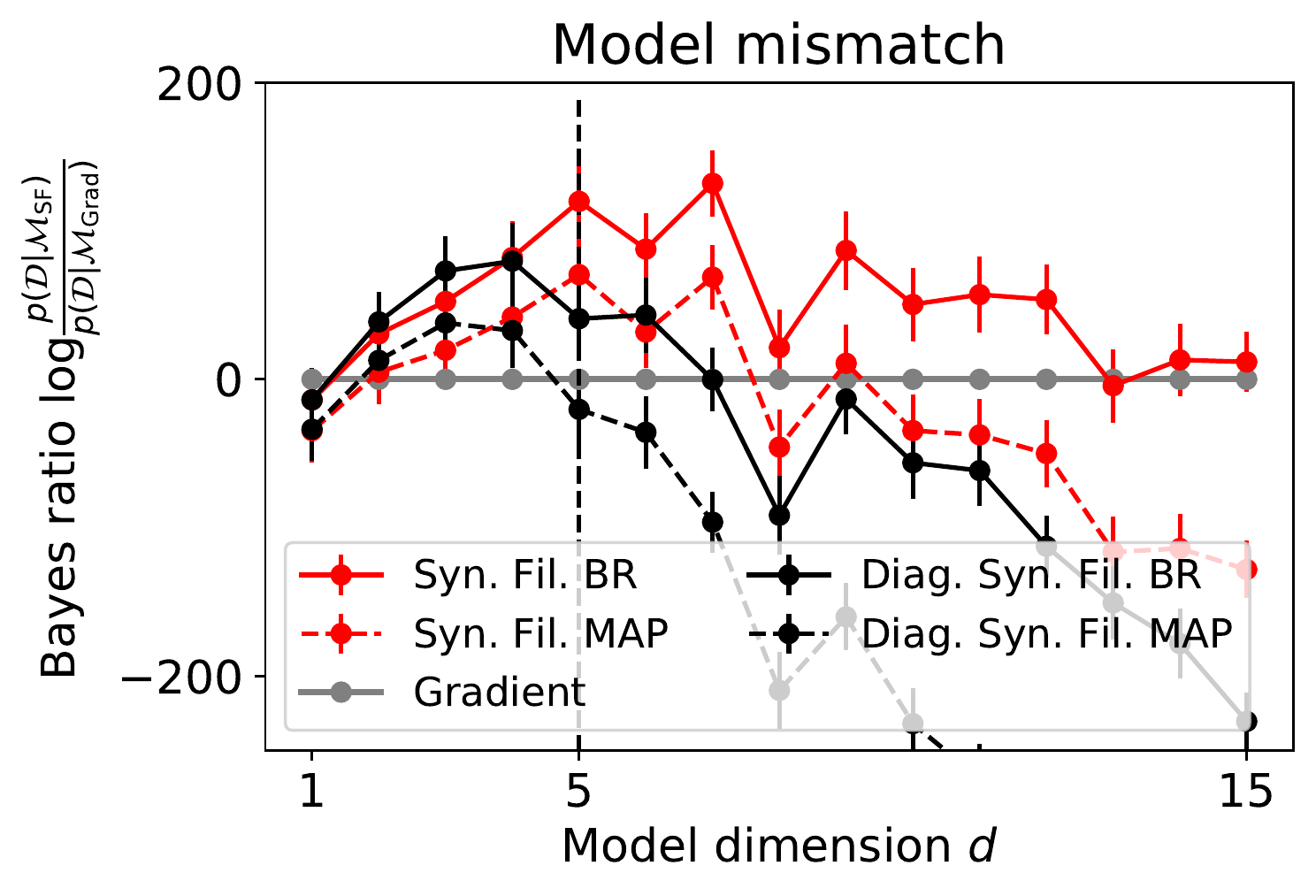}
\end{minipage}
\end{tabular}
\end{minipage}
\caption{ \label{fig:comp-results} % MSE-vs-d
% key message
The Synaptic Filter has the best overall MSE and predictive performance.
% describe plots
\textbf{(A)} The MSE of Synaptic Filter (red line) and the Diagonal Synaptic Filter (black line) decrease as the determinism $\beta_0$ increases. At $\beta_0 = 0$, the MSE corresponds to the equilibrium variance of the prior, $\sigma_{\text{ou}}^2 = 1$. The Diagonal Synaptic Filter performs slightly worse.
\textbf{(B)} The MSE increases as a function of dimension. Again, the Diagonal Synaptic variants (black) perform worse.
\textbf{(C)} The MSE of the Synaptic Filter (red) is lower than the MSE of a gradient learning rule (gray) for a range of learning rates $\eta$. The symbols indicates three combinations of determinism $\beta_0$ and dimension $d$. Consistently with \textbf{(A, B)}, the lowest MSE (black) is obtained at the lowest dimension and highest determinism, i.e. $\beta_0 = 1$ and $d=5$. 
%
%% C2 %\caption{ \label{res:fig:c2}
%The Synaptic Filter has higher predictive performance than the Diagonal Synaptic Filter and a optimal gradient rule.
\textbf{(D)} The predictive performance is measured by the Bayes factor relative to a gradient learning (gray) rule with optimised learning rate. The Synaptic Filter (solid red line) and the Diagonal Synaptic Filter (solid black line) have superior performance and the Synaptic Filter performs best overall. Using Maximum a Posteriori (MAP, dashed lines), which does not include uncertainty during prediction, yields lower performance than using Bayesian Regression (BR, solid lines). 
\textbf{(E)} In the presence of model mismatch, $d_{\rm{tutor}} = 5 \neq d$, the Synaptic Filter has the best overall performance. At high dimensions the optimal gradient rule performs equally well. The Diagonal Synaptic Filter and filters with MAP prediction overfit in the regime 
$d_{\rm{tutor}} \ll d$. 
Dots and errors in \textbf{(A-E)} denote the mean and SEM from 100 simulations. %In \textbf{(D-E)}, the mean and SEM were computed under the assumption of log normally distributed performance data and 
The error bars take the gradient rule's SEM into account via the root mean square.
%Errors represent the SEM across 100 simulations.
% parameters
%The simulated time per run was $10 \tau_{ou} = 1000$s.
See \Cref{met:sec:c012} in Materials and Methods for simulations details.
}
\end{figure}

%
% MSE(b0)
The MSE of both filtering models decreases as function of $\beta_0$, as shown in \cref{fig:comp-results} \textbf{(A)}. The Synaptic Filter (red line) performs slightly better than its diagonalised counterpart (black line). As $\beta_0$ increases the observations convey more information about the ground truth weights $w_t$, hence allowing for a more accurate estimation. At $\beta_0 = 0$, the MSE of all models is close to 1, a value that corresponds to the MSE of the prior.

%
% MSE(d)
For all four filtering models, the MSE increases as a function of the dimension $d$, as depicted in \cref{fig:comp-results} \textbf{(B)}. The Diagonal Synaptic Filter has a slightly higher MSE. Increasing the dimension increases the difficulty of the filtering task because more weights compete for explaining the observation. For instance, at $d=1$, observations can be attributed to $w_{t,0}$ uniquely but at $d=2$ the information in the observations is distributed across both weights, $w_{t,0}$ and $w_{t,1}$. Thus, the scope of learning the weights via filtering is limited to the regime in which the observations convey enough information per dimension and per time.

%
%% Gradient rule
As a benchmark for the Synaptic Filter, we use a gradient rule with a scalar learning rate\footnote{This implicitly defines a Euclidean metric in weight space. A more general learning choice corresponds to a matrix-valued learning rate.} (see \Cref{met:sec:wML} in Material and Methods). As shown in \Cref{fig:comp-results} \textbf{(C)}, the MSE of the gradient rule (gray) is higher than the MSE (red) of the Synaptic Filter for a large range of learning rates. The MSE as a function of learning rate $\eta$ exhibits a minimum when the combined effect of delayed learning and overshooting are minimal. Delayed learning occurs at low learning rates because the gradient does not converge before the ground truth weights change. At high learning rates, the update steps of the gradient rule are too large which leads to overshooting. In contrast, the Synaptic Filter optimally tunes the learning rate for each weight individually and at each moment based on the amount of information available in the data.

%
%% Conclusion
The Synaptic and the Diagonal Synaptic Filter have the highest MSE performance in low dimensions $d$ and with high determinism $\beta_0$. In this regime, observations contain the largest amount of information per weight. The existence of a limited high performance regime is a feature of the dynamic learning task, not a limitation of the filters. In particular, the Synaptic Filter outperforms the optimised gradient rule.

%
%% MSE performance
Despite the fact that the benefit of learning as filtering seems to be limited to a regime of low dimensionality and that biological neuron have up to $10^4$ synaptic inputs, learning as filtering is a relevant framework for modelling synaptic plasticity. First, the sparsity of the neuronal code could limit the effective number of input dimensions at any point in time. Thus it is possible that learning in many brain systems takes place inside the learning regime. Secondly, the framework can be applied to high dimensions by aggregating the majority of synapses and inputs into a single variable, e.g., the bias $w_{t,0}$, and modelling only the remaining small group of synapses explicitly. It would be interesting to investigate whether multiple of these low-dimensional models yield good performance based on observations generated from a high-dimensional tutor network.

\subsection{Predictive performance of the Synaptic Filter}
\label{res:sec:c2}
%
% motivation
The MSE is a function of the hidden variable, in our case the weights $w_t$, but it is not directly sensitive to whether the predictions of the network are useful, e.g., a low MSE does not necessitate poor predictions. The MSE is particularly limiting in presence of a model mismatch, when the student network in \Cref{fig:generative-model} \textbf{(B)} makes false assumptions about how the tutor network maps inputs to outputs. If the model mismatch follows from a differences in weight dimensionality, the MSE cannot be defined. In the case of the nervous system, model mismatch is always present because sensory data is generated from physical processes which cannot be represented in detail by the brain.

%
%% Solution
This motivates the study of predictive performance which measures how well the student network predicts the emission of the next output spike $y_t$. The measure is equivalent with the Bayesian model evidence. Prediction performance can be evaluated even when the generative model is incorrect. %, i.e., when the structure of the student and the tutor match, and in the case that the generative model is wrong, i.e., the student wrongly assumes that its own structure matches the one of the tutor.

%
%% Method
The Synaptic Filters make predictions on the basis of Bayesian regression (see \Cref{met:sec:c012} in Materials and Methods), a method that takes advantage of the filtering distribution to compute the posterior predictive distribution:
\begin{align}
\label{res:eq:BR}
    p(y_t | \mathcal{D}_{t}, x_t^\epsilon) = \int p(w_t | \mathcal{D}_{t}) p(y_t|w_t, x_t^\epsilon) \text{d}w_t.
\end{align}
Intuitively, the marginalisation over the weights $w_t$ represents a weighted average over the probability of an output spike $y_t$. Thus, Bayesian regression takes advantage of parameter uncertainty encoded in the filtering distribution, including correlations between parameters when represented. To study the importance of including parameter uncertainty during prediction, we included a MAP prediction based on replacing the posterior in \Cref{res:eq:BR} with a delta-function around its maximum.
%% model comparison perspective
Based on the posterior predictive and the data $\mathcal{D}_t$, we compute the log evidence for four models $\mathcal{M}$, i.e., the Synaptic Filter, the Diagonal Synaptic Filter and their MAP-version (see \Cref{met:eq:ppd-gamma} in Materials and Methods), from time 0 to $t$:
$\log p(\mathcal{D}_t | \mathcal{M})$. The model evidence is reported relative to a null model given by the baseline firing rate $g_0$. % and in units of the loglikelihood $\mathcal{L}_g$ of the tutor's firing rate $g(w_t x^\epsilon_t)$. 
The model evidence is a testing error in the sense that the prediction has not been informed by the current output $y_t$, i.e., the parameters of the weight distribution have not been updated. As a benchmark, we use a gradient rule with optimal learning rate (\Cref{met:sec:c012} in Materials and Methods).

%
%% Results standard
In the standard setting, i.e., without model mismatch, the Synpatic Filter and the Diagonal Synaptic Filter outperform the gradient rule, as shown in \cref{fig:comp-results} \textbf{(D)}. The Synaptic Filter has the best performance, followed by the Diagonal Synaptic Filter and finally the gradient rule. The performance gain of the Synaptic Filter and Diagonal Synaptic Filter over the gradient method can attributed to two factors. The first factor is their ability to estimate the value of the weight $w_t$ of the tutor network better, consistent with the gain in MSE performance in \cref{fig:comp-results} \textbf{(C)}. The second factor becomes evident by comparing the Bayesian regression prediction with the MAP prediction, i.e., including weight uncertainty via Bayesian regression improves performance.

%
%% MM
Next, we wondered whether the introduction of a model mismatch would affect the model evidence of the Synaptic Filters compared to the benchmark. Specifically, we considered a tutor network with $d_{\rm tutor} = 5$ and a student network with varying dimension $d \in (1, \dots, 15)$. Our rationale for this type of model mismatch was twofold. First, differences in model evidence can no longer be explained in terms of the MSE because the MSE is not defined; thus, the differences in model evidence can be more directly be attributed to the use of Bayesian regression or MAP for prediction. Secondly, a central argument for using parameter uncertainty is that it helps to avoid overfitting. The risk of overfitting is high when the tutor has fewer dimensions than the student, i.e., $d_{\rm tutor} < d$. In this case, the process that generates the data has less degrees of freedom than the model that aims at explain the data.

\Cref{fig:comp-results} \textbf{(E)} shows that the Synaptic Filters perform better than the optimised gradient rule when their dimension is smaller or equal to the tutor $d \leq d_{\text{tutor}}$. However, when the tutor has less dimensions that the student networks, $d > d_{\text{tutor}}$, the performance of the Diagonal Synaptic Filter drops while the Synaptic Filter continues to outperform the gradient method.
This result can be explained by the fact that the Synaptic Filter includes correlations between weights while the Diagonal Synaptic Filter does not. In the case of the Synaptic Filter, weights become negatively correlated as multiple combinations of weights compete for explaining the data. This weight uncertainty cannot be reduced by additional observations because its fundamental cause is the model mismatch, i.e., extra dimensions of the student compared to the tutor. With Bayesian regression (BR) it is possible to include this uncertainty in the predictions. In contrast, not including weight correlations as in the case of the Synaptic Filter with Maximum a Posteriori (MAP) predictions leads to lower performance.
% redo!
%This is the reason why the Synaptic Filter has the best predictive performance. 
%In contrast, the drop in predictive performance of the Diagonal Synaptic Filter can be attributed to overfitting, i.e., the failure of an overly flexible model to account for parameter uncertainty, here encoded as weight correlations, when computing predictions. 

It might seem surprising that the gradient rule, which does not account for any parameter uncertainty, eventually outperforms all Synaptic Filters. The reason is that the optimisation of the loglikelihood with respect to the learning rate adds considerable power to the model. For instance, small learning rates effectively limit the risk of overfitting because weights do not converge quickly enough.
%
% conclusion
Overall, the Synaptic Filter has higher predictive performance than the Diagonal Synaptic Filter and a gradient learning rule with optimal learning rate. This result was obtained in two conditions, with and without model mismatch. In particular, the model mismatch condition showed that the Synaptic Filter in combination with Bayesian regression is robust to overfitting because it accounts for the full posterior, including weight correlations.

\subsection{The Synaptic Filter is consistent STDP}
\label{res:sec:b1}
% background: STDP
Spike-time dependent plasticity (STDP) refers to the property of a synapse to exhibit long-term potentiation (LTP) if the presynaptic spike comes before the postsynaptic spike and long-term depression (LTD) otherwise \citep{markram1997regulation,bi1998synaptic}. The results are usually depicted as STDP curve, i.e., the normalised change in synaptic weight as a function of the time interval between the pre and the postsynaptic spike.

% Question and motivation
Normative models of STDP have explained the LTP lobe, i.e., the weight change for pre- before postsynaptic spiking, in terms of causality reinforcement \citep{booij2005gradient,gutig2006tempotron,urbanczik2014learning,xu2013supervised,bohte2005reducing,pfister2006optimal}. The delay of the postsynaptic relative to the presynaptic spike represents the degree to which the occurrence of the postsynaptic spike can be attributed to the presynaptic activity trace and its decay resembles the shape of LTP lobe.
In contrast, a post-before-pre spike pair has no causal relationship. Indeed, the computational rationale for the LTD lobe has remained a matter debate with proposals including the regulation of the postsynaptic firing rate and temporal locality \citep{pfister2006optimal}.

% method
We wondered whether the Synaptic Filter could reproduce the STDP curve, in particular the LTD part. Our rationale was that postsynaptic spiking predominately affects the bias $w_{t,0}$, which represents the neuronal excitability. From this perspective, STDP is a secondary (differential) effect, i.e., changes in the synaptic weight account for the prediction error left unexplained by the adjustment of bias. Immediately after the occurrence of a postsynaptic spike, the expected firing rate is increased, leading to more LTD (\Cref{eq:dotmu}). Thus, the time scale of the LTD lobe corresponds to the time scale of the transition probability of the prior $\tau_{\rm ou, bias}$. In the simulations, we assumed $\tau_{\rm ou, bias} = \tau_{\rm m}$ but set all other transition time scales to values much larger than the duration of the experiment (\Cref{met:sec:b123} Materials and Methods).

To study the effect of the bias on the STDP curve, we applied a STDP protocol with one spike pair to three versions of the Synaptic Filter, i.e., a single synapse without bias and the Diagonal Synaptic Filter and Synaptic Filter with bias. To avoid effects from transients, the protocol was applied after the bias had reached its steady state. In contrast to biological experiments with up to 60 spike pairs, we simulated a single spike pair to avoid complications from saturation effects and induction times.

%
% mean, bias and negative lobe
In all three experiment, the resulting STDP curve shows an exponentially-shaped LTP lobe while the LTD lobe occurs only when the bias is included, as shown in \Cref{fig:STDP-prediction} \textbf{(A)}). 
% positive lobe
The LTP lobe resembles the exponential shape of the presynaptic activation because when the postsynaptic spike occurs, the weight update is proportional to the current amplitude of the presynaptic trace.
% negative lobe
The LTD lobe is present when the bias is included (black and red lines). Without the bias (gray line), the LTD part of the STDP curve is independent of the spike timing. Moreover, the bias lowers the amplitude of LTP. Both observations, LTD lobe and lower LTP, are caused by the modulation of the expected firing rate $\gamma_t$ due to the bias $w_{t,0}$. The bias acts as low-pass filter of the postsynaptic spike train. Its value is maximal immediately after the occurrence of a postsynaptic spike at $t_{\rm{post}}$ and relaxes back to its equilibrium afterwards. The faster the pre follows the postsynaptic spike, the larger the value of the expected firing rate $\gamma_{t_{\rm pre}}$ when the presynaptic spike occurs. Since LTD is proportional to $\gamma_t$, shorter intervals between pre and postsynaptic spikes lead to more LTD, in correspondence with biological STDP experiments. The dynamics of the variables of the Synaptic Filter are shown in detail %in \Cref{si:fig:STDP-variable-dynamics} in 
the Supplementary Information.

%
%% Conclusion
The Synaptic Filter and Diagonal Synaptic Filter is consistent with the STDP. %This is remarkable because they were derived from the learning as filtering. 
The appearance of the LTD lobe is contingent on inclusion of the bias. Indeed, since the underlying mechanism does not require the inclusion of uncertainty, a similar result could be obtained in a learning as optimisation framework as well, e.g., as the post-only term in the expansion of a generic update function  \citep{gerstner2002spiking}.
% single synapse
The fact that only a single synapse with bias was considered does not impair generality because additional unstimulated synapses do not affect the result.
From an experimental perspective, the proposed mechanism can be tested by comparing the time scale of the negative lobe with the time scale of the excitability of a neuron, which is controlled by the bias.

%
%% Variance STDP
\subsection{The Synaptic Filter predicts spike-timing dependent changes of the variance}
\label{res:sec:b2}
%
% variance
%The prediction of variance STDP is shown in \Cref{fig:STDP-prediction} \textbf{(B)}. In all three cases, the variance decreases. The amplitude is spike timing dependent because of the term $\langle g \rangle_w$ in the update rule. In the limit that $\langle g \rangle_w \approx \text{const}$, the variance decrease independently of the spike timing. Including the bias, smoothes the negative lobe at $t_{\text{pre}} > t_{\text{post}}$ such that it shape becomes more similar to the negative lobe at $t_{\text{pre}} < t_{\text{post}}$.\\ 

%
% overview and setup: same protocol, same rules, cov prediction, measurement
So far, we have assumed that the posterior mean $\mu_t$ can be measured experimentally as average EPSP amplitude. From here on, we assume the sampling hypothesis to be true, i.e., that the posterior variance $\sigma_t^2$ corresponds to the EPSP variance \citep{aitchison2015synaptic}. We wondered whether, in this case, the Synaptic Filter predicts spike-timing dependent changes in the EPSP variance.

\begin{figure}[h!]
\begin{minipage}{\myFigureWidth\linewidth}
\begin{tabular}{ll}
{\bf (A)} & {\bf (B)} \\
\begin{minipage}{0.45\textwidth}
\includegraphics[width=\textwidth,trim={0cm 0cm 0cm 0cm},clip]{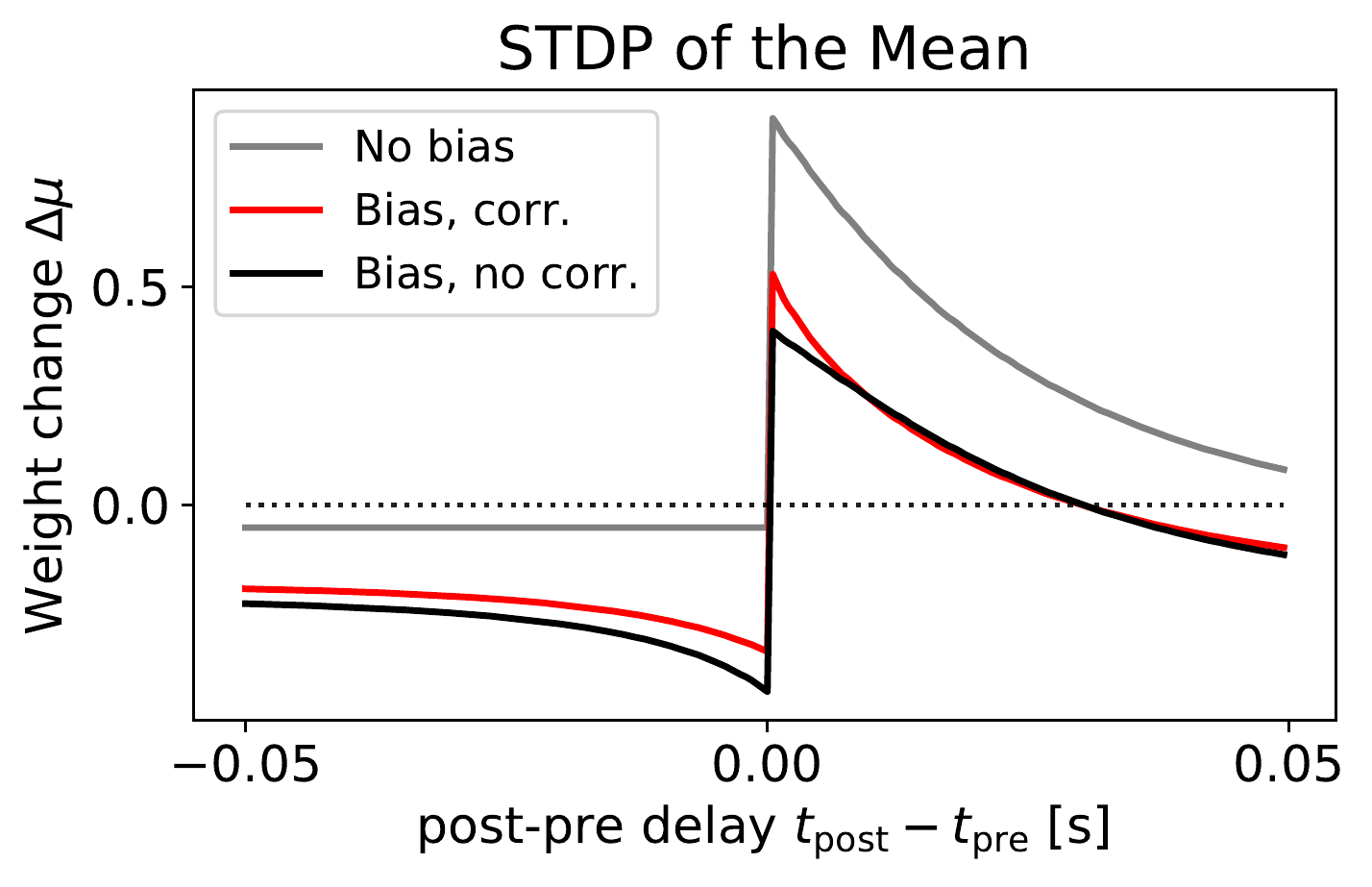}
\end{minipage} &
\begin{minipage}{0.45\textwidth}
\includegraphics[width=\textwidth]{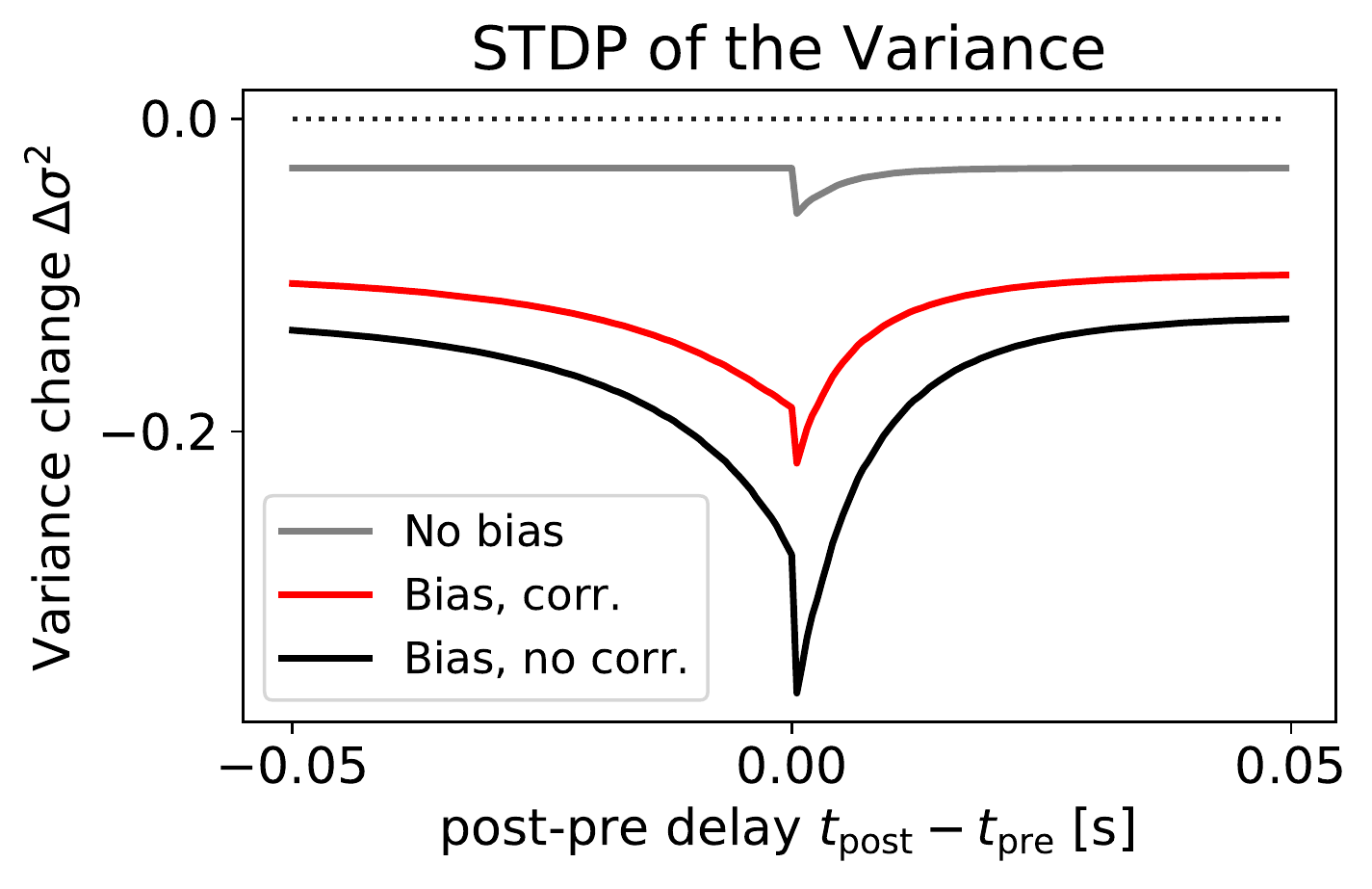}
\end{minipage}
\end{tabular}
\end{minipage}
\caption{ \label{fig:STDP-prediction}
% main message
The Synaptic Filter exhibits STDP of the mean and variance. \textbf{(A)} The change of the mean of the filtering distribution as a function of the temporal difference between a pre-post spike pair. For a single weight (excluding the bias, gray line), the Synaptic Filter produces only the LTP lobe ($t_{\rm{pre}} < t_{\rm{post}}$)
while the LTD lobe ($t_{\rm{post}} < t_{\rm{pre}}$) is independent of spike-timing. Inclusion of the bias, either with off-diagonal covariance elements (red line) or without (black line) yields the LTD lobe; and the magnitude of LTP decreases. \textbf{(B)} For the same protocol and learning rules, the variance $\sigma^2$ of the weight exhibits a spike-timing dependent decreases. When the bias (black and red lines) is included, the the change in variance resembles a symmetrised LTD lobe, i.e., it scales as the inverse of $|t_{\rm{pre}} - t_{\rm{post}}|$. Without the bias, the amplitude of change is reduced and the dependence on spike-timing disappears for $t_{\rm{post}} < t_{\rm{pre}}$. See \Cref{met:sec:b123} in Materials and Methods for simulations details.
%The initial values for the means and diagonal variance elements where set to 1. To ensure comparability, $\beta \equiv 1$ for all three settings.
}
\end{figure}

%
% mean, bias and negative lobe
Studying the same three conditions as in the previous section, we found that the variance decreases for all conditions and all pre-post timings, as shown in \Cref{fig:STDP-prediction} \textbf{(B)}. In a Bayesian framework this is expected because inputs spikes, which represent informative data, decrease uncertainty.
Interestingly, the reduction depends on the spike-timing.
For a single synapse without bias (gray line), the effect is weak and confined to the causal pairings, i.e., $t_{\rm{pre}} < t_{\rm{post}}$. 
Including the bias (black and red lines) increases the amplitude of the variance change. Moreover, it adds a qualitatively new feature: a negative lobe in the regime $t_{\rm{pre}} < t_{\rm{post}}$.
The underlying mechanism is the same as in the case of the LTD lobe of the mean (\Cref{fig:STDP-prediction} \textbf{(A)}). Changes in the synaptic weight and in the bias modulate the expected firing rate $\gamma_t$. %The single synapse model (gray line) is only aware of the modulation caused by its own increase in weight in response to causal spike pairs. The 2-dimensional synapse models (red and black lines) include the posterior distribution over the bias in the computation of expected firing rate. 
In the models with bias, a postsynaptic spike increases the expected firing rate temporally and, hence, the potential for variance reduction when a presynaptic spiking occurs close in time. %When a presynaptic spike occurs shortly before or after a postsynaptic spike, the reduction in variance is stronger due to the increased expected firing rate.
When both spikes coincide $t_{\rm pre} = t_{\rm post}$ the reduction is maximal because the temporally increased bias and the presynaptic activation increase the expected firing rate superlinearly.
%
%% Conclusion and remarks
Thus, with the sampling hypothesis, the Synaptic Filter makes the novel prediction of spikes-timing dependent changes of the EPSP variance. %Experimental evidence in support of this prediction would not only support the Synaptic Filter as model for synaptic plasticity but also the sampling hypothesis. However, if the plasticity of the EPSP variance contradicts the prediction, the Synaptic Filter would remain a valid proposal for plasticity as long as other plausible implementations of the posterior variance exist.
%
%% Figure

\subsection{The Synaptic Filter explains heterosynaptic plasticity}
\label{res:sec:b3}
%
%% motivation for heterosynaptic plastcitity
From a hebbian perspective on plasticity, it is required that the presynaptic neuron's activation takes part in postsynaptic neuron's activation. Heterosynaptic plasticity contradicts a purely hebbian view on learning because plasticity occurs without presynaptic activation \citep{abraham1983asymmetric,royer2003conservation}. 
For example, LTP induction at hippocampal synapses leads to LTD at synapses that did not receive stimulation \citep{lynch1977heterosynaptic}. %Thus, heterosynaptic plasticity defies the central idea of Hebb's postulate, i.e., that pre- and postsynaptic activity jointly gate plasticity. In fact, 
It has been argued that the role of heterosynaptic plasticity is complementary to homosynaptic, hebbian plasticity, which can destabilize neuronal dynamics through run-away weights
\citep{bailey2000heterosynaptic,chistiakova2015homeostatic}.

%
%% intuition for explaining away “theory"
We wondered whether the Synaptic Filter is consistent with heterosynaptic plasticity. Our starting point was that heterosynaptic plasticity could be linked to the explaining-away effect in Bayesian reasoning. Explaining-away occurs when one infers the posterior over multiple causes for an observation. When additional observations provide evidence for only one of the causes, the competing causes are ``explained away". For example, hearing a triggered alarm (an observation) is best explained by a burglar. However, upon learning that an earthquake occurred when the alarm was set off, the posterior probability for the burglar becomes much lower.

\begin{figure}[h!]
\begin{minipage}{\myFigureWidth\linewidth}
%\begin{minipage}{\myFigureWidth\linewidth} % shrink
\begin{tabular}{ll}
{\bf (A)} & {\bf (B)} \\
\begin{minipage}{0.5\textwidth}
\includegraphics[width=\textwidth,trim={0cm 7cm 0cm 3cm},clip]{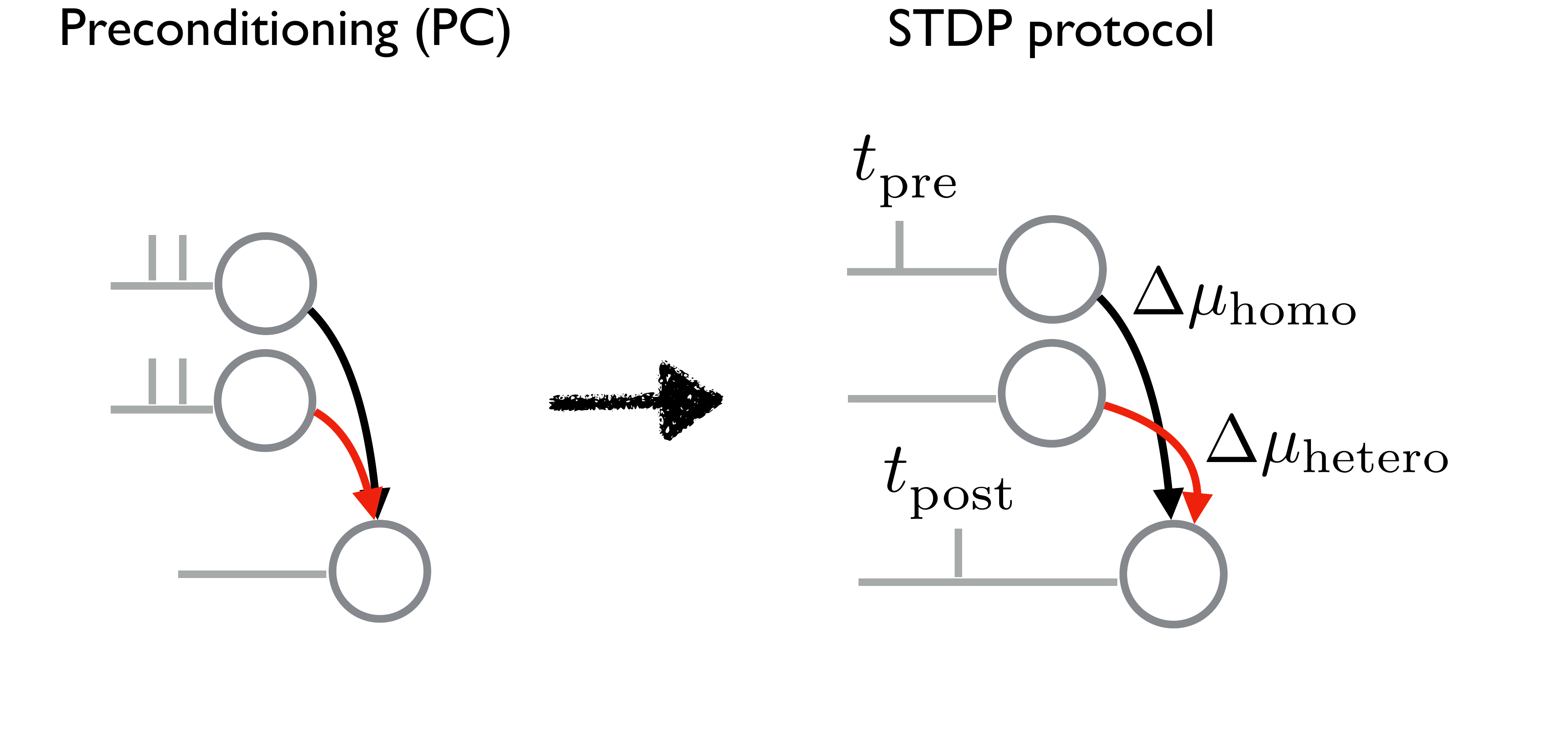}
\end{minipage} &
\begin{minipage}{0.3\textwidth}
\includegraphics[width=\textwidth]{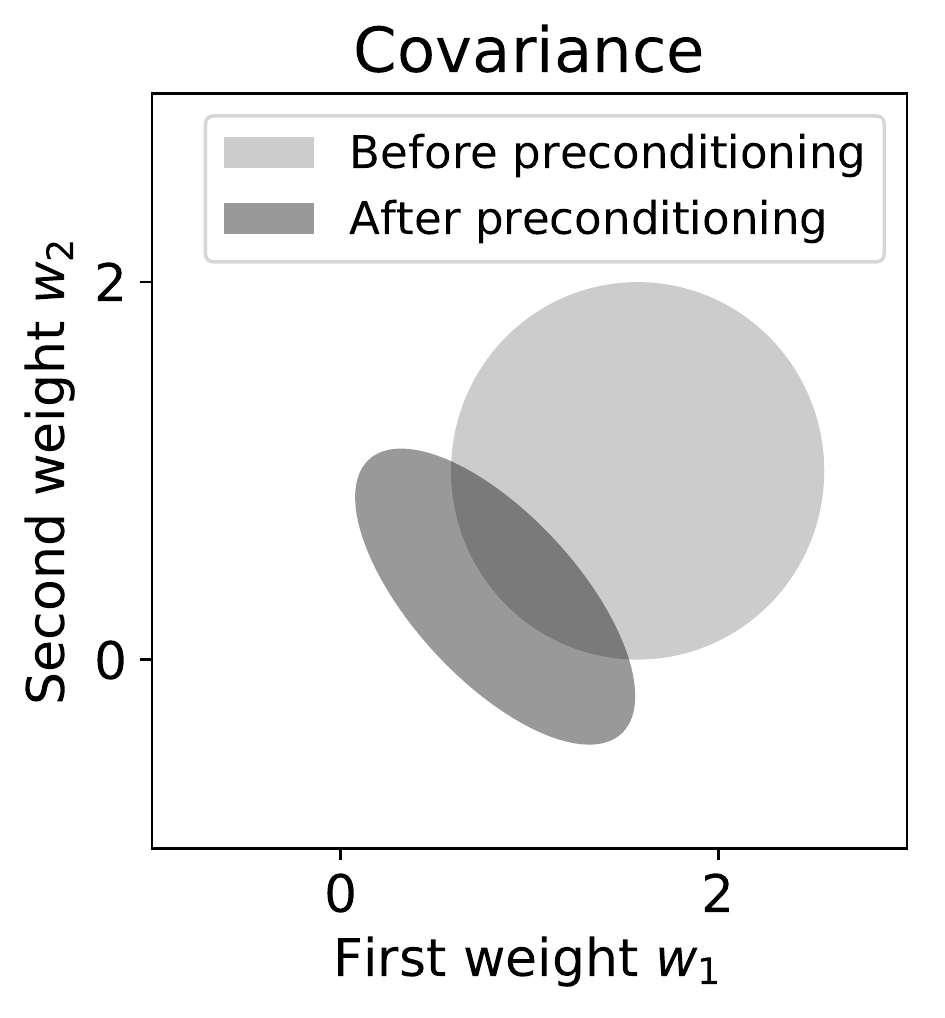}
\end{minipage}\\
{\bf (C)} & {\bf (D)} \\
\begin{minipage}{0.45\textwidth}
\includegraphics[width=\textwidth,trim={0cm 0cm 0cm 0cm},clip]{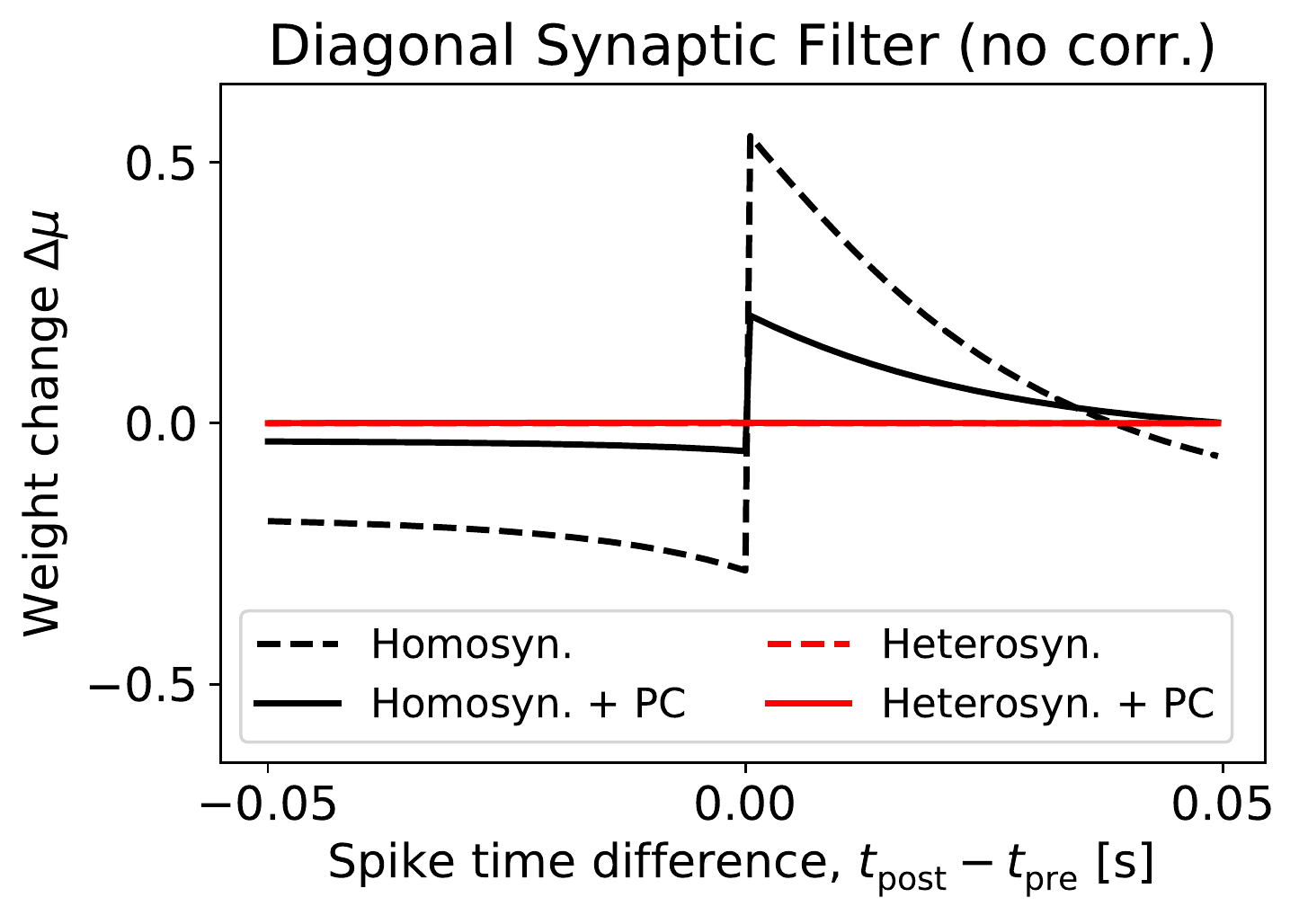}
\end{minipage} &
\begin{minipage}{0.45\textwidth}
\includegraphics[width=\textwidth]{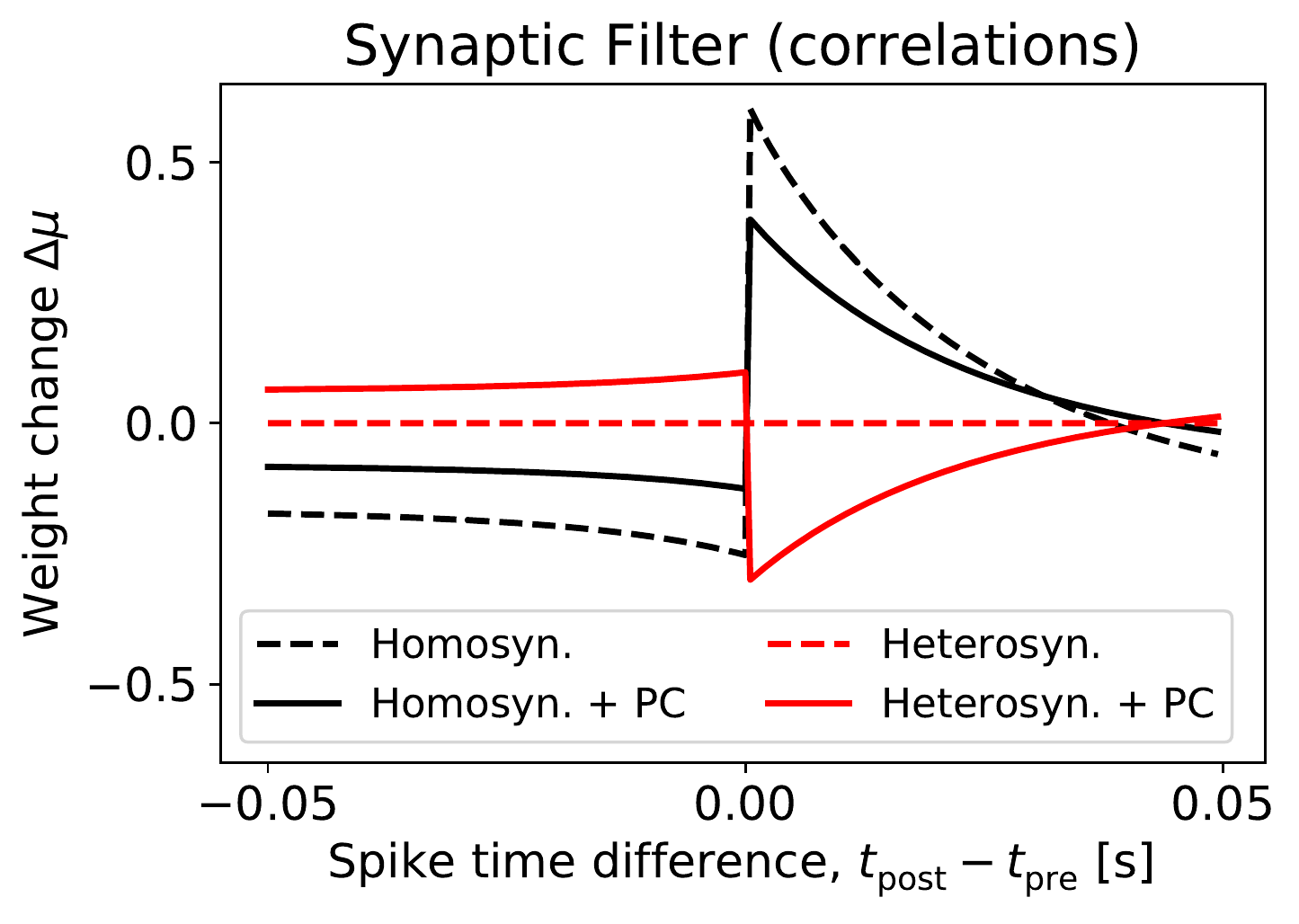}
\end{minipage}\\
{\bf (E)} & {\bf (F)} \\
\begin{minipage}{0.45\textwidth}
\includegraphics[width=\textwidth,trim={0cm 0cm 0cm 0cm},clip]{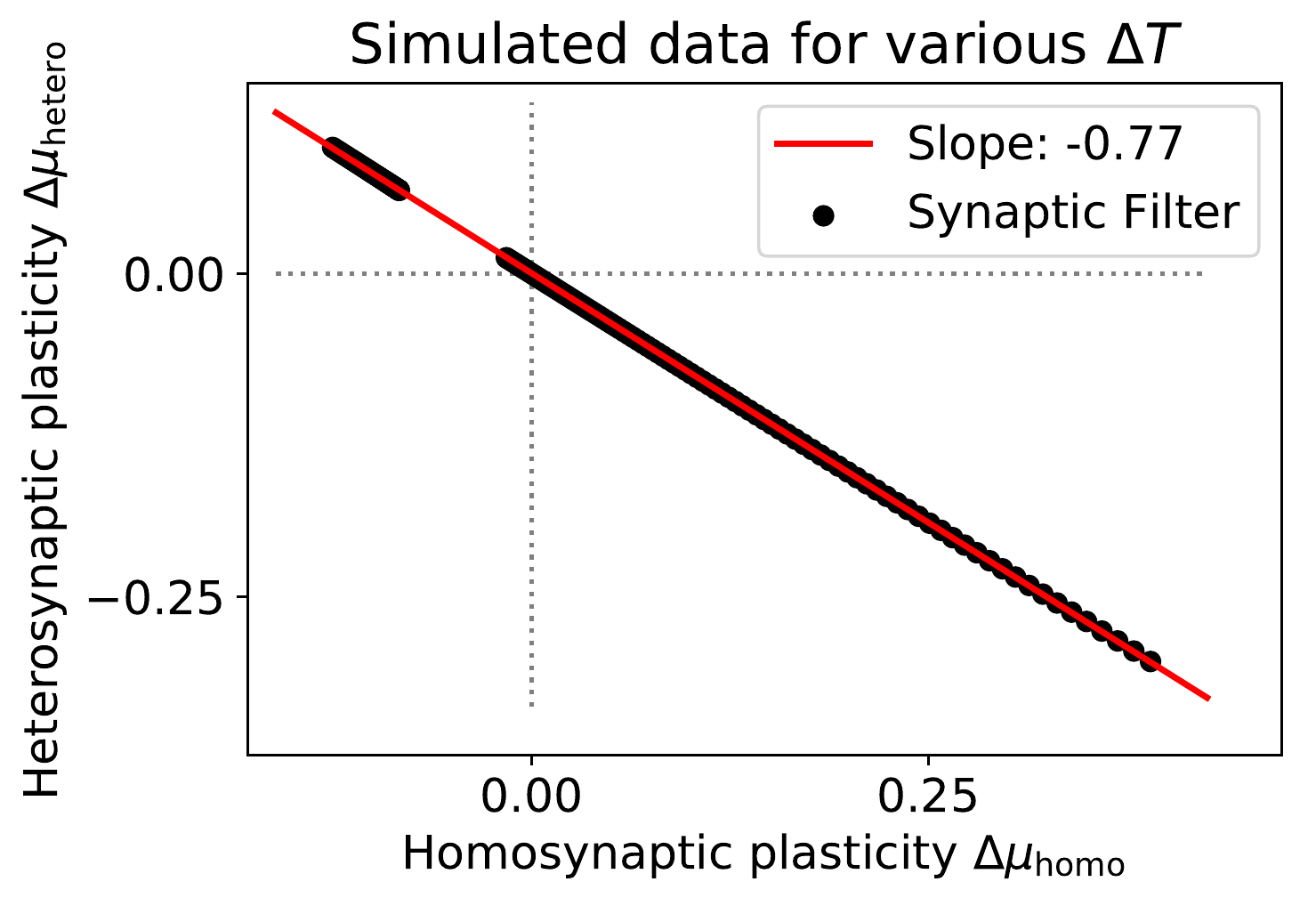}
\end{minipage} &
\begin{minipage}{0.45\textwidth}
\includegraphics[width=\textwidth]{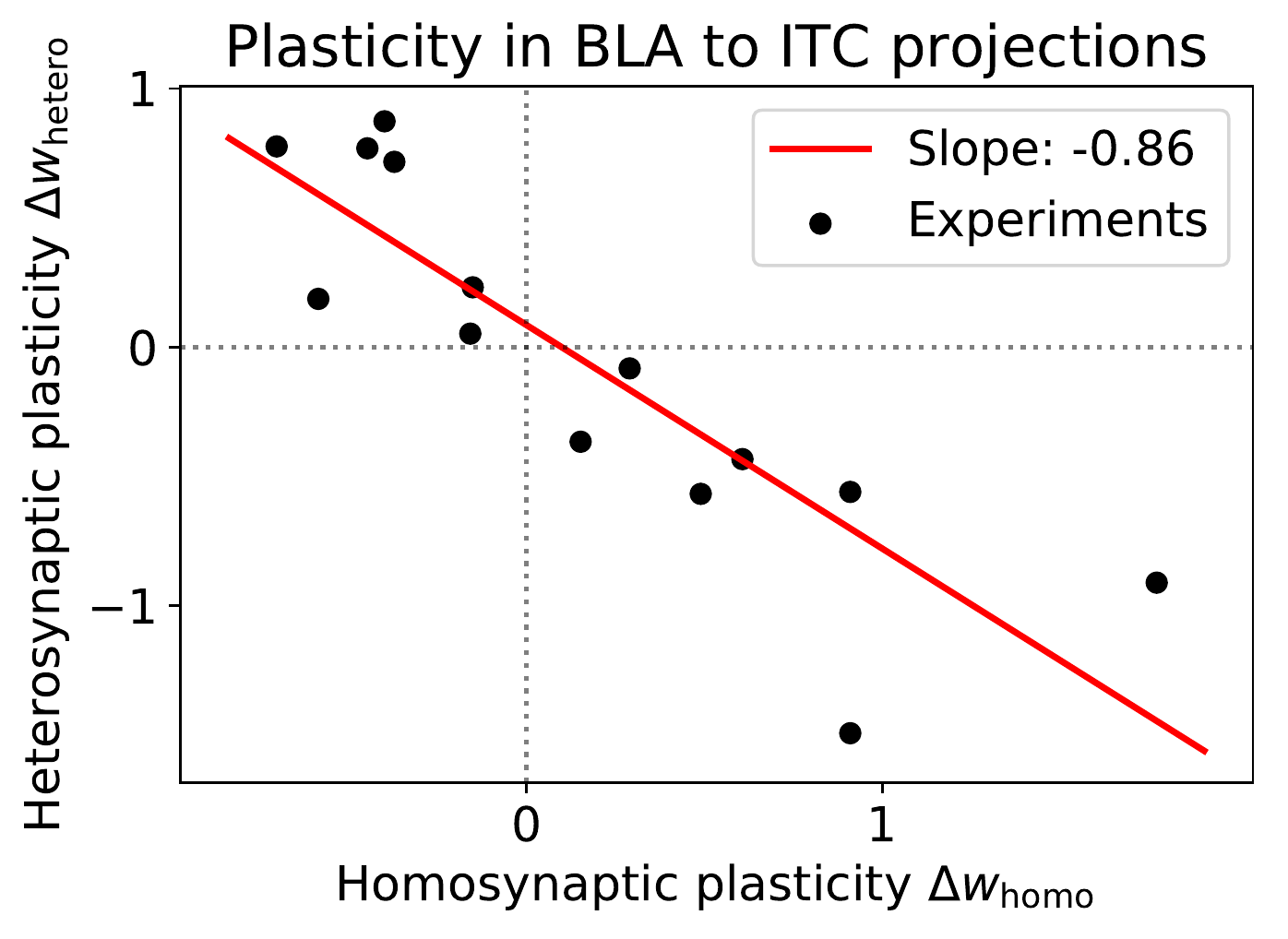}
\end{minipage}
\end{tabular}
\end{minipage}
%\end{minipage} %  shrink
\caption{ \label{fig:hetero-prediction}
The Synaptic Filter explains experimentally observed anticorrelation of homo- and heterosynaptic plasticity.
% protocol
\textbf{(A)} Two inputs drive a neuron. During (optional) preconditioning (PC), two synchronous input spikes are delivered. Changes in the weights in the first and second weight in response to a STDP protocol are reported as homosynaptic (black) and heterosynaptic (red) plasticity respectively. %Transients decay via a delay. 
See \Cref{met:sec:b123} in Materials and Methods for more details.
% Cartoon
\textbf{(B)} The effect of PC on the equilibrium weight distribution, visualised by contours, of the Synaptic Filter (light gray). After PC, the weight distribution (dark gray) has lower mean and weights become anticorrelated.
\textbf{(C)} The Diagonal Synaptic Filter exhibits homosynaptic STDP (black lines) but no heterosynaptic plasticity (red lines). PC reduces plasticity (solid black line).
% SF: yes hetero synaptic plasticity
\textbf{(D)} The Synaptic Filter exhibits homo- and heterosynaptic plasticity (solid black and red lines) after PC. Without PC (dashed lines), the Synaptic Filter behaves like its diagonal counterpart shown in \textbf{C}.
% anti correlation between LTP and LTD
\textbf{(E)} The Synaptic Filter predicts that homo- and heterosynaptic plasticity are anticorrelated. The $x$- and $y$-locations of the black points correspond to the solid black and red lines in \textbf{(C)}.
% data
\textbf{(F)} Anticorrelated homo- and heterosynaptic plasticity was found in synaptic projections from BLA to ITC neurons, Figure redrawn from \citep{royer2003conservation}.
}
\end{figure}

%
%% protocols and methods
In the spirit of explaining-away, we designed a preconditioning protocol to set up two synapses as competing causes for observations, i.e., the postsynaptic activity. The preconditioning protocol consists of synchronous spikes trains at both inputs without postsynaptic spiking. 
In a second step, a STDP protocol was applied to the first synapse but plasticity was reported from both. %used to study whether additional evidence in the form of presynaptic activity at one of the two synapses would, consistently with heterosynaptic plasticity, affect the synapse that did not directly receive further stimulation.
The weight change at the first and second synapses is our prediction for homo- and heterosynaptic plasticity respectively.
Both protocols are shown in \Cref{fig:hetero-prediction} \textbf{(A)}.
% two step protocol
%We were particularly interested in how the preconditioning protocol inspired by the explaining-away effect would influence heterosynaptic plasticity. Therefore, 
We obtained the homo- and heterosynaptic STDP curves with and without preconditioning. The effect of preconditioning is illustrated in \Cref{fig:hetero-prediction} \textbf{(B)}: the equilibrium weight distribution, which is the initial condition of the STDP-step, becomes negatively correlated.
% system to study
We simulate a 3-dimensional Synaptic Filter with two synapses and bias. To test whether correlations between weights are important for heterosynaptic plasticity, we also include the Diagonal Synaptic Filter. The same time constants and initial conditions are assumed as in the STDP experiment (\Cref{met:sec:b123} Materials and Methods).

%
%% experimental results for hetero synaptic
% homo
The homosynaptic STDP curve (black lines) appear robustly in all experiments, as shown in \Cref{fig:hetero-prediction} \textbf{(C, D)}, i.e. with and without preconditioning (dashed vs solid lines) and in both models, the Synaptic Filter and Diagonal Synaptic Filter. Preconditioning lowers the overall amplitude of the STDP curve, which was to be expected because presynaptic activity reduces the variance which acts as learning rate.
% hetero
Only a single experiment yields the heterosynaptic STDP curve: the Synaptic Filter with preconditioning (solid red line), shown in \Cref{fig:hetero-prediction} \textbf{(D)}. The heterosynaptic curve has the same shape as the homosynaptic STDP curve but with opposite sign. In contrast, the Diagonal Synaptic Filter exhibits no heterosynaptic STDP, i.e., the red curves in \Cref{fig:hetero-prediction} \textbf{(C)} are flat; and the Synaptic Filter without preconditioning has a flat heterosynaptic STDP curve (dashed red line, \textbf{(D)}) as well.
% explain mechanism
These results make clear that the mechanism of heterosynaptic plasticity in Synaptic Filter requires weight correlations. The Diagonal Synaptic Filter cannot represent weight correlations, which is why it never exhibits heterosynaptic plasticity. The Synaptic Filter, in contrast, can represent weight correlations but they have to be induced by the preconditioning protocol (because we made the idealised assumption that the equilibrium distribution has strictly uncorrelated weights). %A preconditioning protocol is required to induce non-zero correlations. 
The correlations are encoded in the covariance matrix $\Sigma_t$ as off-diagonal elements. The covariance matrix acts as inverse metric of the parameter space. It scales the update of the mean weight via $\Sigma_t x^\epsilon_t$ \citep{surace2020choice}. Thus, activity at one of the weights can lead to plasticity at correlated weights. Therefore the Synaptic Filter exhibits heterosynaptic plasticity only in combination with the preconditioning protocol.

%
%% anti correlations
The observation that homo- and heterosynaptic STDP curves have opposite signs is explained by the negativity of the off-diagonal entries in the covariance matrix. 
%Negative correlations makes sense from a mathematical and an intuitive perspective.
From a mathematical perspective, the sign of the updates of the off-diagonal elements are negative when correlations are present. This follows from using non-negative inputs (see \Cref{met:sec:negCorr} Material and Methods). Intuitively, the negative correlation between two weights (with same-sign inputs) encodes how much they can explain-away each other.

% 
%% Link to explaining away and negative correlation and data
Next, we wondered whether the negative correlation between homo- and heterosynaptic plasticity was consistent with experimental data. We quantified their relation by first plotting the values of the homo- and heterosynaptic STDP curves against each other and subsequently fitting a line, shown in \Cref{fig:hetero-prediction} \textbf{(E)}. The negative slope means that homosynaptic LTP is correlated with heterosynaptic LTD. The amplitude of heterosynaptic plasticity is around three quarters of the amplitude of homosynaptic plasticity. While the value of the slope depends on the number of spikes in the preconditioning protocol and other model parameters, the negativity of the slope is a robust feature of the Synaptic Filter caused by negativity of the weight correlations.
% data
A similar relation between homo- and heterosynaptic plasticity was reported in projections from basoateral amygdala (BLA) to intercalated cells (ITC) of the amygdala \citep{royer2003conservation}. The authors used extracellular low- and high-frequency stimulation to induce LTD and LTP respectively. They associated the induced weight change in the stimulated connection with homosynaptic plasticity and weight changes at other connections with heterosynaptic plasticity. Their main result aggregates plasticity results from multiple recorded ITC cells in a single figure, replotted in \Cref{fig:hetero-prediction} \textbf{(F)}. The data confirms a robust negative correlation between homo- and heterosynaptic plasticity, consistent with the prediction of the Synaptic Filter. Points in \Cref{fig:hetero-prediction} \textbf{(E, F)} are comparable in the sense that they originate from the variability in the induction mechanism. In the case of the Synaptic Filter, only one parameter, the spike-timing differed between points. In a biological plasticity experiment, the number of sources of variability is much higher, including somatic properties, initial conditions of synapses, speed of signal transmission and effectiveness of extracellular stimulation. On the premise that the aggregate outcome of this variability modulates the effectivity of biological plasticity in a similar way as spike-timing in the stimulated experiment, \Cref{fig:hetero-prediction} \textbf{(E, F)} provides evidence for the Synaptic Filter. Moreover, the Synaptic Filter explains the inverse correlation between homo- and heterosynaptic plasticity in terms of the explaining-away effect. The inverse correlation has also been found in Hippocampus \citep{lynch1977heterosynaptic} and more generally, see \citep{chistiakova2009heterosynaptic} for a review.
%Heterosynaptic depression: a  postsynaptic correlate of long-term potentiation \citep{lynch1977heterosynaptic}. Commissoral and Schaffner fibres are stimulated with 15s LTP protocol (tetanus stimulus). Plasticity in both is measured through population spike. Table is procuded (67\% LTD and 380 LTP \% 10min after stimulation). 

\section{Discussion}

%\red{Summary \\}
In this study, we showcased the framework of learning as filtering through the Synaptic Filter, an Assumed Density Filter for the weights in a spiking network. The main advantage of learning as filtering is that it accounts for the dynamics of the environment and weight uncertainty in a mathematically principled way. 
% C1,2
In a dynamic learning task, the Synaptic Filter outperforms a gradient rule with optimised learning rate in weight space. In combination with Bayesian regression, the Synaptic Filter also has better prediction performance. The representation of weight correlations proved particularly important to prevent overfitting in the presence of a model mismatch.
% B1-3
The relevance of the Synaptic Filter to biological plasticity is threefold. It exhibits STDP of the mean weight, including the negative lobe; it predicts STDP of the weight variance; and based on weight correlations, it predicts heterosynaptic LTD and homosynaptic LTP consistently with experimental evidence. Thus, the Synaptic Filter combines computational benefits with biological insights into plasticity.

%\red{Framework extensions \\}
% Framework
The framework learning as filtering can be used to derive additional learning rules. Here, we considered the simple case of a Gaussian weight distribution, an OU-process as prior and an exponential gain function in combination with spike-based observations. Alternative choices yield new learning rules. Parametrising the weight distribution through the binomial model of stochastic release represents an exciting possibility to study dynamics of EPSP variability and quantal parameter plasticity. The case of log-normally distributed weights has been studied in discrete time \citep{aitchison2015synaptic}. From a biological perspective, log-normal or binomial models have the advantage of obeying Dale's law. %, i.e., biological cannot change the sign of their weight. 
However, this advantage comes at the cost of intractability, the requirement of additional approximation or more complicated update-rules. To avoid these complications, we chose Gaussian synapses in this work.
Another option is to study gated plasticity through a hierarchical weight distribution in which an additional hidden variable infers whether a synapse should be plastic or not. Moreover, the framework can encompass different types of observations, for example continuous rates instead of spikes. %Instead of spikes, one can consider continuous observations or increase their dimensionality. % either rm or expand
Closely related to the observed variable is the choice of gain function. While the exponential offers simplicity, the sigmoidal and soft-max yield analytically tractable learning rules under additional approximations. Thus, learning as filtering offers a rich set of options to study learning and synaptic learning in particular.

%
%%
%\red{Extended to RNN, when all neurons are visible \\}
The generalisation of the single neuron analysis to the case of a recurrent neuronal network is straightforward as long as the output neurons' activities are conditionally independent of each other. In a recurrent neuronal network with only visible neurons, this condition is satisfied because past spiking affects the current output spikes only through the presynaptic kernels in the membrane potential. Formally, the joint probability distribution of output spikes conditioned on past spiking must factorise. Thus, the Synaptic Filter can be used to gain insights into learning dynamic weight distributions not only in single neuron but also in the more complex setting of a recurrent neuronal network model.

%\red{Implementation of testing g vs training y \\}
For the derivation of the learning rules, we make the assumption that the output neuron receives an external and neuron-specific supervision signal. Recent studies have addressed the question of how such a signal could be computed in biological networks \citep{sacramento2018dendritic} and in continuous time \citep{scellier2017equilibrium}. In our model, the supervision signal takes the form of spikes of the output neuron. This assumption does not exclude the possibility that biological neurons receive one type of spike as supervision signal to guide plasticity, and generate another type of spike to make predictions \citep{urbanczik2014learning,schiess2016somato}. Experimental findings in cerebellum and cortex are compatible with this idea. Indeed, complex and simple spikes in Purkinjee cells, and bursts and normal firing in cortical pyramidal neurons play distinct roles for plasticity \citep{yang2014purkinje,jacob2012regular}. Therefore, the assumption of a continuously provided, event based supervision signal does not impair the biological relevance of the Synaptic Filter.

%\subsection{Weight correlations: further predictions and implementation}
The Synaptic Filter represents correlations between weights. From a biological perspective, this suggest two directions for future research. First, how can correlations between synaptic weights be implemented by biological synapses? On the premise, that EPSP samples represent weight uncertainty, non-linear summation at dendrites is a potential candidate. However, non-linear dendritic summation would be confined to spatially close synapses, e.g., ones located on the same dendritic branch. Thus, this mechanism cannot account for the full covariance matrix which scales as $\mathcal{O}(d^2)$ with the number of synapses $d$. One possibility to address this is to approximate the covariance matrix with a limited number off-diagonals. Another approximation aggregates the effect of spatially distant synapses on the membrane potential in a single aggregation variable, e.g., the bias. %In particular, a more sophisticated version of the Sampling Synaptic Filter should be developed to predict dendrite summation properties explicitly.
Secondly, the Synaptic Filter was derived under the assumption of positive presynaptic inputs while the sign of the weights could either be positive or negative. As a consequence, weight correlations are never positive. Based on the negative weight correlations the Synaptic Filter could explain the negative correlations between homo- and heterosynaptic plasticity (shown in \Cref{fig:hetero-prediction}). As an extension of this result, it would be interesting to generalise the Synaptic Filter to the case of positive and negative inputs, representing excitatory or inhibitory neurons. We hypothesise that a generalised Synaptic Filter would make the prediction of positively correlated homo- and heterosynaptic plasticity between synapses from inhibitory and excitatory neurons.

%
%% Metaplasticity
Because uncertainty controls the speed of learning the Synaptic Filter can in combination with the sampling-hypothesis link synaptic variability to synapse specific metaplasticity, which has been observed experimentally \citep{chistiakova2009heterosynaptic}. The Synaptic Filter predicts that synaptic variability and learning speed reduce upon presynaptic stimulation but relax back to maximal value on the time scale of the OU-prior. Indeed, consistent with an OU time scale of hours, plasticity experiments have shown that LTP saturates temporally but recovers within hours \citep{abraham1996metaplasticity}. 
% TODO: age vs plasticity
%Moreover, there is evidence that a release probability of 0.5, which is necessary condition for maximal synaptic variability, is an attractor for synapses in layer 2/3 of rat neocortex \citep{hardingham2007presynaptic}. Interestingly, postsynaptic stimulation drives release probabilities at presynaptic synapses towards this value \citep{volgushev2000retrograde}. The Synaptic Filter can mechanistically explain this observation as potentiation of the bias in the presence of negative correlations between bias and weights, which results in heterosynaptic plasticity of the variances.

%
%% know thy neighbour
% recap
The presented work is closely related to the Know-Thy-Neighbour (KTN) theory  \citep{pfister2009know}. It assumes that synapses solve the problem of estimating the presynaptic membrane potential from the arriving spike train within the filtering framework. Consequently, it links the dynamics of short-term synaptic depression to the mean and variance updates of the filtering distribution.
% Similarities
The Synaptic Filter and the KTN theory formalise plasticity via Assumed Density Filtering with an Ornstein-Uhlenbeck prior. At any point in time, the synapses encode a posterior distribution over the hidden variable given past observations, i.e., membrane potential and presynaptic spikes in the KTN model and ground truth weight with pre- and postsynaptic spikes in our model. In both models the classically defined synaptic efficacies, the averaged postsynaptic responses, corresponds to the mean of the posterior; and the variance plays the role of learning rate in the update equations. 
% Differences
Our novel contributions are the extension of the KTN framework to multiple and potentially correlated hidden variables and a more complex emission process, i.e., the emissions are generated by the sum of the hidden variables weighted by the presynaptic trace. The KTN model is formally equivalent to the one-dimensional Synaptic Filter when the bias is the only hidden variable. On the level of the biological interpretation, KTN focuses on short-term plasticity while our work makes a connection to experiments in the context of long-term plasticity.
%%

%\red{Instant feedback, SSF, integrals and spike sampling \\}
A previous study has addressed the computational role of synaptic uncertainty \citep{kappel2015synaptic}. The authors propose that spine turnover implements samples from the posterior distribution over synaptic weights via Langevin sampling. Their work differs from ours because they consider a static inference task (bottom right in \Cref{fig:generative-model} \textbf{(A)}), not filtering. One consequence of the static nature of their task is that the online version of their learning rule includes a fixed data set size as external parameter. Compared to previous learning rules in the context of filtering \citep{aitchison2015synaptic}, we make four additional contributions. First, we connect the learning task to the rich literature of filtering. In particular, this facilitates a simple, rigorous and continuous-time treatment. Secondly, we go beyond the assumption of a diagonal Gaussian Assumed Density Filtering by including weight correlations; and show that their importance for filtering performance. Thirdly, we show that the filtering distribution can be used in combination with Bayesian regression to improve predictive performance, a more relevant performance measure for learning than the MSE, which is computed in weight space. Finally, based on the spiking, continuous-time analysis, the Synaptic Filter recovers the phenomenon of spike-timing dependent plasticity, i.e., the mean synaptic increases if the postsynaptic spike follows the presynaptic spikes closely, and decreases if the spike order is reversed. Moreover, the Synaptic Filter predicts spike-timing dependent changes of the EPSP variability. Finally, it explains the negative correlation between homo- and heterosynaptic plasticity in terms of the Bayesian explaining-away effect.

%\red{Conclusion \\}
Overall this article provides evidence that learning as filtering is a serious candidate for a computational principle underlying plasticity and provides testable predictions.

\section{Materials and Methods}
\subsection{The generative model and learning rules}
In this section, we define the filtering problem in terms of a generative model for plasticity. %The generative model describes an output neuron that receives input from several neurons. The goal of filtering is to compute the distribution over the synaptic weights that connect the inputs to the output neuron. 
%The goal of filtering corresponds to learning the distribution over the weights. 
In addition, the update equations of the learning models are introduced, i.e., the Synaptic Filter, the Diagonal Synaptic Filter %the particle filter 
and the gradient learning rule.

\subsubsection{The generative model}
%
% motivation
Learning as static optimisation has the goal of finding a parameter value that minimizes a cost function of given training data. Here, we consider a different framework for learning: learning as filtering. The goal of filtering is to continuously compute the probability distribution over dynamic, hidden variables based on continuously emitted observations. In contrast to optimisation, filtering includes time and parameter uncertainty in a principled manner on the level of the task.

%
%% GM needed
A generative model specifies an observation process that relates the hidden parameters to observations and a transition probability that represents assumptions about the dynamics of the hidden parameters. To apply the framework of learning as filtering to synaptic plasticity, we consider the following generative model.

The fundamental assumption is that $d$ time dependent parameters, the synaptic weights $w_t \in \mathbb{R}^{d}$, govern how input spikes $x_t$ give rise to output spikes $y_t$ via the observation process $p(y_t | x_{0:t}, w_t)$ (explained in more detail below). The mapping between inputs and outputs is not static but changes due to the dynamics of $w_t$. For the transition probability, we assume that the hidden weights evolve according to an Ornstein–Uhlenbeck (OU) process with time scale $\tau_{\rm{ou}}$, mean $\mu_{\rm{ou}}=0$ and diagonal covariance matrix $\Sigma_{\rm{ou}} = \sigma^2_{\rm ou} \mathbb{1}$ (with non-zero elements) $\sigma_{\rm{ou}}^2=1$:
\begin{align}
\label{met:eq:dw}
\text{d}w_t = \tau_{\rm ou}^{-1} (\mu_{\rm ou}-w_t) \text{d}t + \sqrt{2 \sigma_{\rm ou}^2 \tau_{\rm ou}^{-1}} \mathbb{1} \text{d}V_t,
\end{align}
where $V_t$ is a $d$-dimensional Wiener process. The process in \Cref{met:eq:dw} can be represented as Gaussian transition probability: $p(w_t | w_{t - \text{d}t}) = \mathcal{N}(w_{t - \text{d}t} + \tau_{\rm{ou}}^{-1} (\mu_{\rm{ou}}-w_{t - \text{d}t}) \text{d}t, 2 \sigma^2_{\rm{ou}}\tau_{\rm{ou}}^{-1} \text{d}t)$.

%
%% likelhood / observations
For the observations, we assume a stochastic leaky-integrate-and-fire neuron, also called Spike-Response Model \citep{gerstner2002spiking}. The output spikes $y_t$ are generated stochastically from a membrane potential $u_t$ via an inhomogeneous Poisson point process. Thus, the output spikes represent a sum of delta distributions with spike times $\{ t^f_j \}$: $y_t = \sum_{f} \delta(t - t^f)$.
To connect the membrane potential to the firing rate of the Poisson process, we choose an exponential gain function:
\begin{align}
    y_t  &\sim \text{PoissonProcess}(g(u_t)) \\
    \label{met:eq:gut}
    g(u_t) &= g_0 \exp(\beta u_t).
\end{align}
The determinism parameter $\beta$ controls how strongly changes in the membrane affect the firing rate, and $g_0$ is the baseline firing rate. An exponential gain function represents a neuron close to the onset of firing but excludes saturation effects of the firing rate at high values of the membrane potential. The exponential gain function has been established as a phenomenological model of neocortical pyramidal neurons \citep{jolivet2006predicting}.
% membrane
The membrane potential is a leaky integrator, with time constant $\tau_{\rm m} = 25$, of the weighted sum of input spikes $x_t$. The leaky integration is represented by an exponential kernel $\epsilon_t = e^{-t/\tau_{\rm m}}\Theta(t)$ where $\Theta(\cdot)$ represents the Heavyside function:
\begin{align}
    u_t &= (w^\top x * \epsilon)_t \approx w_t^\top (x * \epsilon)_t,
    \label{eq:u_t tau_m << tau_ou approximation},
\end{align}
% Markovian GM
The approximation in \Cref{eq:u_t tau_m << tau_ou approximation} is valid in the regime $\tau_{\rm{m}} \ll \tau_{\rm{ou}}$, i.e., that the membrane dynamics is much faster than the dynamics of the ground truth weights assumed by the generative model. This assumption simplifies the generative model by casting it as a Markov process. Otherwise, the current observation $y_t$ would dependent on the entire history of hidden weights through the low-pass filtered membrane potential. The current observation does, however, dependent on the history of input spikes via the convolutions $x^\epsilon_t := (x * \epsilon)_t$. This type of history dependence does not complicate the analysis because it can be straightforwardly taken into account in the spiking probability:
$p(y_t | x_{0:t}, w_t)$ can be replaced by $p(y_t | x_t^\epsilon, w_t)$. Moreover, it has the biological interpretation of a presynaptic trace.

%
%% Bias
The notion of a bias can be in included in the generative model by setting one of the inputs to one permanently. A bias stabilises the performance simulations and yields interesting biological insights. Thus, we adopt the convention that weight $w_{t,0}$ represents the bias and the input with index $i=0$ is permanently set to $x^\epsilon_{t,0} = 1$. The generative model remains $d$-dimensional in terms of hidden variables but has only $d-1$ presynaptic inputs.

\subsubsection{The Synaptic Filter: update equations and prediction}
The goal of learning as filtering is to compute the distribution over the hidden weights $p(w_t | \mathcal{D}_t)$ given all previously observed input and output spikes $\mathcal{D}_{t} := \{x_t, y_t \}_{\tau=0}^t$.
On a formal level, the Markovian structure of the generative model, enables a recursive solution of the filtering problem:
\begin{align}
\label{met:eq:forward}
    p(w_t | \mathcal{D}_{t}) \propto p(y_{t} | x_t^\epsilon, w_{t})  \int  p(w_t | w_{t - \text{d}t}) p(w_{t-{\text{d}t}} | \mathcal{D}_{0:t-{\text{d}t}}) \text{d}w_{t-\text{d}t}.
\end{align}
% (see Supplementary Information)
The Kushner-Stratonovich Equation \citep{kushner1967dynamical} gives a formal solution for all moments of the filtering distribution. However, for most generative models the solution is intractable because of the closure problem, i.e., the evolution of lower moments depends on higher moments of the filtering distribution. 
One way to address the closure problem is Assumed Density Filtering. The central idea is to replace the exact filtering distribution with a proposal distribution $q$ parameterised by $\theta_t$:
\begin{align}
\label{met:eq:ADF}
    p(w_t | \mathcal{D}_{t}) \approx q_{\theta_t}(w_t).
\end{align}
When substituting the approximation \Cref{met:eq:ADF} into the right-hand side of \Cref{met:eq:forward}, the resulting posterior will generally not lay in the family of the proposal distribution $q_\theta$. Therefore, one has to decide how to best approximate the result with a member of the proposal family \citep{sugiyama2012density}.

Here, we derive the \textit{Synaptic Filter}, an approximate solution based on a Gaussian proposal density with mean $\mu_t \in \mathbb{R}^d$ and covariance matrix $\Sigma_t \in \mathbb{R}^{d \times d}$. The results for the \textit{Diagonal Synaptic Filter} are identical with exception that off-diagonal elements of $\Sigma_t$ are set to zero. The evolution of the distribution parameters $\theta_t = (\mu_t, \Sigma_t)$ can be computed from the Kushner-Stratonovich Equation. To remain in the Gaussian family, higher moments are omitted (see Supplementary Information). For the generative model specified above, the resulting update equations for $\mu_t$ and $\Sigma_t$ are \Cref{eq:dotmu,eq:dotsigma2}.

The expected firing rate $\gamma_t$ is a central quantity in the Synaptic Filter. It represents the predicted firing rate based on the filtering distribution:
\begin{align}
    \label{met:eq:gamma}
    \gamma_t := \int_{w} q_{\theta_t}(w_t) g(u_t) \text{d}w_t = g_0 \exp \left(  \beta \mu_{t}^\top x_{t}^{\epsilon} + \tfrac{1}{2}\beta^2 (x_t^{\epsilon})^\top \Sigma_{t} x_{t}^{\epsilon} \right).
\end{align}
The expected firing rate depends not only on the mean of the filtering distribution but also on the covariance. As expected from a convex gain function, the covariance increases the expected firing rate compared to a scenario where only the mean is taken into account.
The expected firing rate enters the computation of the error signal $y_t - \gamma_t$ in the update \Cref{eq:dotmu} of the mean and controls the reduction of the covariance in \Cref{eq:dotsigma2}. 
In addition, the expected firing rate appears naturally when combining the Synaptic Filter with Bayesian regression. Bayesian regression is a method for making predictions in the presence of parameter uncertainty. The central idea is to marginalise over the parameters, as in \Cref{res:eq:BR}. For the Synaptic Filter, \Cref{res:eq:BR} can be evaluated, by rewriting the probability of an output spike $p(y_t | x^\epsilon_t, w_t)$ as Bernoulli probability. The probability that a output spike occurs in the infinitesimal interval $\text{d}t$, indicated by $\text{d}N_t := y_t \text{d}t \in \{0,1\}$ is:
\begin{align}
\label{met:eq:SF-pdN}
    \langle p(\text{d}N_t = 1 | x^\epsilon_t, w_t) \rangle = \langle g(u_t) \text{d}t \rangle = \gamma_t \text{d}t,
\end{align}
where we used \Cref{met:eq:gamma} in the last step. From normalisation over both states of $\text{d}N_t$ and converting the Bernulli probability back to a point emission process, we obtain the posterior predictive distribution of the Synaptic Filter: 
\begin{align}
    \label{met:eq:ppd-gamma}
    p(y_t | x^\epsilon_t, \theta_t)  &= \text{PoissonProcess}(\gamma_t).
\end{align}

%
%% Gradient
\subsubsection{Gradient learning rule}
\label{met:sec:wML}
As a performance benchmark, we use the gradient learning rule. Assuming updates are proportional to the gradient of the log output probability yields:
\begin{align}
    \label{met:eq:dotwML}
    \dot{w}_{t}^{\rm{ML}} &= 
    \eta \beta x_t^{\epsilon}\beta (y_t - g_t^{\rm{ML}}),
\end{align}
where $\eta$ is the learning rate parameter and $g_t^{\rm{ML}} = g_0 \exp(\beta (w_{t}^{\rm{ML}})^\top x_t^\epsilon)$. We did not absorb $\beta$ in the learning rate $\eta$ to make the values of $\eta$ more comparable to values of the variance in the Bayesian learning rule and to use the same scaling of $\beta$ with the dimension as in the Synaptic Filters (see \Cref{met:sec:c012} in Materials and Methods).

%
%% negative correlations
\subsection{The weight correlations of the Synaptic Filter are mostly negative}
\label{met:sec:negCorr}
The weight correlations in the Synaptic Filter are represented by the off-diagonal elements of the covariance matrix $\Sigma_t$. 
In the 2-dimensional case and for positive inputs ($x^\epsilon_t \geq 0$) these elements $\Sigma_{t,i \neq j}$ are always negative. This follows from the following two observations. First, the change of weight correlations is negative when the initial weight distribution is diagonal, i.e., $\Sigma_{0} = \sigma_0^2 \mathbb{1}$. Secondly, a negatively correlated weight distribution cannot evolve into a positively correlated weight distribution without assuming a diagonal form in between.

The change of the covariance is given by \Cref{eq:dotsigma2}. Omitting the temporal index for clarity and assuming $i \neq j$, the update of an off-diagonal element is:
\begin{align}
    \dot{\Sigma}_{ij} = - \gamma (\Sigma x^\epsilon)_i (\Sigma x^\epsilon)_j - 2 \tau_{\rm ou}^{-1} \Sigma_{ij},
\end{align}
where we used that $(\Sigma_{\rm ou})_{ij} = 0$. For the initial condition given by a diagonal covariance matrix, this expression simplifies to:
\begin{align}
\label{met:eq:sigma_ij}
    \dot{\Sigma}_{ij} = - \gamma \Sigma_{ii} x^\epsilon_i \Sigma_{jj} x^\epsilon_j.
\end{align}
Since all factors in this expression are positive but the overall sign is negative, an initially diagonal weight distribution can only evolve towards a negatively correlated distribution. Because in 2 dimensions a transition from a negatively correlated to a positively correlated weight distribution is not possible without a diagonal state in between, positive correlations cannot occur. Conditions under which this result holds for $d>2$ are discussed in the Supplementary Information Section 6.

%
%% C0:2, Simulation details
\subsection{Computational performance: hyperparameters and simulation details}
\label{met:sec:c012}

% parameters like + transition
Here, we describe how the computational performance discussed in \Cref{res:sec:c1,res:sec:c2} in Results was evaluated. In \Cref{met:sec:u-scaling}, the scaling of the membrane potential with dimensions is introduced. The following sections describe how the predictive performance and the model mismatch were implemented and the technical details of the simulations.

%
%% membrane scaling
\subsubsection{Scaling of the membrane potential with dimensions}
\label{met:sec:u-scaling}
The performance of the learning rules is reported as a function of the dimension $d$ of the generative model. Varying $d$, influences the statistics of the firing rate $g(u_t)$ and, hence, the amount of information available for learning. %In biological synapses homeostatic mechanisms compensate for changes in the number of input synapses by regulating the neuronal gain and, thus, the firing rate \citep{turrigiano2004homeostatic}. 
In our model we scale the determinism $\beta$ with the number of inputs such that the expected firing rate $\langle g(u_t) \rangle$ becomes (approximately) independent of the number of inputs $d$. In this section, the expected value is taken with respect to the statistics of the generative model, not the filtering distribution. %For example, $\langle x_t \rangle$ denotes the expected firing rate vector of the input neurons.

To compute $\langle g(u_t) \rangle$, we first approximate the membrane potential statistics as Gaussian. With $\langle w_t \rangle = \mu_{\rm ou} = 0$, the mean and variance of the membrane are: 
\begin{align}
    \langle u_t \rangle &= \langle w^\top_t \rangle \langle x^\epsilon_t \rangle = 0 
    \label{met:eq:u-scale}
    \\ 
    \text{var}[u_t^2] &= 
    \sum_{i=0}^{d-1} \text{var}[ w_{t,i} x^\epsilon_{t,i} ]
    = 
    \sigma_{\rm ou}^2 ( 1 + (d - 1) \langle (x_{t,i}^\epsilon)^2 \rangle ) \approx d \frac{\sigma_{\rm ou}^2 \tau_{\rm m} \nu_0}{2}
    \label{met:eq:u2-scale}
\end{align}
where we used that for two independent random variables $a$, $b$ with $a$ being zero-mean: $\text{var}[ab] = \text{var}[a]\mathbb{E}[b^2]$ and dropped the special treatment of the bias in the final step. We also assumed that all presynaptic neurons fire at the same rate $\langle x_{t,i} \rangle = \nu_0$.
Approximating the membrane statistics by their first and second moment, given by \Cref{met:eq:u-scale,met:eq:u2-scale}, we used the Gaussian expectation of an exponential (see Supplementary Information) to obtain the expected firing rate in the generative model as a function of the dimension $d$:
\begin{align}
    \label{met:eq:g-scaling}
    \langle g(u_t) \rangle = g_0 \exp( \tfrac{1}{4}d \beta^2 {\sigma_{\rm ou}^2 \tau_{\rm m} \nu_0}  ).
\end{align}
To ensure a comparable observation rate across dimensions, we remove the dependence on $d$ in \Cref{met:eq:g-scaling} by making the determinism $\beta$ parameter a function of the dimension:
\begin{align}
    \beta \rightarrow \beta(d) = c \beta_0 d^{-1/2}.
\end{align}
The proportionality constant $c \beta_0$ is split into a factor of order $\beta_0 = \mathcal{O}(1)$ that is varied in the experiments and a constant $c$ that ensures that the neuron's firing rate only rarely exceeds $g_{\rm max} = 50$Hz. Specifically, the 5-sigma environment of the membrane statistics, $\sigma_u^{(5)} := \langle u_t \rangle \pm 5\sqrt{ \text{var}[u_t^2]}$ (given by \Cref{met:eq:u-scale,met:eq:u2-scale}), defines the condition for $c$:
\begin{align}
    g_{\rm max} \overset{!}{=} g_0 \exp \left( \beta(d) \sigma_u^{(5)}  \right) 
    \Rightarrow 
    c \equiv \frac{\log\left(\tfrac{g_{\rm max}}{g_0}\right)}{5\sqrt{\tfrac{1}{2} {\sigma_{\rm ou}^2 \tau_{\rm m} \nu_0}}}
\end{align}
In the simulations, we vary the proportionality constant $\beta_0$, which we refer to as determinism parameter for simplicity.

%
%% loglikelihood
\subsubsection{Measuring predictive performance via the model evidence }
In \Cref{res:sec:c2}, the predictive performance of a model $\mathcal{M}$, e.g., the Synaptic Filter, is measured in terms of the evidence $p(\mathcal{M}|\mathcal{D}_t)$. The log evidence is directly related to the predictive distribution evaluated on the data $\mathcal{D}_t$. Assuming a flat model prior, we have:
\begin{align}
    \label{met:eq:model-evidence}
    \log p(\mathcal{D}_t| \mathcal{M})
    & = \int^t_0 \left( \log \int p(y_\tau | x_\tau^\epsilon, w_\tau) p(w_\tau |\mathcal{D}, \mathcal{M}) \text{d}w_\tau \right) \text{d}\tau.
\end{align}
The model dependent parameter distribution $p(w_\tau | \mathcal{D}, \mathcal{M})$ corresponds to the filtering distribution for the Synaptic Filter and to a delta-distribution centered on the current estimator $w^{(\rm ML)}_\tau$ for the gradient rule (see \Cref{met:sec:wML}). Thus, the argument of the log in \Cref{met:eq:model-evidence} is the predictive distribution evaluated on the data. 
%and the Diagonal Synaptic Filter is measured via the loglikelihood $\mathcal{L}(\theta_{0:t})$, where $\theta_{0:t}$ represents the sequence of distribution parameters of the filters. The loglikelihood compares the observed output spike $y_t$ to the prediction of the filter obtained via the posterior predictive distribution, given by \Cref{met:eq:ppd-gamma,met:eq:gamma}, in response to an input $x_t^\epsilon$:
%\begin{align}
%    \mathcal{L}(\theta_{0:T}) := \int_{0}^T \log  %p(y_t|x_t^\epsilon,\theta_t) \text{d}t 
%    \label{met:eq:loglike}
%\end{align}
In practise, we use the same discretisation as in \Cref{met:eq:SF-pdN} to evaluate \Cref{met:eq:model-evidence} based on the size of the simulation time steps $\Delta t$. For instance, the log Bayes factor between the Synaptic Filter and a null-model $\mathcal{M}_0$ based on the baseline firing rate $g_0$ is:
\begin{align}
    \log \frac{p(\mathcal{D}_t | \mathcal{M}_{\rm SF}) }{p( \mathcal{D}_t | \mathcal{M}_0)}  \approx 
    \sum_{k=0}^{T/\Delta t} \log  \left( \tfrac{\gamma_{t_k}}{g_0} \right)^{\Delta N_{t_k}} 
    + \log \left(\tfrac{1 - \gamma_{t_k}  \Delta t}{ 1- g_0 \Delta t} \right)^{1-\Delta N_{t_k}},
    \label{met:eq:loglike-discrete}
\end{align}
where $t_k := k \Delta t $ and $\Delta N_{t_k} = 1$ if a spike occurred in the interval $[t_k, t_k + \Delta t]$ and zero otherwise.

% errors
%The results for predictive performance in \Cref{fig:comp-results} \textbf{(D, E)} represent the difference of the Bayes factor relative to the optimised gradient rule. The Bayes factor $f$ is given by \Cref{met:eq:loglike-discrete}. We compute $f$ and its SEM from 100 simulations $i \in (1,\dots 100)$. Because second term in \Cref{met:eq:loglike-discrete} is heavy-tailed under the dynamics of the model, the mean and variance estimators of $f$ are susceptible to outliers. To address this, we assume that $f$ approximately follows a log normal distribution: $\log f \sim \mathcal{N}(m, s^2)$. 
%This is a reasonable assumption because an exponential gain function with normally distributed argument follows a log normal distribution. With the standard estimators $\hat{m} = \langle \log f_i \rangle_i$ and $\hat{s}^2 = \langle \log^2 f_i \rangle_i - \hat{m}^2$, the Bayes factor's mean and variance (from which the SEM is computed) become: $\mu_f = \exp(\hat{m} + \hat{s}^2/2)$ and $\sigma^2_f = \mu_f^2 (\exp(\hat{s}^2) - 1)$. Differences in $\mu_f$ with respect to the models and the root mean square of the SEM's are shown in \Cref{fig:comp-results} \textbf{(D, E)}. Including the log normal assumption has the effect of smoothing the plots without affecting the performance results qualitatively.

%
%% loglikelihood of the gradient rule
\subsubsection{Predictive performance of the optimised gradient rule}
In \Cref{res:sec:c2}, we use the log predictive distribution to evaluate learning rules. Because the gradient rules does not include parameter uncertainty, this metric is equivalent to the loglikelihood. % of a gradient rule with optimal learning rate as benchmark. 
For the optimisation of the loglikelihood with respect to the learning rate, we obtained the loglikelihood performance of the gradient rule for 11 log-spaced values of the learning rate in an interval $\eta \in [0.05, 2]$, which contains the optimal learning rate, and interpolated with a 3rd order polynomial (see Supplementary Information). Based on the fit, we selected the maximal loglikelihood value. The same procedure was used to compute the SEM of the loglikelihood.

The intuition behind the fact that $[0.05, 2]$ contains the optimal learning rate is that the learning rate replaces the variance in the update of the Diagonal Synaptic Filter, as eminent when comparing \Cref{eq:dotmu} and \Cref{met:eq:dotwML}. The variance values of the Diagonal Synaptic Filter are generally below their equilibrium value $\sigma^2_t \leq \sigma^2_{\rm ou} = 1$ but well above 0.05. Because the Diagonal Synaptic Filter adjusts the variance of each synapse optimally, it is expected that the optimal value of $\eta$ has the same order of magnitude. As shown %can be seen in %\Cref{si:fig:grad-fit} of 
in the Supplementary Information, the optimal learning rate was indeed located in the interval $[0.05, 2]$.

%
%% model mismatch
\subsubsection{Model mismatch}
Bayesian regression takes advantage of the filtering distribution to compute the posterior predictive distribution. A central advantage of the posterior predictive is that it alleviates overfitting. In \Cref{res:sec:c2}, we introduce a model mismatch between the model that generates the observations, i.e., the tutor network in \Cref{fig:generative-model} \textbf{(B)}, on the one hand, and, on the other hand, the generative model on which the learning rules are based. We study model mismatch in \Cref{res:sec:c2} as follows.

When the generative model includes excess dimensions compared to the tutor, $d > d_{\text{tutor}}$, we provide the student network with additional $d - d_{\text{tutor}}$ Poissonian input with firing rate $\nu_0$ that have no effect on the generation of the observations. When the generative model has less dimensions compared to the tutor, $d < d_{\text{tutor}}$, tutor and generative model share the first $d$ input neurons but the tutor included additional $d_{\text{tutor}} - d$ ones, which are used to generate observations. % input neurons in the generation of the observations. 
Both, generative model and tutor, share the same value of the determinism parameter $\beta = \beta(d_{\rm tutor})$. In the simulations, we kept the dimension of the tutor at a constant value $d_{\rm tutor} = 5$ and varied the dimension of the generative model.

%
%% Hyperparameters
\subsubsection{Hyperparameters and simulated time}
% hyper pars likelihood
The membrane time constant and baseline firing rate are $\tau_{\rm m} = 25$ms and $g_0 = 1$Hz. The $d-1$ input spikes at dimensions $0 < i < d$ were drawn from Poisson neurons with $\nu_0 = 40$Hz. 
Thus, the expected synaptic activation is $\langle x^\epsilon_t \rangle = \tau_{\rm m} \nu_0 = 1$, i.e., on average each input neuron emits a spike per membrane time constant.
The first dimension denotes the bias and does not receive spiking input.

For the MSE simulations %$of the normalised moments and the MSE, 
discussed in \Cref{res:sec:c1}, we used 100 simulations with an OU time scale of $\tau_{\rm ou} = 100$s and duration $T_{\rm sim} = 10 \tau_{\rm ou}$. Shorter time scales yielded overall higher and less dynamic MSE curves for the hyperparameters studied. 

For the evaluation of the predictive performance, reported in \Cref{res:sec:c2}, a shorter time scale was used $\tau_{\rm ou} = 5$s. Because the predictive distribution is a noisier metric than the MSE and our computational resources were limited, we reduced the $\tau_{\rm ou}$ in order to be able to run more $\tau_{\rm ou}$-periods and, consequently, reduce the SEM. We chose $T_{\rm sim} = 200 \tau_{\rm ou}$. As before, 100 simulations we used. A second rationale for reducing the time scale was that the optimisation of the learning rate $\eta$ was computationally demanding.

%
%% simulation dt, run time, cutoff
\subsubsection{Simulation details: time steps and error tolerance}
For the Synaptic Filters and gradient rule in \Cref{res:sec:c2} we used a time step of $\Delta t = 0.5$ms and for the particle filter $\Delta t = 1$ms. These time steps were a compromise between minimization of the frequency with which discretisation errors occurred and the requirement to average over sufficiently many $\tau_{\rm ou}$-period to improve the errors. When a discretisation in the firing rate occurred, i.e., $g(u_t) {\Delta}t > 1$, we corrected it by enforcing a value of 1. From all conditions, this problem occurred most frequently for $\beta_0 = 2$ (irrespectively of the dimension). However, even for $\beta_0 = 2$, only $10^{-4}$ of the time steps needed a correction, which we regarded within error tolerance.
In the case of the particle filter, discretisation errors can lead to negative particle weights, which we corrected by setting them to 0 and renormalising all particle weights afterwards. 
Again, high values of $\beta_0$ caused the highest frequency of discretisation errors. For $\beta_0 = 2$ and $\beta_0 = 1.67$ negative particle weights occurred in $0.5 \%$ and $0.1 \%$ of the time steps. %We tolerated these errors because the values of the negative importance weights were always close to zero compared to the remaining ones and thus did not substantially affect the accuracy of the particle filter.

%
%% initial conditions
\subsubsection{Initial conditions}
The initial value of the tutor's weights was $w_{t=0} = \mu_{\rm ou}$. The distributional parameters were initialised as $\Sigma_{t=0} = \Sigma_{\rm ou}$ and $\mu_{t=0} \sim \mathcal{N}(\mu_{\rm ou},\Sigma_{\rm ou})$. For the particle filter, the initial positions of the particles, indexed with $l$, were drawn from the prior as well: $v^{(l)}_{t=0} \sim \mathcal{N}(\mu_{\rm ou},\Sigma_{\rm ou})$. After initialisation, a burn-in period of $\tau_{\rm ou}$ was simulated.

%
%% B1:3, Simulation details
\subsection{Biological predictions: parameters and technical details}
\label{met:sec:b123}
In this section, we specify the values of the hyperparameters, protocols, initial conditions and simulation parameters used in the simulated STDP experiments in  \Cref{res:sec:b1,res:sec:b2,res:sec:b3}.
%\\
%\\
\subsubsection{Hyperparameters}
% hyper pars likelihood
For the simulated experiments, the membrane time constant and baseline firing rate were set to their standard values, $\tau_{\rm m} = 25$ms and $g_0 = 1$Hz and the determinism parameter was always set to $\beta \equiv 1$ (unlike in the performance simulations in \Cref{fig:comp-results}%fig:z-plot,
). Scaling $\beta$ with dimensions as we did in the study of computational performance would have made it difficult to compare the weight change observed in the single synapse model ($d=1$), the synapse with bias model ($d=2$) and the heterosynaptic experiments ($d=3$).

% OU parameters details
For the simulated experiments, we assume that the bias $w_{t,0}$ changes on the same time scale as the membrane potential, $\tau_{\rm ou, bias} = \tau_{\rm m} = 25$ms. 
%The same time scale is assumed for correlations between bias and other weights. 
The time scale of the weights and weight correlations is set to $\tau_{\rm ou} = 10 000$s. This implies that the correlations between the bias and other weights decay on the order of $\tau_{\rm m}$ (see %\Cref{si:eq:two-timescales} in
Supplementary Information).

With the exception of the bias, the equilibrium values of the transition probability are set to $\mu_{\rm ou} = 0$ for the weights and $\sigma^2_{\rm ou} = 1$ for the diagonal covariance elements and off-diagonal covariance elements are set to zero: $\Sigma_{{\rm ou}, i \neq j} = 0$. The bias is treated differently because it represents the neuronal excitability, as explained in the main text. To prevent run-away dynamics of the bias in the absence of spikes, we chose $\mu_{\rm ou, 0} = 1$. The prior variance $\sigma^2_{\rm ou, 0}$ determines how strongly the bias changes in response to input and output spikes. For the STDP experiments in \Cref{fig:STDP-prediction}, we chose  $\sigma^2_{\rm ou} = 2$ and 
%To increase the amplitude of the LTD lobe in the STDP experiments, $\sigma^2_{\rm ou} = 2$ was chosen. 
for the heterosynaptic experiments in \Cref{fig:hetero-prediction}, we chose $\sigma^2_{\rm ou} = 1$. %fulfills this requirement and was used in the simulated experiments on heterosynaptic plasticity. To increase the amplitude of the LTD lobe in the STDP experiments, $\sigma^2_{\rm ou} = 2$ was chosen. See the SI for a direct comparison to the case of $\sigma^2_{\rm ou, 0} = 1$.\\
%\\
%\\ 
% waiting times
\subsubsection{Protocols and initial conditions}
The STDP protocol consisted of pre- and postsynaptic spikes with 200 different values for the delay (shown on the $x$-axis of the STDP curve). The preconditioning protocol consists of a presynaptic spike pair with $5$ms delay simultaneously applied to both presynaptic inputs.
Prior to applying any protocol, the Synaptic Filter was simulated without any input or output spikes for $T_{\rm wait} = 6 \tau_{\rm m}$ such that 
%Visual inspection of the time series confirmed that $T_{\rm wait}$ was sufficiently long for 
the mean and variance values of the bias converge to their equilibria. The same waiting time was simulated after the preconditioning protocol. The simulated STDP curve was computed based on the value of the weight directly before the application of the STDP protocol and the value of the weight after $2 T_{\rm wait}$. %, where the larger time window was chosen because the difference between pre- and postsynaptic spike can represent a significant fraction of $T_{\rm wait}$.
%
% init cond
The initial conditions for the distribution parameters of the Synaptic Filter were chosen %either one or zero: 
$\mu_{t=0,i} = 1$ with $i \in (0, \dots, d-1)$ for the weights and $\Sigma_{t=0} = \mathbb{1}\sigma_0^2$ with $\sigma_0^2 = 1$ for the covariance. % and $\Sigma^2_{t=0,ij} = 0$ for $i \neq j \in \{0,\dots d-1\}$.
%\\
%\\
\subsubsection{Technical details: simulations and fits}
%
% simulation parameters
We solved the ODEs of the Synaptic Filters with the Euler method. The time step was $\Delta t = 0.1$ms for the STDP experiments presented in \Cref{res:sec:b1,res:sec:b2}. Since the preconditioning protocol induces sharp decreases of the variance, a time step of $\Delta t = 0.01$ms was used for the simulations in \Cref{res:sec:b3} to ensure that the variance remained positive.

% Fit
The slopes reported in \Cref{fig:hetero-prediction} \textbf{(E, F)} correspond to a linear fit with a least-squares objective and two free parameters, slope and offset. The data shown in \Cref{fig:hetero-prediction} \textbf{(F)} were extracted manually.
 % end methods environment
%\showmatmethods{} % Display the Materials and Methods section

\section{Supplementary Information}

The Supplementary Information addresses three main questions. \Cref{sec:si:ssf} asks whether the sampling hypothesis is compatible with MSE performance. In \Cref{sec:si:c0-all}, we analyse whether the Synaptic Filter is a faithful solution filtering problem. Thirdly, in \Cref{sec:si:derivation-all}, we answer how the update equations of the Synaptic Filter (Equations (1) and (2) in the Main Text, here \Cref{eq:si:dot-mu,eq:si:dot-sigma2}) are derived. 

Additional short sections provide additional explanations for the results in the Main Text. \Cref{si:sec:grad-fit} details how the predictive performance was evaluated for the gradient rule and \Cref{si:sec:SF-variable-dynamics} shows how the variables of the Synaptic Filter evolve during the plasticity protocols. In the final Section, we discuss when weight correlations are negative in the Synaptic Filter with more than two dimensions.

%
%% 1 SSF and C1
\subsection{The sampling hypothesis is compatible with MSE performance}
\label{sec:si:ssf}

%\subsubsection{The Sampling Synaptic Filter}
%
%% SSF and diagonal
The sampling hypothesis states that EPSP are samples from the filtering distribution \citep{aitchison2015synaptic}. We wondered whether the sampling hypothesis could be used to inspire a version of the Synaptic Filter that maintains good filtering performance while at the same time computing the expected firing rate through EPSP samples.

%
%% sampling and smoothign
Based on two heuristic modifications of the update equations of the Synaptic Filter \Cref{eq:si:dot-mu,eq:si:dot-sigma2}, we derive the Sampling Synaptic Filter. The first modification is that the expected firing rate $\gamma_t$ is replaced with the sampling based firing rate $\gamma^{\rm s}_t := g(u^{\rm s}_t)$. The sampling based membrane potential $u^{\rm s}_t$ is directly motivated by the sampling hypothesis, i.e., it is a low-pass filtered sum of spike-triggered samples from all synapses. %Upon the arrival of a spike at the $i^{\rm th}$ synapse, a sample is drawn from the marginal filtering distribution:
%$w^{\rm s}_{t,i} \sim \mathcal{N}(\mu_{t,i},\Sigma_{t,ii})$. Because no two spikes occur in the same moment, all samples in ${\gamma}^{\rm s}_t$ come from marginals of the filtering distribution. Thus, ${\gamma}^{\rm s}_t$ is not sensitive to weight correlations. 
The second modification aims at slowing down learning to the sampling time scale $\tau_{\rm s}$. This modification is motivated by the empirical observation that good filtering performance requires the sampling based firing rate ${\gamma}^{\rm s}_t$ to change on the same time scale as other spike-related terms in the update \Cref{eq:si:dot-mu,eq:si:dot-sigma2}. In the following, both modifications are discussed in more detail.

%
%% Modifications from Methods
%\subsubsection{Methods: the Sampling Synaptic Filter}
%\label{met:sec:ssf}
The sampling based firing rate $\gamma_t^{\rm s} := g(u^{\rm s}_t)$ is computed on the basis of a sampling based membrane potential $u^{\rm s}_t$, which is modelled as leaky-integrator of spike-triggered EPSPs $w^{\rm s}_t$:
\begin{align}
    u^{\rm s}_t = ((w^{\rm s})^\top x * \epsilon)_t.
\end{align}
The sum  $(w^{\rm s})^\top x$ takes a value different from zero only when a spike arrives at synapse $0<i<d$ and a sample from the marginal filtering distribution is drawn:
\begin{align}
    w^{\rm s}_{t,i} \sim \mathcal{N}(\mu_{t,i},\Sigma_{t,ii}).
\end{align}
Since no two spikes occur in the same moment, the Sampling Synaptic Filter does not include correlations between weights.
The bias is treated as a special case because it does not receive spikes and, thus, the idea of spike-triggered sampling cannot be applied. We make the assumption that Sampling Synaptic Filter includes the bias in the sampling based firing rate $\gamma_t^{\rm s}$ in the same way as the Synaptic Filter, i.e., we adopt the following convention for the first component of the sum:
\begin{align}
\label{met:eq:SSF:bias}
    (w^{\rm s}_0 x_0 * \epsilon)_t \equiv \mu_{t,0} + \tfrac{1}{2} \beta \sigma_{t,00}^2.
\end{align}
With \Cref{met:eq:SSF:bias} and in the case of $d=1$, the expected firing rate of the Synaptic Filter and sampling based firing rate in the Sampling Synaptic Filter are identical: $\gamma_t^{\rm s} = \gamma_t$.

%
%% smoothing 4 \tau_{\rm m}
The second modification aims at slowing down learning in order to suppress fast feedback dynamics between weights and expected firing rate. Without this suppression, filtering performance deteriorates because the first modification prevents the fast feedback from working properly. Consider how the fast feedback works in the case of the Synaptic Filter. A large expected firing rate $\gamma$ can reduce its own value indirectly by reducing the mean and variance values of all weights with active inputs through the update \Cref{eq:si:dot-mu,eq:si:dot-sigma2}. This reduction can be faster than the membrane time scale $\tau_{\rm m}$. In the Sampling Synaptic Filter, this fast feedback mechanism does not work anymore. The sampling based firing rate $\gamma^{\rm s}_t$ depends on samples of past filtering distributions and it cannot influence the value of these samples through the update \Cref{eq:si:dot-mu,eq:si:dot-sigma2}. The sampling based firing rate $\gamma^{\rm s}_t$, responds to changes in the filtering distribution with a delay of $\tau_{\rm m}$. Thus, the weight updates can overshoot which leads to poor performance. %because  equations are derived under the assumption that the expected firing rate $\gamma_t$ is part of a fast feedback loop but its substitute, 

One way to address the broken fast feedback is by introducing a delay in the updates in \Cref{eq:si:dot-mu,eq:si:dot-sigma2} such that the delayed response of $\gamma^{\rm s}_t$ to changes in the filtering distribution becomes irrelevant. Specifically, we consider spike related quantities as fast and replace them by their low-pass filtered version. With the exponential kernel $\alpha$ with \textit{sampling time scale} $\tau_{\rm s}$, the update \Cref{eq:si:dot-mu,eq:si:dot-sigma2} of the Sampling Synaptic Filter are defined as follows:
\begin{align}
    % mean
    \dot{{\mu}}_t &= \beta {\Sigma}_t \left( (\alpha *  x^{\epsilon} y )_t - (\alpha * x^{\epsilon} \gamma^{\rm s} )_t \right) - \tau_{\rm ou}^{-1}(\mu_t - \mu_{\rm ou}) \label{met:eq:SSF-mean} \\
    % var update
    \dot{{\Sigma}}_t &=
    - \beta^2 {\Sigma}_t (\alpha * x^{\epsilon} (x^{\epsilon})^\top \gamma^{\rm s} )_t {\Sigma}_t - 2 \tau_{\rm ou}^{-1} (\Sigma_t - \Sigma_{\rm ou}).
    \label{met:eq:SSF-cov}
\end{align}
In all simulations, we chose $\tau_{\rm s} = 4 \tau_{\rm m} = 100$ms.
Additionally, we test the Diagonal Sampling Synaptic Filter which is updated according to \Cref{met:eq:SSF-mean,met:eq:SSF-cov} with the only difference that a diagonal covariance matrix is enforced. While both versions of the Sampling Synaptic Filter compute the sampling membrane based on the marginal filtering distribution, i.e., without considering correlations between samples, the Sampling Synaptic Filter includes the off-diagonal elements of the covariance matrix during learning while the Diagonal Sampling Synaptic Filter uses a diagonal covariance matrix.

Because the model of the membrane potential differs between the sampling filters and their deterministic counterparts, \Cref{met:eq:SSF-mean,met:eq:SSF-cov} do not converge to \Cref{eq:si:dot-mu,eq:si:dot-sigma2} in a non-trivial limiting case.
Furthermore, the history of the weight samples $w^{\rm s}_t$ must be included in the data $\mathcal{D}^{\rm s}_t := (w_{0:t - \text{d}t}^{\rm s}, \mathcal{D}_t)$. 
The key advantage of the Sampling Synaptic Filter over the Synaptic Filter is that it does not require the instantaneous evaluation of the expected firing rate $\gamma_t$ but instead uses a biologically more plausible expected firing rate ${\gamma}^{\rm s}_t$ inspired by the sampling hypothesis. 

%
%% C1 results
%\subsubsection{Performance of the various Synaptic Filters}
%\label{res:sec:c1}
%
% outlook
In the following, the MSE performance of the Synaptic Filter, the Sampling Synaptic Filter, the Diagonal Synaptic Filter and the Diagonal Sampling Synaptic Filter are evaluated for the same range of values for the determinism parameter $\beta_0$ and the dimension $d$ as in the Main Text. The results for the Synaptic Filter and the Diagonal Synaptic Filter are replotted.

\begin{figure}[htb!]
\begin{minipage}{\myFigureWidth\linewidth}
\begin{tabular}{ll}
{\bf (A)} & {\bf (B)} \\
\begin{minipage}{0.45\textwidth}
\includegraphics[width=\textwidth,trim={0cm 0cm 0cm 0cm},clip]{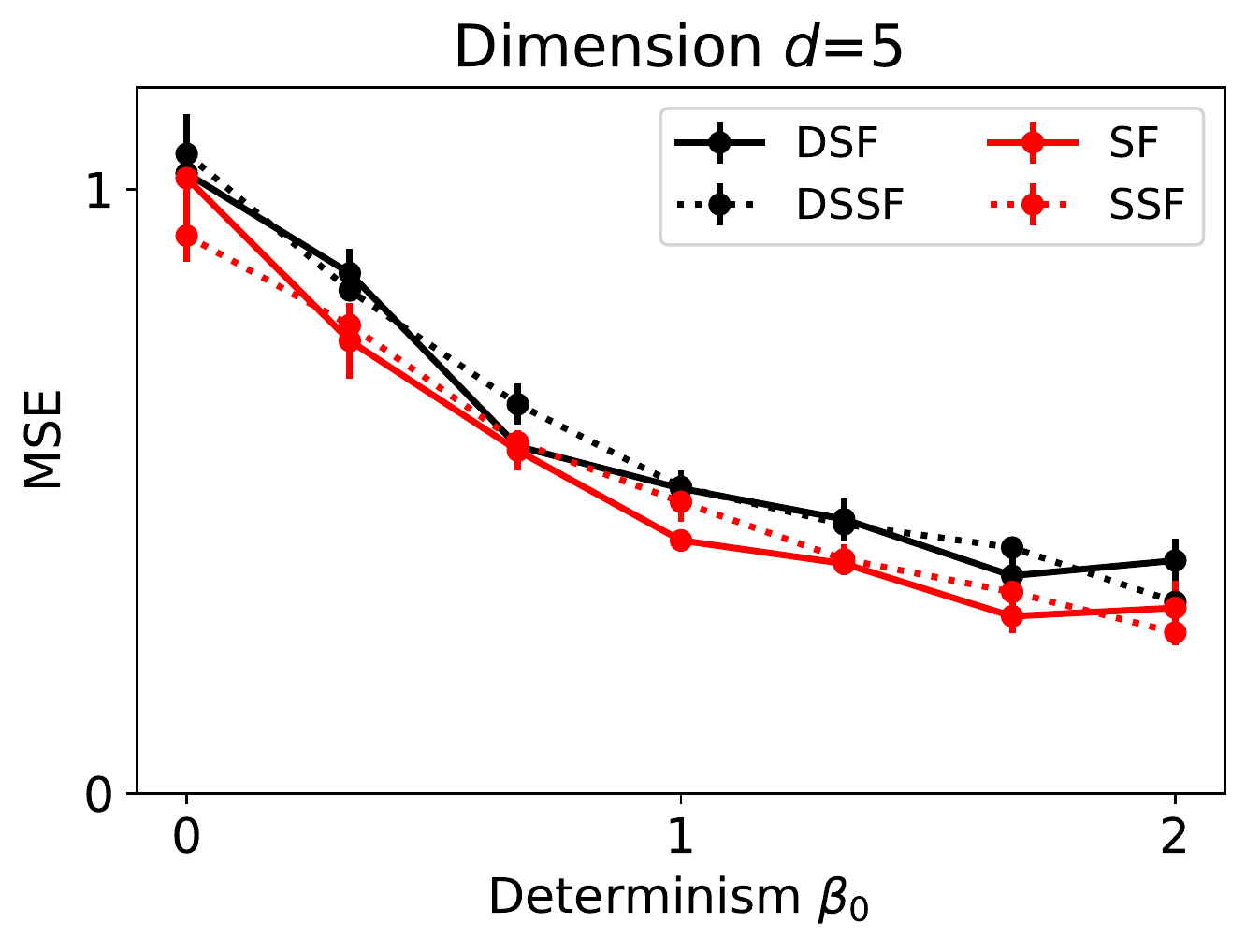}
\end{minipage} &
\begin{minipage}{0.45\textwidth}
\includegraphics[width=\textwidth]{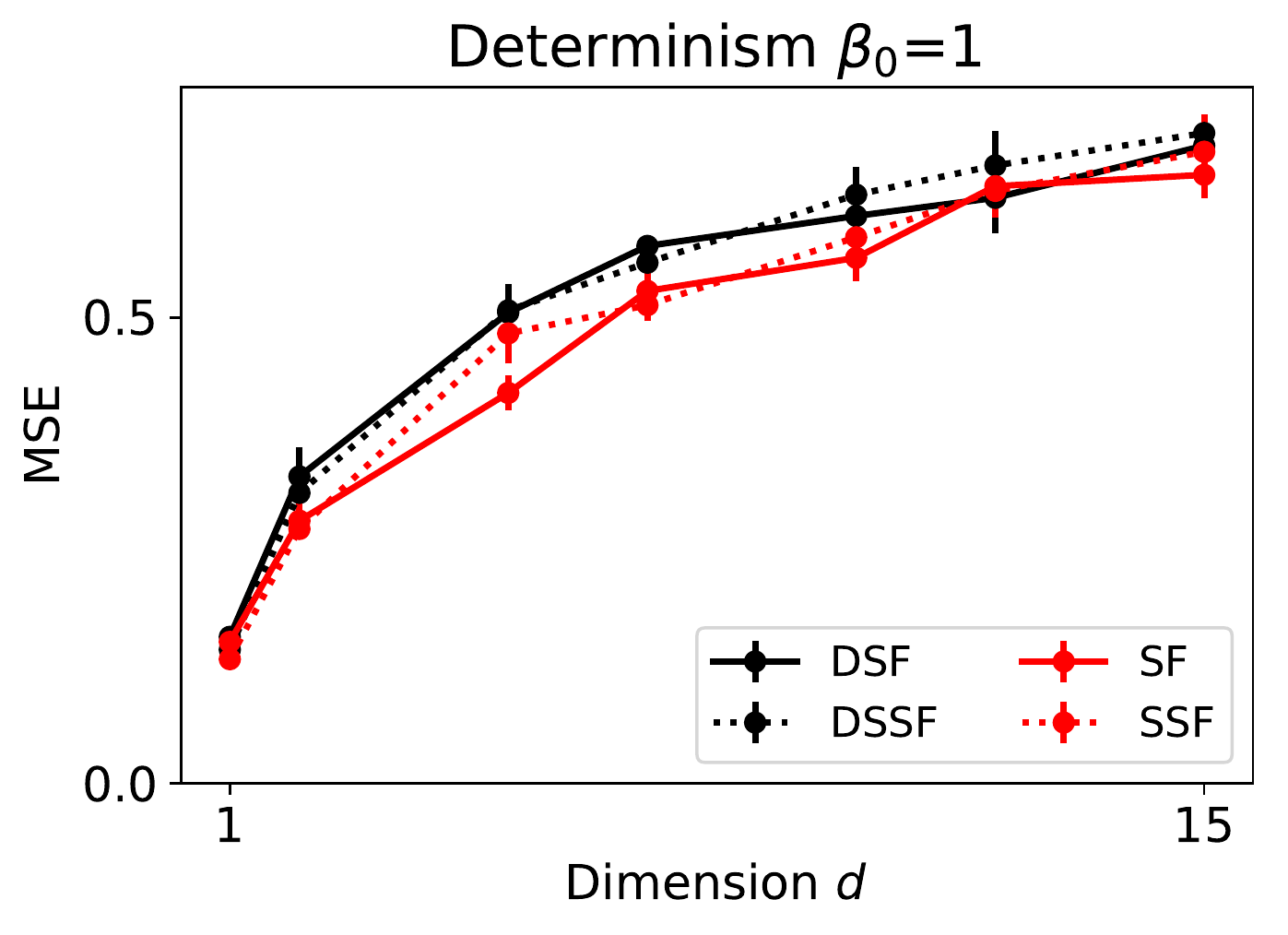}
\end{minipage}
\end{tabular}
\end{minipage}
\caption{ \label{fig:MSE-vs-d}
% key message
The Sampling Synaptic Filters have similar MSE performance to their deterministic counterparts for all values of the dimension $d$ and determinism $\beta_0$.
% describe plots
\textbf{(A)} The MSE of Synaptic Filter (red solid), Sampling Synaptic Filter (red dashed) and their diagonal counterparts (black) decrease as the determinism $\beta_0$ increases.
\textbf{(B)} For all filtering models, the MSE increases as a function of dimension. Again, the diagonal variants (black) perform worse.
% parameters
Dots and errors denote the mean and SEM from 100 simulations. The simulated time per run was $10 \tau_{ou} = 1000$s.
}
\end{figure}

%
% MSE(b0)
Taking into account the sampling hypothesis does not substantially impair MSE performance, as shown in in \Cref{fig:MSE-vs-d}. The sampling filters behave similarly to their deterministic counterparts, including a small performance gain from using the covariance matrix (red lines) in the update equations. Still, the Synaptic Filter (red solid line) remains the best model overall. 

%
%% Conclusion
%%% Sampling Synaptic Filter only MSE: \footnote{The Sampling Synaptic Filter and Diagonal Sampling Synaptic Filter are not tested here because the sampling time scale $\tau_{\rm s}$ causes delayed updates of the distributional parameters. To account this delay when evaluating predictive performance would introduce considerable complexity in defining a metric for predictive performance.}

\subsection{The Synaptic Filter solves the filtering problem}
\label{sec:si:c0-all}
%
%% results
\subsubsection{The normalised moments of the Synaptic Filter correspond to the moments of the exact filtering distribution}
\label{res:sec:c0}
%
% motivation
The Synaptic Filter is an approximated solution to the filtering problem. Thus, we need to determine the level of accuracy of the approximation. Specifically, we check if the moments of the Synaptic Filter are consistent with the corresponding moments of the exact filtering distribution. We test a range of values for the determinism parameter $\beta_0$ with fixed dimension $d=5$, and a range of dimensions $d$ for a fixed value of the determinism $\beta_0 = 1$.

%
% method
To compare the moments of the Synaptic Filter $q_{\theta_t}(w_t)$ with the moments of the exact filtering distribution $p(w_t | \mathcal{D}_t)$, we compute the normalised estimators $z^{(1)}$ and $z^{(2)}$ of the first and second moment respectively. If the moments of $q$ converge to the first two moments of the exact distribution, the  normalised estimators converge $z^{(1)} \rightarrow 0$ and $z^{(2)} \rightarrow 1$ (see \Cref{met:sec:z_z2}). We use these conditions to measure the quality of the Synaptic Filters.
%
%% pf
Additionally, we obtain an approximation to the exact filtering distribution $p(w_t | \mathcal{D}_t)$ with a particle filter (see \Cref{met:sec:pf}). The particle filter converges to the exact solution of the filtering problem in the limit of a large number of particles.

\begin{figure}[h!]
\begin{minipage}{\myFigureWidth\linewidth}
\begin{tabular}{ll}
{\bf (A)} & {\bf (B)} \\
\begin{minipage}{0.49\textwidth}
\includegraphics[width=\textwidth,trim={0cm 0cm 0cm 0cm},clip]{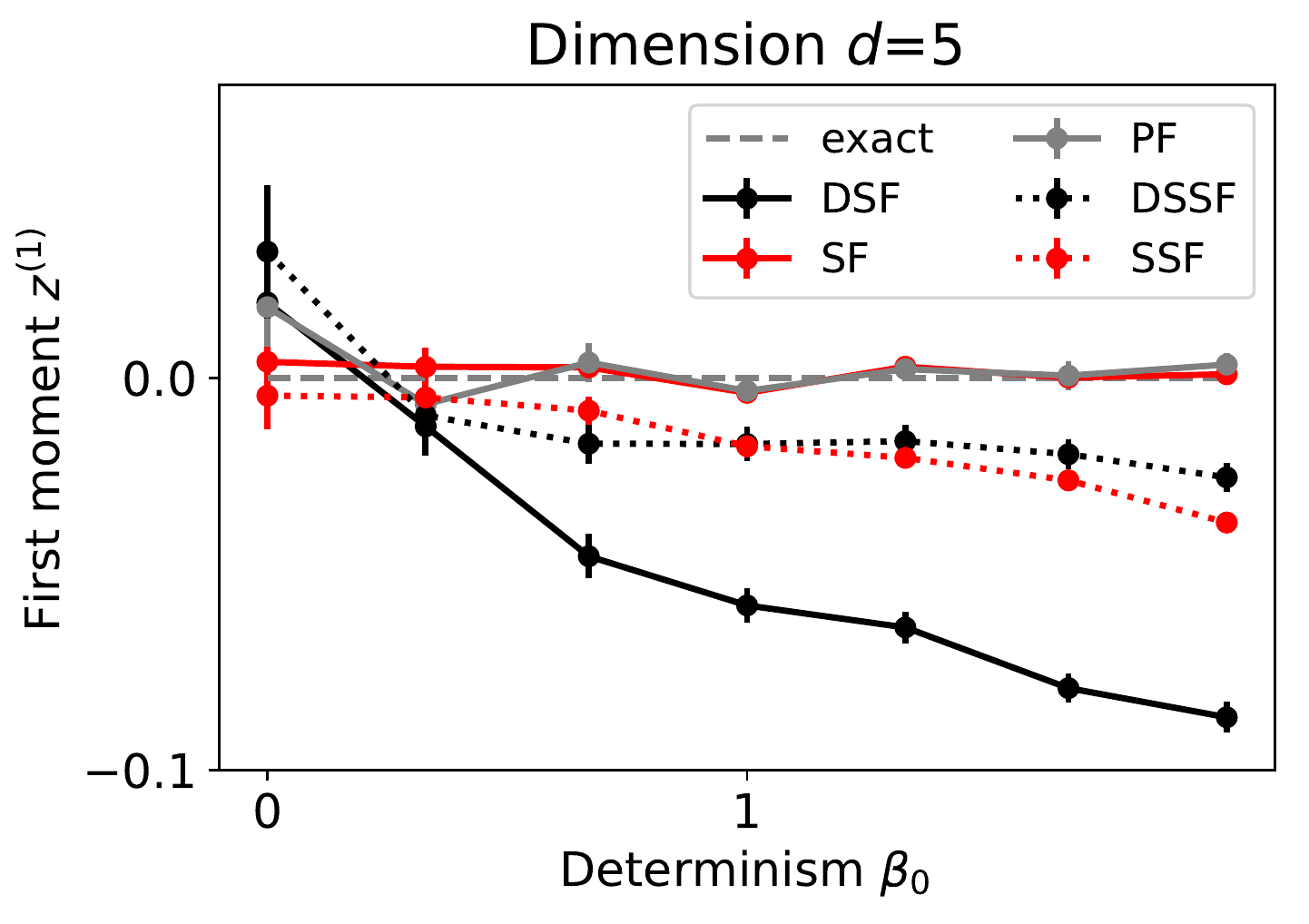}
\end{minipage} &
\begin{minipage}{0.49\textwidth}
\includegraphics[width=\textwidth]{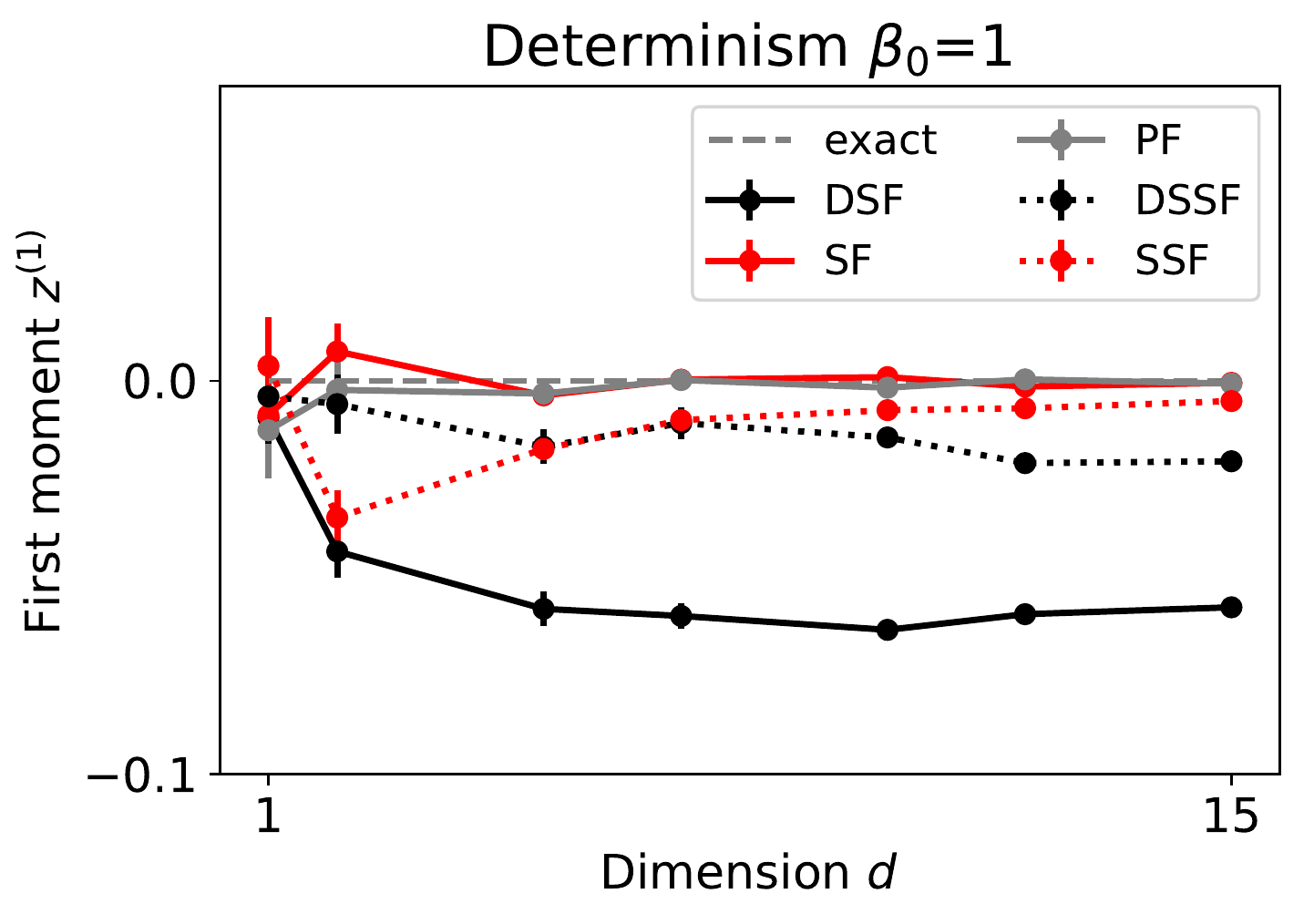}
\end{minipage}\\
{\bf (C)} & {\bf (D)} \\
\begin{minipage}{0.46\textwidth}
\hspace{0.4cm}
\includegraphics[width=\textwidth]{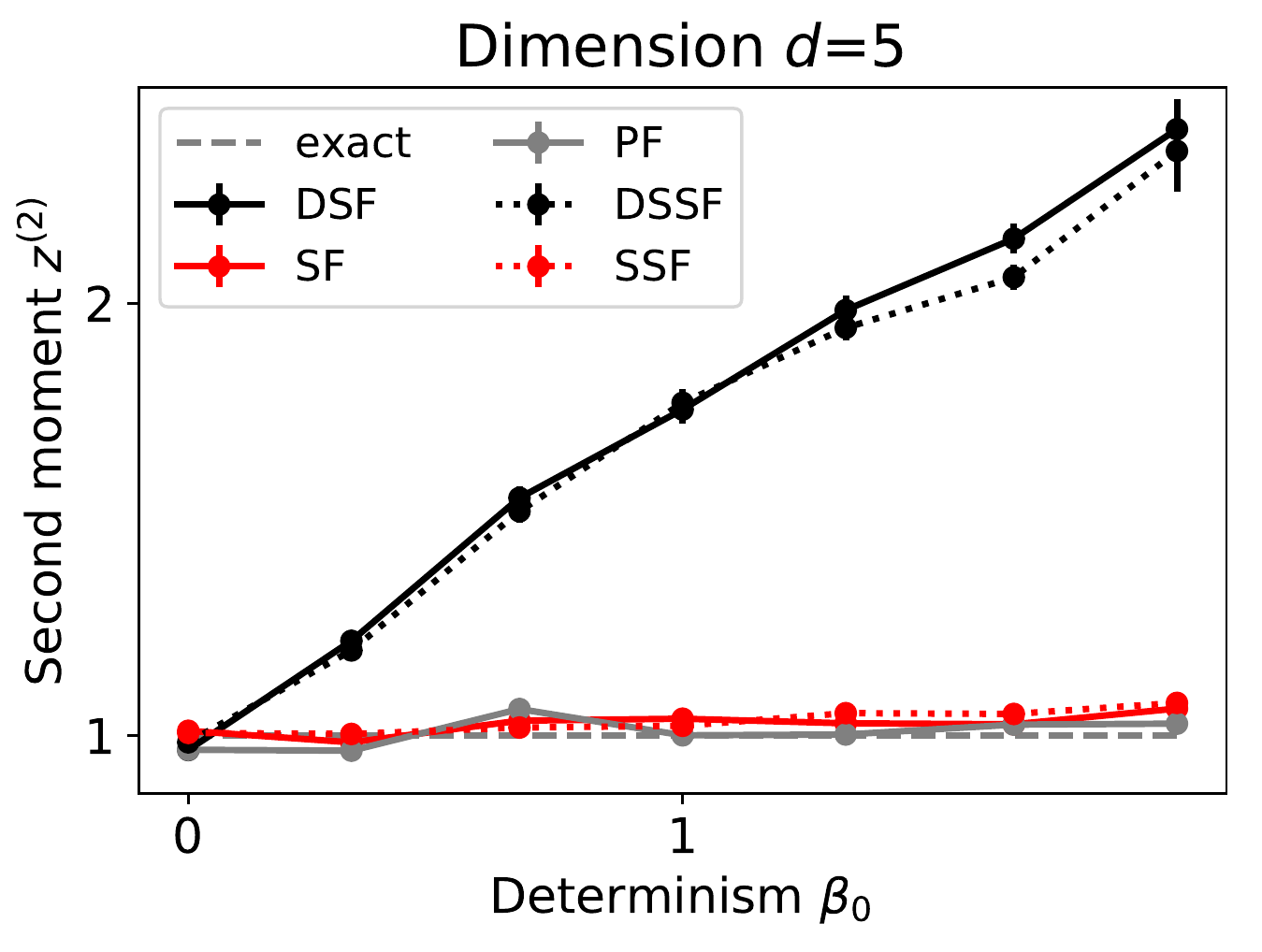}
\end{minipage} &
\begin{minipage}{0.46\textwidth}
\hspace{0.4cm}
\includegraphics[width=\textwidth]{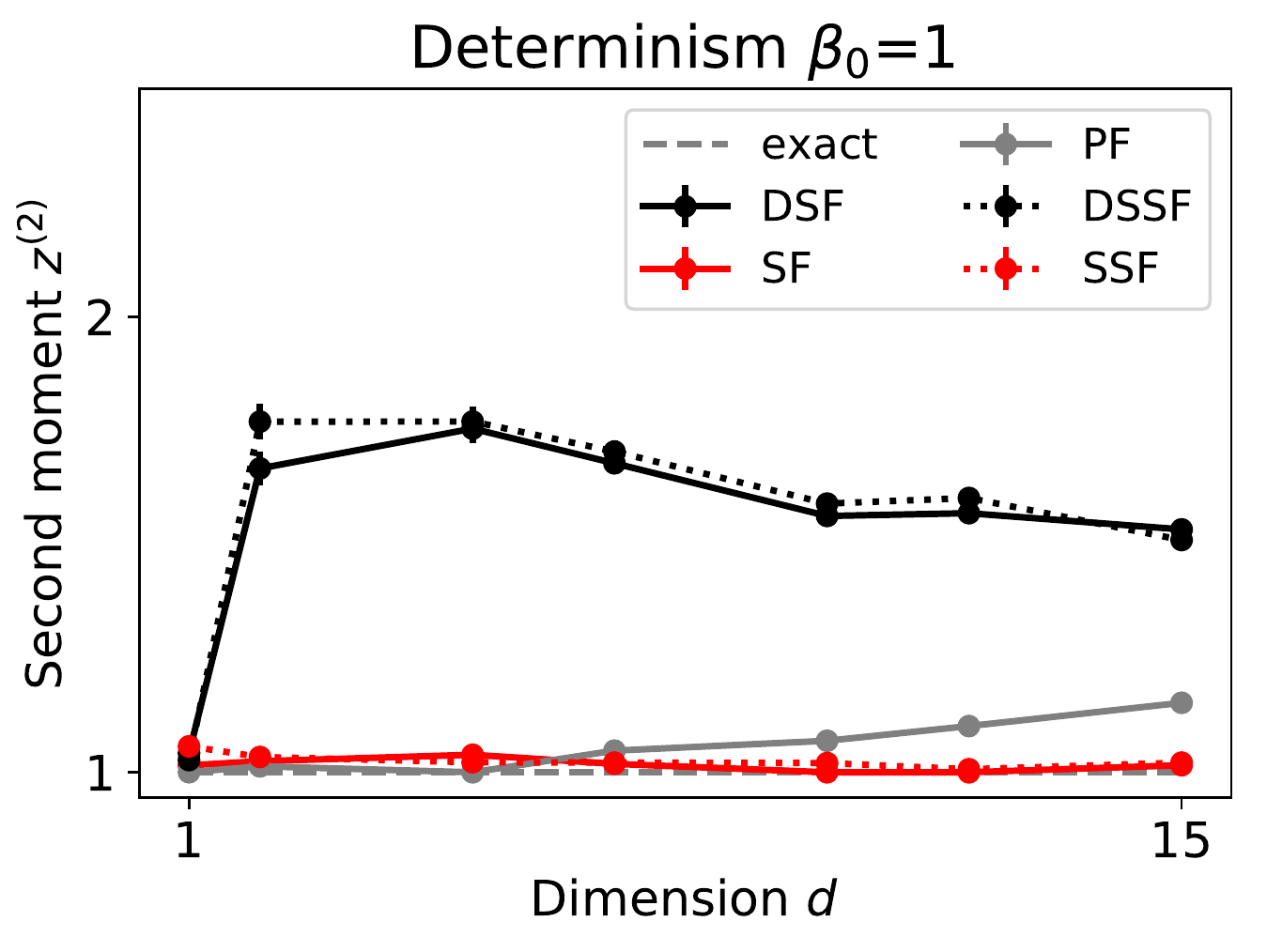}
\end{minipage}
\end{tabular}
\end{minipage}
\caption{ \label{fig:z-plot}
% key message
The first and second moments of the Synaptic Filter (SF, solid red line) match the corresponding moment of the exact filtering distribution, i.e. the normalised moments of the Synaptic Filter satisfy $z_{\rm SF}^{(1)} \approx 0$ and $z^{(2)}_{\rm SF} \approx 1$.
% describe plots
\textbf{(A)} For $d=5$ and a $0 \leq \beta_0 \leq 2$, the first normalised moment $z^{(1)}$ of the Synaptic Filter and the particle filter (PF, gray line) are close to 0 while the Sampling Synaptic Filter (SSF, solid black), Diagonal Synaptic Filter (DSF, dashed red line), and Diagonal Sampling Synaptic Filter (DSSF, dashed black line) deviate from 0. At $\beta_0 = 0$, all filtering distribution resemble the prior. As the determinism $\beta_0$ increases, the deviation increases as well.
\textbf{(B)} For $\beta_0=1$ and $1 \leq d \leq 15$, the first normalised estimator $z^{(1)}$ of all models excepct for the Diagonal Synaptic Filter are close. The dimension $d$ corresponds to the number of presynaptic inputs plus one (for the bias) so when $d = 1$, the Synaptic Filter and Sampling Synaptic Filter (red lines) are equivalent to their diagonalised (black). However, the diagonalised versions perform worse for $d>2$.
\textbf{(C)} The second normalised estimator $z^{(2)}$ of the Synaptic Filter, the Sampling Synaptic Filter and the particle filer are close to 1 while the diagonalised versions (black) overestimate $z^{(2)}$. The deviation is linear in $\beta_0$.
\textbf{(D)} As before, the $z^{(2)}$ is close to one for Synaptic Filter and Sampling Synaptic Filter and deviates for the diagonal models. %The deviation for the diagonalised models jumps when the first spiking input is included at $d = 2$ and decreases slowly for higher $d$.
The particle filter is consistent with $z^{(2)}=1$ at low dimensions but deviates systematically for increasing $d$.
% parameters
Dots and error bars denote the mean and SEM from 100 simulations. The simulated time per run was $10 \tau_{ou} = 1000$s.
}
\end{figure}

%
% main finding
The results in \Cref{fig:z-plot} \textbf{(A-D)} show that, for a range of dimensions $d$ and values of the determinism $\beta_0$, the normalised estimators of the Synaptic Filter (SF) confirm consistency between the approximated and the exact filtering distribution, i.e. $z^{(1)}_{\rm SF} \approx 0$ and $z^{(2)}_{\rm SF} \approx 1$. The Synaptic Filter (red solid line) and the Diagonal Synaptic Filter (black solid line) show similar performance at $d=1$ (\Cref{fig:z-plot} \textbf{(B, D)}) because in this case the covariance matrix is a scalar and, hence, both models are identical. However, at higher dimensions the Diagonal Synaptic Filter exhibits a small deviation from $z^{(1)} = 0$ and a strong positive deviation from $z^{(2)} = 1$. For the later, the deviation grows linearly with the determinism $\beta_0$, as shown in \Cref{fig:z-plot} \textbf{(C)}.

%% REDO
The reason is that due to its omission of correlations, the Diagonal Synaptic Filter underestimates the overall weight uncertainty, and hence overestimates the overall precision and $z^{(2)}$, which is proprotional to the precision matrix, i.e., the inverse of the covariance matrix. %Because we consider positive inputs, i.e., $x^\epsilon_t \geq 0$, weight correlations are always negative, i.e., $\Sigma_{i \neq j} \leq 0$ (see \Cref{met:sec:negCorr} Material and Methods).
%Thus, the omission of weight correlations leads to less negative contributions to $z^2$ and results in an overestimation of $z^2$. %the expected firing rate $\gamma_t$ (see \Cref{met:eq:gamma}), which is proportional to the variance decrease (see \Cref{eq:si:dot-sigma2}).
Correlations arise from the likelihood term in the update \Cref{eq:si:dot-sigma2}, which is proportional to $\beta_0$. This explains the scaling of the deviation with $\beta_0$.
The superior performance of the Synaptic Filter shows that the off-diagonal elements of the covariance matrix are important to obtain a good approximation to the filtering distribution.

%
%% PF
The estimators computed based on the particle filter (gray) are generally consistent with the exact distribution and with the Synaptic Filter, i.e., they satisfy $z_{\rm PF}^{(1)} \rightarrow 0$ and $z^{(2)}_{\rm PF} \rightarrow 1$. This was expected since particle filters are asymtotically exact in the limit of infinitely many particles. The systematic deviation $z^{(2)}_{\rm PF} > 0$ in (\Cref{fig:z-plot} \textbf{(D)}) arises because particle filters suffer from the curse of dimensionality, which leads to underestimates of the covariance in higher dimension.

%
% second finding
The Sampling Synaptic Filter (red dashed line) and the Diagonal Sampling Synaptic Filter (black dashed line) estimate the second moment $z^{(2)}$ with similar accuracy as their counterparts without sampling (solid lines), as shown in \Cref{fig:z-plot} \textbf{(C, D)}. The Sampling Synaptic Filter performs well, i.e., $z^{(2)}_{\rm SSF} \rightarrow 1$, while the Diagonal Sampling Synaptic Filter deviates strongly from $z^{(2)}=1$. Both sampling filters perform well in terms of the first normalised moment, shown in \Cref{fig:z-plot} \textbf{(A, B)}. The fact that both sampling filters estimate the moments of the exact filtering distribution with comparable accuracy as their deterministic counterparts implies that the explicit inclusion of the sampling hypothesis in the updates does not impair filtering performance. For this conclusion to hold, we had to assume a sampling time scale $\tau_{\rm s}$ that was many orders of magnitude smaller than the time scale of the weight evolution $\tau_{\rm ou}$ (see Main Text).

%
% Conclusion
The analysis of the first and second normalised moment estimators, $z^{(1)}$ and $z^{(2)}$, shows that the mean and covariance computed by the Synaptic Filter correspond closely to the mean and covariance of the exact filtering distribution. A particle filter solution to the filtering problem confirms this. The fact that the Diagonal Synaptic Filter and the Diagonal Sampling Synaptic Filter perform poorly shows that off-diagonal elements in the covariance matrix must be included to match the moments of the exact filtering distribution.

%
%% 2 methods
\subsubsection{Time averaged normalised moment estimators}
\label{met:sec:z_z2}

The naive approach to quantify whether the moments of a proposal distribution, i.e., an approximation to the exact filtering distribution, match the moments of the exact filtering distribution, is computing the moments of both. However, this requires the potentially expensive computation of the moments of the exact filtering distribution. To avoid this problem, we use an implicit comparison via the normalised moments. In particular, we calculate the first two normalised moments $z^{(1)}$ and $z^{(2)}$, which form the basis of the analysis in \Cref{res:sec:c0}.

Consider a proposal distribution $q_\theta(w_t)$ with mean $\mu_t$ and covariance $\Sigma_t$. The moments of the proposal can be compared to the respective moments $\mu_t^\star$ and $\Sigma_t^\star$ of the exact filtering distribution $p(w_t | \mathcal{D}_t)$ by averaging over many realisations of the generative model, i.e., over the history of the hidden weight $w_{0:t}$ and data generated $\mathcal{D}_t$. To see this, we rewrite the distribution over the realisations of the generative model using the definition of conditional probabilities:
\begin{align}
    p(w_{0:t}, \mathcal{D}_t) = p(w_{t}, w_{0:t-\text{d}t} | \mathcal{D}_t) p(\mathcal{D}_t).
    \label{met:eq:prob_of_GM}
\end{align}
Now, we define the time dependent, normalised moments $z_t^{(1)}$ and $z^{(2)}_t$ as a function of the hidden weight $w_t$ and the moments of the proposal:
\begin{align}
    \label{met:eq:z_t}
    z^{(1)}_t &:=  d^{-1} \sum_{i=1}^d (\Sigma_t^{-\frac{1}{2}} (w_t - \mu_t))_i \\
    \label{met:eq:z2_t}
    z^{(2)}_t &:= d^{-1} (w_t - \mu_t)^\top \Sigma_t^{-1} (w_t - \mu_t).
\end{align}
Since \Cref{met:eq:z_t,met:eq:z2_t} do not dependent on $w_{0:t - \text{d}t}$, the expectation with respect to realisations of the generative model (\Cref{met:eq:prob_of_GM}) is equivalent to the expectation with respect to the posterior and the data, i.e. $p(w_{t} | \mathcal{D}_t) p(\mathcal{D}_t)$:
\begin{align}
    \label{met:eq:Ez_t}
    \mathbb{E}[ z^{(1)}_t | w_{0:t},\mathcal{D}_{t}]  &=  d^{-1} \sum_{i=1}^d \mathbb{E}[ (\Sigma_t^{-\frac{1}{2}} (\mu_t^\star - \mu_t))_i | \mathcal{D}_{t}] \\
    \label{met:eq:Ez2_t}
    \mathbb{E}[ z^{(2)}_t | w_{0:t},\mathcal{D}_{t}] &= d^{-1} \text{Tr}\left( \mathbb{E}[ \Sigma_t^{-1} (\mu^\star_t - \mu_t) (\mu^\star_t - \mu_t)^\top + \Sigma_t^{-1} \Sigma^\star_t | \mathcal{D}_t ]
    \right),
\end{align}
where we evaluated $w_t$ using the mean and covariance of the exact posterior. To obtain \Cref{met:eq:Ez2_t}, we further used that the trace operator leaves a scalar quantity invariant and permits cyclic permutation. Note that the moments of the exact posterior and the proposal dependent on the history of the data $\mathcal{D}_t$, which is why the expectation on the right-hand side of \Cref{met:eq:Ez_t,met:eq:Ez2_t} remains. However, in the special case that the proposal is equivalent to the exact posterior, we have $\mu_t^\star = \mu_t$ and $\Sigma_t^{-1} \Sigma_t^\star = \mathbb{1}$, and consequently:
\begin{align}
    \label{met:eq:Ez_t_0}
    \mathbb{E}[ z_t^{(1)} | w_{0:t},\mathcal{D}_{t}]  &=  0 \\
    \label{met:eq:Ez2_t_1}
    \mathbb{E}[ z^{(2)}_t | w_{0:t},\mathcal{D}_{t}] &= 1.
\end{align}
Thus, \Cref{met:eq:Ez_t,met:eq:Ez2_t} can be compared to \Cref{met:eq:Ez_t_0,met:eq:Ez2_t_1} to test whether the moments of a proposal distribution match the moments of the exact posterior without explicitly calculating the latter.

The quantities shown in the Main Text are time averaged estimates of \Cref{met:eq:Ez_t,met:eq:Ez2_t} computed from 100 realisations of the generative model. The time average runs over 10 $\tau_{\rm ou}$ time constants. The average over time is justified because the generative model is stationary, i.e., there is no explicit dependence on time. Indexing realisations by $k$, we define the normalised moments by:
\begin{align}
    \label{met:eq:Ez_t_final}
    z^{(1)} & :=  \tfrac{1}{100}\sum_{k=1}^{100} \langle z_{t,k}^{(1)} \rangle_t \\
    \label{met:eq:Ez2_t_final}
    z^{(2)} & :=  \tfrac{1}{100}\sum_{k=1}^{100} \langle z^{(2)}_{t,k} \rangle_t.
\end{align}

%
%% PF
\subsubsection{Particle filter}
\label{met:sec:pf}
To compute an asymptotically correct approximation to the posterior distribution (without the approximation of the Assumed Density Filter) we use a particle filter with $L = 2^{13}$ particles. %$l \in (1, \dots,L)$ particles, with $L = 2^{13} = 8192$. 
The update of the position $v^{(l)}$ and importance weight $a^{(l)}$ of the $l^{\rm th}$ particle are given by:
\begin{align}
    \text{d}v^{(l)}_t &= \frac{ \mu_{\text{ou}} - v_t^{(l)} }{\tau_{\text{ou}}}  \text{d}t + \sqrt{2 \sigma_{\text{ou}}^2 \tau_{\text{ou}}^{-1}} \mathbb{1} \text{d}V_t \\ 
    \text{d}a^{(l)}_t &= a_t^{(l)} \left( \frac{g(\beta  (v_t^{(l)})^\top x^{\epsilon}_t )}{\langle g_t \rangle}  -  1 \right)( y_t - \langle g_t \rangle)  \text{d}t,
\end{align}
where the average $\langle g_t \rangle = \sum_{l=1}^L a^{(l)}_t g(\beta (v_t^{(l)})^\top x_t^\epsilon )$ is computed based on the particle weights and their location. When the effective particle number $N_t = (\sum_{l =1}^L (a^{(l)}_t)^2 )^{-1} < \tfrac{3}{4}L$, we resample the particle location and set all weights to $1/L$. Additional details about particle filtering are found in the literature, e.g. \citep{doucet2000sequential,kutschireiter2020hitchhiker}.

%
%% Derivation
\subsection{Derivation of the Synaptic Filter}
\label{sec:si:derivation-all}
The derivation of the Synaptic Filter extends the work of Pfister et al. \citep{pfister2009know} to multidimensional hidden variables and a more complex observation process. The starting point of the derivation is the general framework of filtering with point observations and hidden diffusion dynamics. Then, the assumed density filter is introduced as a strategy to solve the filtering problem. The next two sections specify the generative model of the Synaptic Filter and show how it yields the update equations used in the Main Text.

%% Derivation of SF -> SI
%\subsubsection{Derivation of the update equations of the Synaptic Filter}
\subsubsection{Filtering with point process observations}
Given the continuous time spiking observations $y_t = \sum_f \delta(t - t^{(f)})$ from hidden weights $w_t \in \mathbb{R}^d$, our goal is to derive the update equations for the parameters $\theta_t$ of the proposal distribution $q_{\theta_t}(w_t)$ of an assumed density filter. The generative model is specified in terms of a prior transition probability and point emission process. For the transition, we consider diffusion processes of the form:
\begin{align}
\text{d}w_t = a(w_t)\text{d}t + b(w_t) \text{d}V_t,
\label{eq:transition-probability-general}
\end{align}
where $V_t$ is a $d$-dimensional Wiener process and $a \in \mathbb{R}^d \rightarrow \mathbb{R}^d$ and $b \in \mathbb{R}^d \rightarrow \mathbb{R}^{d \times d}$ are deterministic functions. The observation $\text{d}N_t \in \{0,1\}$ indicates whether a spike is present in the infinitesimal interval $\text{d}t$:
\begin{align}
\label{si:eq:dN}
\text{d}N_t \sim \text{Poisson}(g_t(w_t)\text{d}t),
\end{align}
where the (deterministic) gain function $g_t(w_t)$ relates the hidden weights to observations.

% KS-eq
This non-linear filtering problem has a general solution given by a Kushner-Stratonovic type of equation for point processes \citep{kushner1964differential,kushner1967dynamical,bremaud1981point}. Using the Laplacian $\mathcal{L}(\cdot) = a^\top \nabla (\cdot) + \tfrac{1}{2} \text{Tr}( b b^\top \nabla \nabla^\top (\cdot))$ all moments $\phi_t$ of the posterior obey the following formal solution:
\begin{align}
\label{eq:KS-eq-general}
\text{d} \langle \phi_t \rangle  = \langle \mathcal{L}(\phi_t) \rangle \text{d}t + \frac{\text{cov}(\phi_t, g_t)}{\gamma_t} \text{d}\delta_t,
\end{align}
where the expectation $\langle \cdot \rangle$ is taken with respect to the filtering distribution and we introduced the error signal $\text{d}\delta_t = \text{d}N_t - \gamma_t \text{d}t$ based on the expected firing rate $\gamma_t := \langle g_t \rangle$.

Generally, \Cref{eq:KS-eq-general} is intractable due to the closure problem, i.e., the evolution of $n$-th moment depends on the $n+1$-th moment. For instance, for the choice $g_t \equiv w_t \in \mathbb{R}$, the evolution of the first moment ($\phi_t \equiv w_t$) depends on the variance $\text{cov}(w_t^2)$. However, when \Cref{eq:KS-eq-general} is used to compute the evolution of the variance ($\phi_t \equiv w_t^2$), third order terms appear: $\text{cov}(w_t^2,w_t)$.

% ADF
\subsubsection{Assumed density filter with Gaussian proposal}
One strategy to apply \Cref{eq:KS-eq-general} is assumed density filtering (e.g. \citep{minka2001family}). The central idea is to replace the exact filtering distribution $p(w_t | \mathcal{D}_t)$ with a more tractable proposal distribution $q_{\theta_t}(w_t)$ with parameters $\theta_t \in \mathcal{S} \subset \mathcal{R}^m$. The fact that the proposal distribution belongs to a parametric family, limits its degrees of freedom. Thus, (in the absence of degeneracy) the number of parameters $m$ determines how many moments have to be computed in order to fully specify the evolution of the proposal distribution. The closure problem can be avoided.

The Synaptic Filter and the Diagonal Synaptic Filter are assumed density filters with a Gaussian proposal distribution:
\begin{align}
\label{si:eq:q}
    q_{\theta_t}(w_t) := \mathcal{N}(w_t ; \mu_t, \Sigma_t).
\end{align}
In the case of the Diagonal Synaptic Filter $\Sigma_t$ contains only diagonal elements. The general notation $\Sigma_t$ (with or without off-diagonal elements) allows for the simultaneous derivation of both filters.
To relate the parameters of the proposal $\theta_t = (\mu_t, \Sigma_t)$, we follow a simple moment-matching strategy, i.e., we use
\Cref{eq:KS-eq-general} to directly compute the updates for $\theta_t$. In the case of Gaussian distributions, moment matching is optimal in the sense that it minimizes the Kullback-Leibler divergence: $\mathcal{D}_{KL}(p(w_t | \mathcal{D}_t) | q_{\theta_t}(w_t))$. However, the question of how to optimally project the exact filtering distribution $p(w_t | \mathcal{D}_t)$ onto the proposal is an active research area in information geometry \citep{sugiyama2012density}.

To compute the evolution of $\theta_t$, we specify \Cref{eq:KS-eq-general} for the first two central moments $\mu_t$ and $\Sigma_t$ (\citep{kutschireiter2020hitchhiker}, Equations (101) and (102)):
\begin{align}
\text{d} \mu_t = \ & \langle a_t \rangle \text{d}t + \frac{\text{cov}(w_t,g_t)}{ \gamma_t} %^{-1}
\text{d}\delta_t
\label{eq:update-mean}
\\
\text{d}\Sigma^2_t = \ & (\text{cov}(a_t,w_t^\top) + \text{cov}(w_t,a_t^\top) + \langle b_t b_t^\top \rangle) \text{d}t
\notag
\\
\label{eq:update-cov}
&
+ \frac{\text{cov}(w_t w_t^\top,g_t) -  \mu_t \text{cov}(w_t^\top,g_t) -  \text{cov}(w_t,g_t) \mu_t^\top } {\gamma_t} %^{-1}
\text{d}\delta_t
\\ \notag \ \  &
- \frac{\text{cov}(w_t,g_t)\text{cov}(w_t^\top,g_t)}{\gamma_t^2} %^{-2}
\text{d}N_t.
\end{align}
Expectations $\langle \cdot \rangle$ are now evaluated with respect to the proposal density $q_{\theta_t}(w_t)$ rather than with respect to the exact filtering distribution $p(w_t | \mathcal{D}_t)$.
For the Diagonal Synaptic Filter, one considers only the updates of the diagonal elements in \Cref{eq:update-cov}. In general, the updates of the diagonal elements can have complex dependencies on off-diagonal elements, e.g., if we had assumed a non-diagonal matrix $b$, the term $b b^\top$ would have introduced such mixing of components. However, since $b$ was assumed to be a diagonal matrix and $w$ and $a$ a vectors, no mixing occurs in \Cref{eq:update-cov}. This is why in our case the updates of diagonal covariance elements in Diagonal Synaptic Filter and the Synaptic Filter are identical.
%Generally, \Cref{eq:update-mean,eq:update-cov} are intractable due to the closure problem, i.e., the evolution of $n$-th moment depends on the $n+1$-th moment. For instance, for the choice $g_t \equiv w_t$, the evolution requires the evaluation of the third-order moment $\text{w_t w_t^\top, w_t}$. The central idea of assumed density filtering is to avoid the closure problem by truncating the higher moments.
%Up to this point, the recited results are general. We have not yet used the fact that the proposal density  Gaussian. If we had chosen a proposal density with  Higher moment of the filtering distribution can be computed via \Cref{eq:KS-eq-general}, and
%Fortunately, this problem does not arise in the case of Gaussian (and log-normal) proposal densities because all moments can be expressed as a function of the first two.
%
%First, we evaluate the contributions from the transition probability:
%To compute the evolution of $\mu_t$ moments, we have to evaluate the following expressions:
%\begin{align}
%\langle a_t \rangle, \  \text{cov}(w_t^n,g_t), \ \text{cov}(a_t,w_t), \ \langle b_t^2 \rangle \ \text{and} \ \langle g_t \rangle, \label{eq:required-quantities}.
%\end{align}
%
%%
\subsubsection{OU-prior and exponential gain function}
%Here, we assume that the synapses normally distributed (not obeying Dale's law). To be consist with the treatment of log-normal synapses, we reserve the symbol $\lambda$ for this type of synapse and use $w$ for sign-preserving synapses.
%\begin{align}
%q(\lambda_t) = \prod_i \mathcal{N}(\lambda_{t,i};\mu_{t,i},\sigma_{t,i}^2) := \prod_i \frac{1}{\sigma_{t,i} \sqrt{2 \pi}} \exp \left( - \frac{ 1 }{2 \sigma^2_{t,i}} ( \lambda_{t,i} - \mu_{t,i} )^2 \right).
%\end{align}
To evaluate the terms in the moment evolution \Cref{eq:update-mean,eq:update-cov}, we must make specific choices for the transition probability, often simply referred to as \textit{prior} and the gain function $g_t$ in the point emission process \Cref{si:eq:dN}.

In our work, the emission probability reflects the output spiking of a neuron with membrane potential:
\begin{align}
u_t = w_t^\top (x * \epsilon)_t
\end{align}
where $\epsilon$ denotes the spike response kernel to given input spike trains $x_{t,i} = \sum_f \delta(t - t_i^{(f)})$.
Since the zero-th weight $w_{t,0}$ acts as bias, we adopt the convention $(x_0 * \epsilon)_t = 1$. The assumption of an exponential gain function has analytical advantages and corresponds to a neuron close to the onset of a sigmoidal gain function:
\begin{align}
\label{si:eq:gain}
g_t(w_t) \equiv g_0 \exp( \beta w_t^\top x^\epsilon_t) \equiv g(u_t).
\end{align}
The determinism parameter $\beta$ can be absorbed in the units of $x_t^\epsilon$ and is ommited from the rest of the derivation. In addition, we drop the temporal index, since the right-hand side in \Cref{eq:update-mean,eq:update-cov} is evaluated exclusively at time $t$. For brevity, we refer to the proposal distribution of the assumed density filter $q_{\theta_t}(w_t)$ as filtering distribution from here on.

For the transition probability of the hidden weights, we use an OU-process with equilibrium values $\mu_{\rm ou}$ and $\Sigma_{\rm ou} = \mathbb{1} \sigma^2_{\rm ou}$, and relaxation time scale $\tau_{\rm ou}$:
\begin{align}
\label{si:eq:ou-process}
a(w) \equiv -\tau^{-1}_{\rm{ou}} ( w - \mu_{\rm{ou}}), \hspace{2cm} b(w) \equiv \sqrt{ \frac{2 \sigma^2_{\rm{ou}}}{\tau_{\rm{ou}}}} \mathbb{1}.
\end{align}
With \Cref{si:eq:ou-process,si:eq:gain} and the update \Cref{eq:update-mean,eq:update-cov} and the proposal density \Cref{si:eq:q}, the assumed density filter is fully specified. The rest of derivation is dedicated towards the explicit computation of the terms in the update \Cref{eq:update-mean,eq:update-cov}.

\subsubsection{Explicit computation of terms in the update equations}
\subsubsection*{Prior}
First, the terms related to the transition probability in \Cref{eq:update-mean,eq:update-cov} are evaluated based on \Cref{si:eq:ou-process}:
\begin{align}
    \label{si:eq:Ea}
    \langle a(w) \rangle &= a(\mu) \\
    \label{si:eq:Ebb}
    \langle b b^\top \rangle &= 2\sigma_{\rm ou}^2 \tau_{\rm ou}^{-1} \mathbb{1} \\
    \label{si:eq:covaw}
    \text{cov}(a,w^\top) &= \langle a w^\top \rangle - \langle a \rangle \langle w^\top \rangle \\ \notag
    &= -\tau^{-1}_{\rm ou}(\mu \mu^\top + \Sigma - \mu_{\rm{ou}} \mu^\top) + \tau^{-1}_{\rm ou}(\mu - \mu_{\rm{ou}}) \mu^\top =  -\tau^{-1}_{\rm ou}\Sigma,
\end{align}
For the stimulated STDP experiments in the Main Text, we consider a fast time scale $\tau_{\rm m}$ for the bias and a slow time scale $\tau_{\rm ou}$ for the remaining weights. The update of the covariance element $\Sigma_{0i}$, which represents the correlations between the bias and the $i^{\rm th}$ weight, contains both time scales because of the first two terms on the left-hand side of \Cref{eq:update-cov}:
\begin{align}
\label{si:eq:two-timescales}
    \text{cov}(a_i,w_0) + \text{cov}(w_i,a_0) &= - (\tau_{\rm m}^{-1} + \tau_{\rm ou}^{-1}) \Sigma_{0i}.
\end{align}
However, since the values of the time scales differ by nine orders of magnitude in the STDP simulations, the contribution of $\tau_{\rm ou}$ can be safely ignored.
%implying that trivially $\langle b b^\top \rangle = 2\sigma_{\rm ou}^2 \tau_{\rm ou}^{-1} \mathbb{1}$ and $\langle a(w_t) \rangle = a(\mu_t)$. The final term associated with the transition probability reads:
%\begin{align}
%\text{cov}(a_t,w_t) &= \langle a_t w_t \rangle - \langle a_t \rangle \langle w_t \rangle = -\tau^{-1}(\mu_t^2 + \sigma^2_t - \mu_{\rm{ou}} \mu_t - (\mu^2_t - \mu_t \mu_{\rm{ou}})) =  -\tau^{-1}\sigma_t^2.
%\end{align}
\subsubsection*{Observations}
Next, we evaluate the terms in \Cref{eq:update-mean,eq:update-cov} that depend on the gain function. We begin by showing that the membrane potential is Gaussian under the statistics of the filtering distribution. Then we evaluate the Gaussian expectation of the gain function. Finally, the expectation of the gain function is used to compute the covariance-terms in \Cref{eq:update-mean,eq:update-cov}.

The weighted sums of a Gaussian random variables yields another Gaussian random variable. With the input kernels $x^\epsilon$ given and $q_\theta(w) = \mathcal{N}(w ; \mu, \Sigma)$, the membrane potential is therefore Gaussian with mean and covariance:
\begin{align}
    \mathbb{E}[u | \theta] &= \mu^\top x^\epsilon := \bar{u} \\
    \text{Var}[u | \theta] &= (x^\epsilon)^\top \Sigma x^\epsilon := \sigma_u^2.
\end{align}
Thus, for any function $g(u)$, the expectation over the $d$-dimensional filtering distribution can be replaced by the 1-dimensional expectation over the membrane distribution. For the exponential gain function this yields:
\begin{align}
\label{si:eq:gamma}
    \mathbb{E}[g | \theta] &= g_0 \int e^{u} \mathcal{N}(u;\bar{u},\sigma_u^2) \text{d}u
    = g_0  \exp ( \bar{u} + \tfrac{1}{2}\sigma_u^2 ) \underbrace{\int \mathcal{N}(u;\bar{u} + \sigma_u^2,\sigma_u^2) \text{d}u}_{= 1} := \gamma
\end{align}
where the second equality follows from completion of the square in the exponent of the Gaussian. The expected firing rate $\gamma:= g_0  \exp ( \bar{u} + \tfrac{1}{2}\sigma_u^2)$ plays a central role in the update equations and for making predictions with the filtering distribution in Bayesian regression.

To compute the covariance terms in \Cref{eq:update-mean,eq:update-cov} we use the fact that the expectation over the filtering distribution commutes with derivatives with respect to $x^\epsilon$:
\begin{align}
    \label{si:eq:covgw}
    \text{cov}[g,w] := \langle w g \rangle - \langle w \rangle \langle g \rangle = \langle \nabla_x g \rangle - \mu \gamma = \nabla_x \gamma - \mu \gamma  =
    (\mu + \Sigma x^\epsilon) \gamma - \mu \gamma = \Sigma x^\epsilon \gamma
\end{align}
The result $\nabla_x \gamma = (\mu + \Sigma x^\epsilon) \gamma$ can be reused to compute the higher order covariance term in \Cref{eq:update-cov}:
\begin{align}
    \label{si:eq:covgww}
    \text{cov}[g,w w^\top] &:= \langle w w^\top g \rangle - \langle g \rangle \langle w w^\top \rangle
    \\ \notag
    &= \nabla_x \nabla_x^\top \gamma - \gamma (\mu \mu^\top + \Sigma)  \\ \notag
    &= \nabla_x  (\gamma (\mu + \Sigma x^\epsilon) )^\top - \gamma (\mu \mu^\top + \Sigma)
    \\ \notag
    &= \gamma (\mu + \Sigma x^\epsilon) (\mu + \Sigma x^\epsilon)^\top  + \gamma \Sigma
     - \gamma (\mu \mu^\top + \Sigma) \\ \notag
    &= \gamma \mu (\Sigma x^\epsilon)^\top + \gamma (\Sigma x^\epsilon) \mu^\top +  \gamma (\Sigma x^\epsilon) (\Sigma x^\epsilon)^\top.
\end{align}

\subsubsection*{Combining results to obtain update equations of the Synaptic Filter}
With the results for the observation terms, i.e.,
\Cref{si:eq:covgw,si:eq:covgww}, the variance update \Cref{eq:update-cov} simplifies considerably. Most terms related to gain function, i.e., terms proportional to $\text{d}t\delta$ and $\text{d}N$, cancel:
\begin{align}
\text{d}\Sigma &\propto
\frac{\text{cov}(w w^\top,g) -  \mu \text{cov}(w^\top,g) -  \text{cov}(w,g) \mu^\top } {\gamma} %^{-1}
\text{d}\delta
- \frac{\text{cov}(w,g)\text{cov}(w^\top,g)}{\gamma^2}
\text{d}N \\ \notag
 &= (\Sigma x^\epsilon) (\Sigma x^\epsilon)^\top \text{d}\delta - (\Sigma x^\epsilon) (\Sigma x^\epsilon)^\top \text{d}N
 =
 - \gamma (\Sigma x^\epsilon) (\Sigma x^\epsilon)^\top \text{d}t.
\end{align}
We obtain the update equations for the Synaptic Filter and the Diagonal Synaptic Filter (for which $\Sigma$ is a diagonal matrix), by substituting the results of prior and observations, i.e., \Cref{si:eq:Ea,si:eq:covaw,si:eq:Ebb,si:eq:covgw,si:eq:covgww}, into the update \Cref{eq:update-mean,eq:update-cov} for the moments. In addition, we use $y := \text{d}N/\text{d}t$ to switch to standard ODE-notation:
\begin{align}
\label{eq:si:dot-mu}
\dot{\mu} &=
\Sigma x^{\epsilon} (y - \gamma) + \tau_{\rm{ou}}^{-1} (\mu_{\rm{ou}}-\mu),
\\
\label{eq:si:dot-sigma2}
\dot{\Sigma} &= - \gamma (\Sigma x^{\epsilon})(\Sigma x^{\epsilon})^\top +
2 \tau_{\rm{ou}}^{-1} (\Sigma_{\rm{ou}}-\Sigma),
\end{align}
where $y := \text{d}N/\text{d}t$. In the Main Text, the determinism parameter $\beta$ scales the input variable $x^\epsilon$.

%
%% 3 C2 gradient
\subsection{Evaluating the predictive of the gradient rule}
\label{si:sec:grad-fit}
To obtain the predictive performance of the gradient rule, we must maximise the loglikelihood as a function of the learning rate. This is challenging because the loglikelihood is noisy and due to limited computational resources, we could only evaluate few values of the learning rate. 

\begin{figure}[h!]
\begin{minipage}{\myFigureWidth\linewidth}
\begin{tabular}{ll}
{\bf (A)} & {\bf (B)} \\
\begin{minipage}{0.45\textwidth}
\includegraphics[width=\textwidth,trim={0cm 0cm 0cm 0cm},clip]{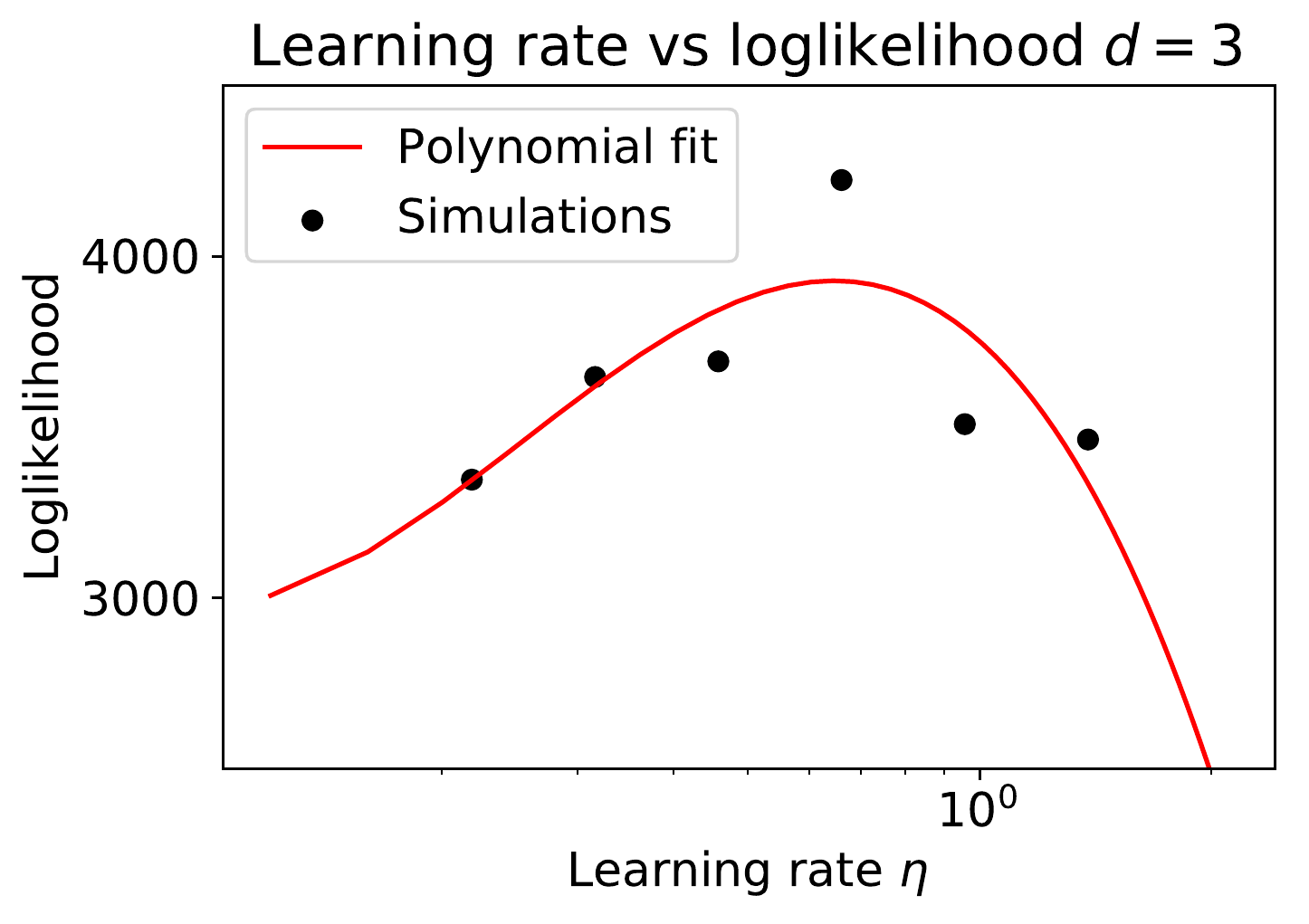}
\end{minipage} &
\begin{minipage}{0.45\textwidth}
\includegraphics[width=\textwidth]{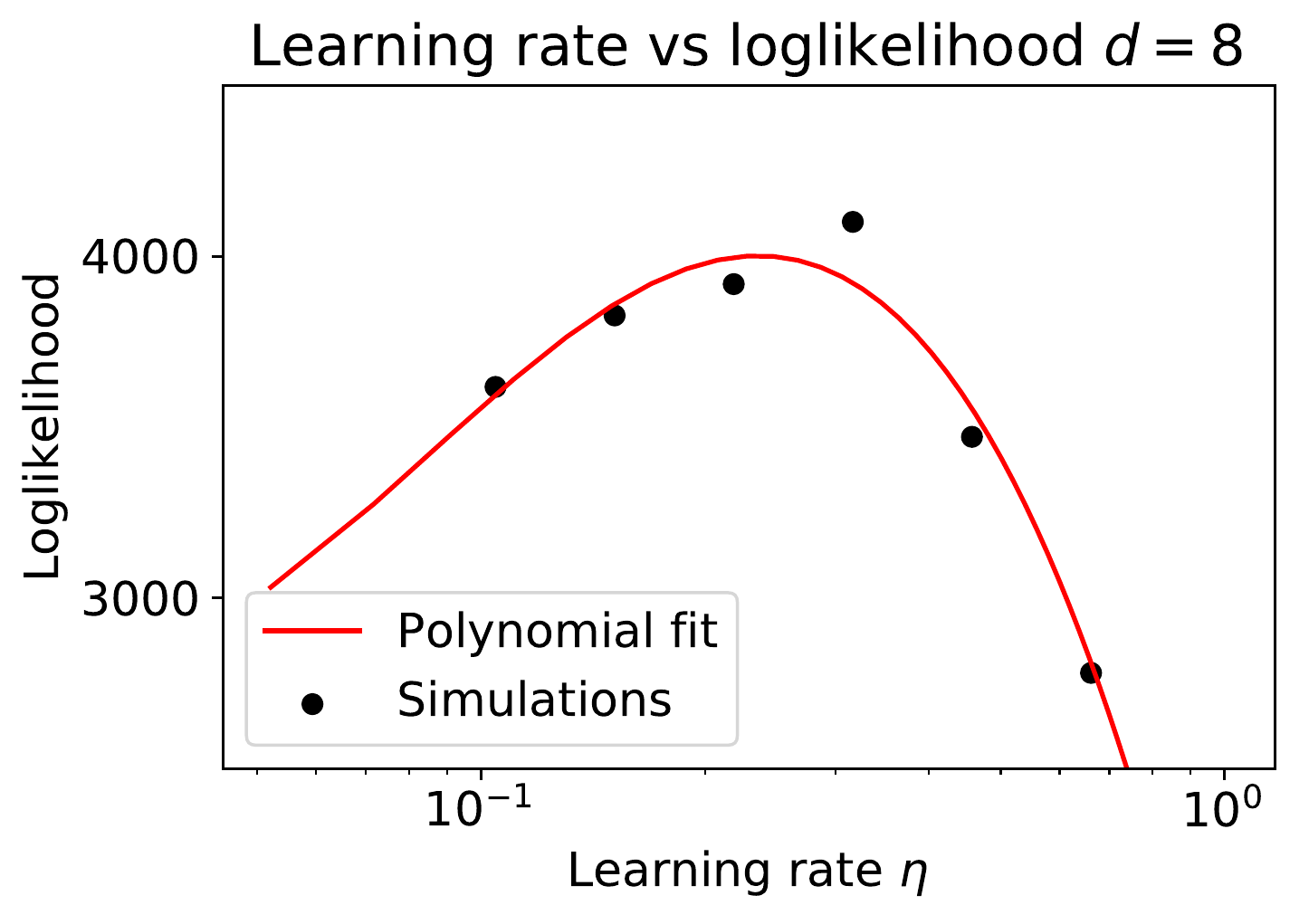}
\end{minipage}
\end{tabular}
\end{minipage}
\caption{ \label{si:fig:grad-fit}
% main message
Two typical examples of the polynomial fit to the loglikelihood of the gradient learning rule. \textbf{(A)} The simulation at $d = 3$ showcases the denoising effect of the fit. Compared to the outlier (central black dot), the maximal loglikelihood based on the fit is lower. \textbf{(B)} The example at $d=8$ show that the optimal learning rate based on the fit shifts is smaller compared to the optimal learning rate obtained by selecting the simulation (black dot) with the highest loglikelihood value.
%The initial values for the means and diagonal variance elements where set to 1. To ensure comparability, $\beta \equiv 1$ for all three settings.
}
\end{figure}

Our strategy was to evaluate the loglikelihood at 11 log-spaced values for the learning rate in the interval $[0.05,2]$. The choice of the interval is motivated in %\Cref{met:sec:c012}
Main Text by noting that the learning rate replaces the variance in the Diagonal Synaptic Filter. \Cref{si:fig:grad-fit} shows that the interval contains the optimal learning rate indeed.

To maximise the loglikelihood with respect to the learning rate, we fit a 3rd order polynomial to a 7 point neighbourhood around the maximum and selected its maximum value of the loglikelihood. Compared to selected the maximal value of the simulated loglikelihoods, using the fit has two advantages. First, it reduces the risk of selecting a statistical outlier as the maximum loglikelihood, as shown in \Cref{si:fig:grad-fit} \textbf{(A)}. Secondly, it compensates for the sparseness of our evaluation of the loglikelihood by interpolation, as seen in \Cref{si:fig:grad-fit} \textbf{(B)}. The predictive performance benchmark shown in the Main Text %in \Cref{res:fig:c2} 
corresponds to maximum value of the polynomial, obtained for each dimension separately.

%
%% 4 Time series
\subsection{Variables of the Synaptic Filter during simulated biological experiments}
\label{si:sec:SF-variable-dynamics}
The Synaptic Filter can explain the biological phenomena of STDP and the negative correlation between homo- and heterosynaptic plasticity. The goal of this section is to show how the interaction of the variables of the Synaptic Filter produces the aforementioned biological observations. In the following, we show the time series of the mean $\mu_t$ and covariance matrix $\Sigma_t$, alongside the simulated protocols.

\begin{figure}[h!]
\begin{minipage}{\myFigureWidth\linewidth}
\begin{tabular}{ll}
{\bf (A)} & {\bf (B)} \\
\begin{minipage}{0.45\textwidth}
\includegraphics[width=\textwidth,trim={0cm 0cm 0cm 0cm},clip]{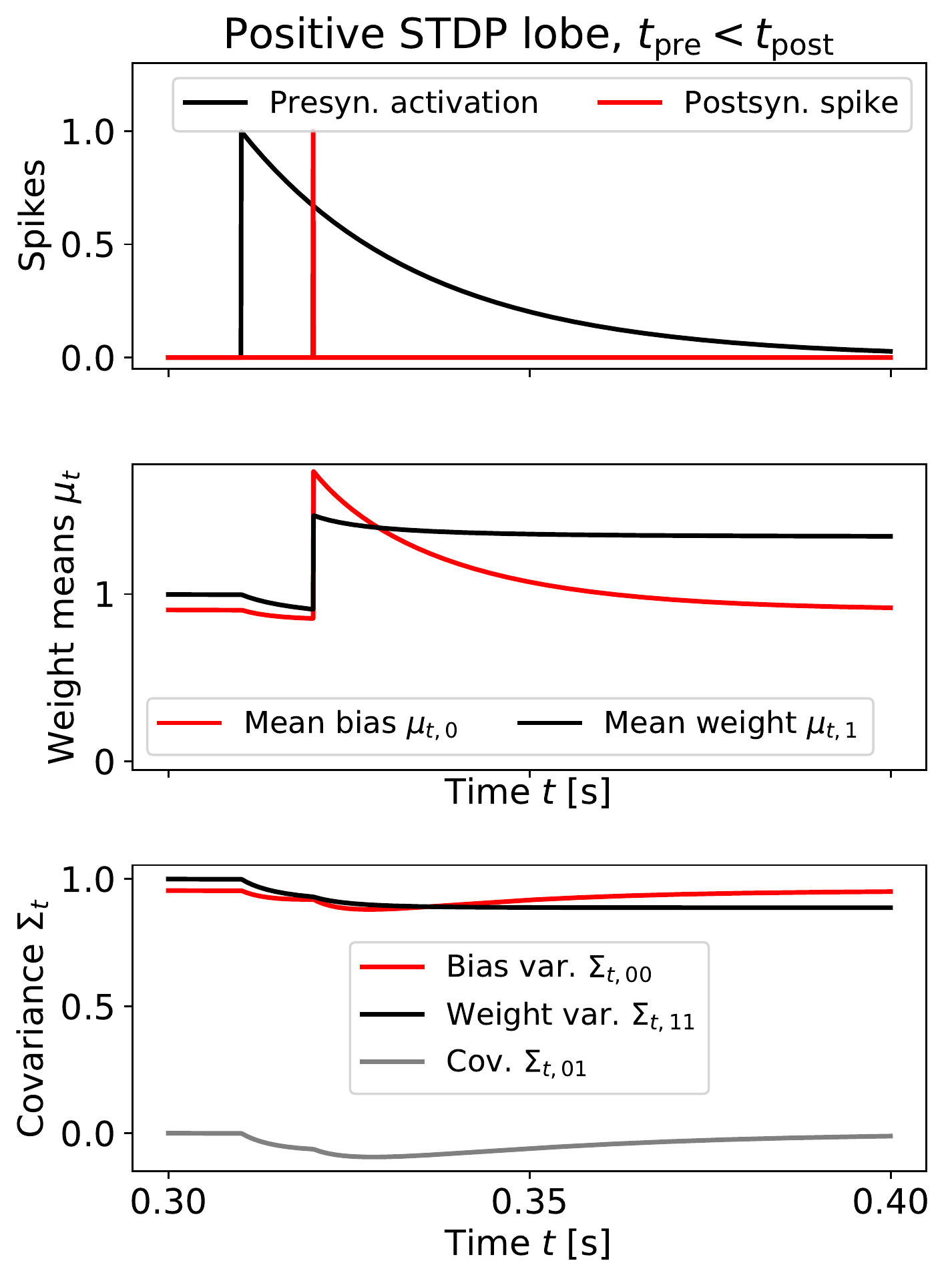}
\end{minipage} &
\begin{minipage}{0.45\textwidth}
\includegraphics[width=\textwidth]{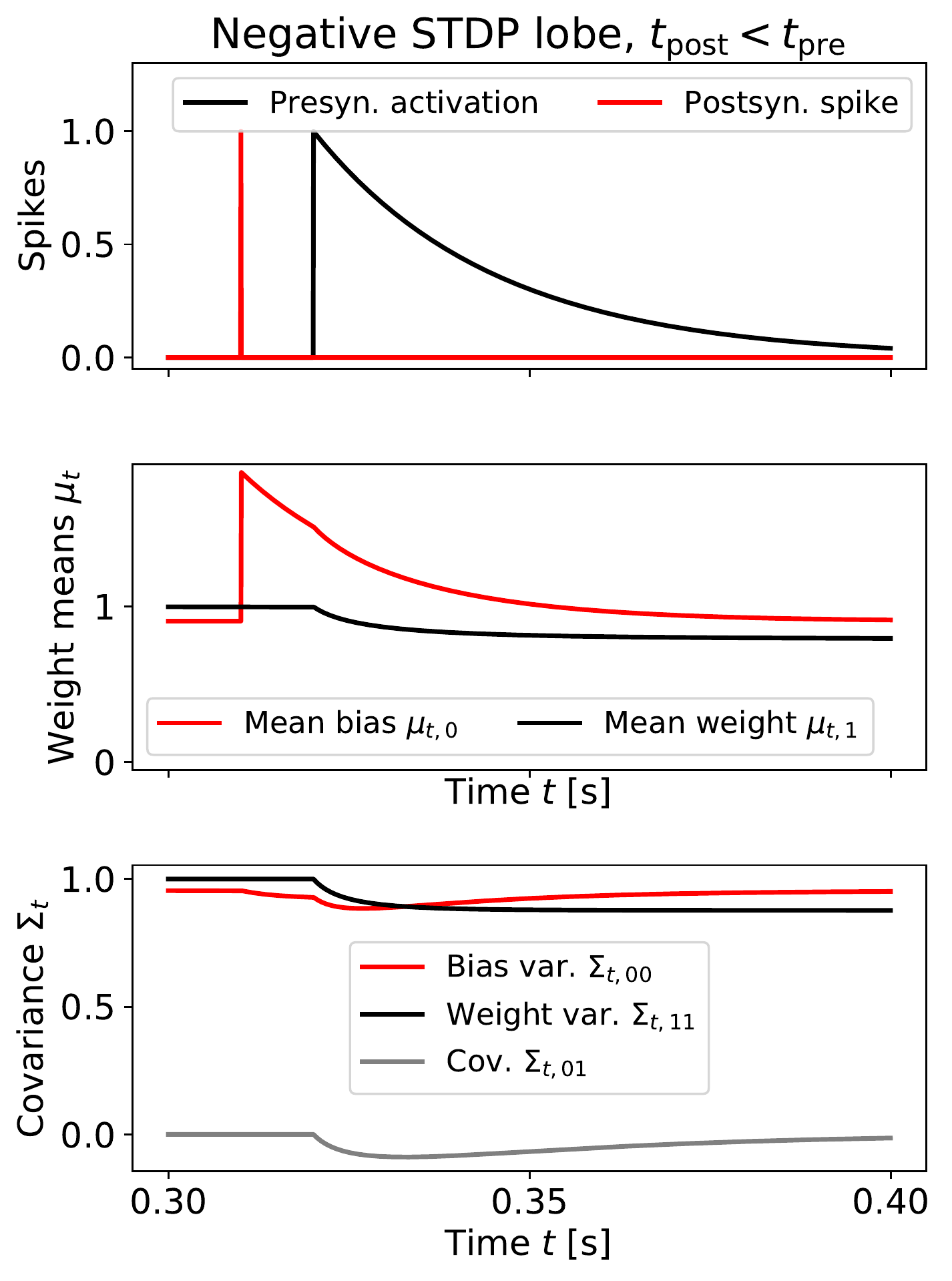}
\end{minipage}
\end{tabular}
\end{minipage}
\caption{ \label{si:fig:STDP-variable-dynamics} The dynamics of the variables of the Synaptic Filter during the STDP protocol. \textbf{(A, top)} shows a protocol for the positive lobe with a presynaptic spike (black) followed by a postsynaptic spike (red) with 10ms delay. The presynaptic activation paired with a postsynaptic spikes increases the mean value of the bias (red) and synaptic weight (black), shown in \textbf{(A, middle)}. The returns to its equilibrium value on a timescale of $\tau_{\rm ou,bias} = 25$ms. \textbf{(A, bottom)} shows that spiking activities reduces the elements of the covariance matrix. The variance of the bias returns quickly to its equilibrium value.
\textbf{(B, top)} shows a protocol for the negative lobe. Importantly the depression of the mean weight (black) in \textbf{(B, middle)} is modulated by mean value of the bias (red). The smaller the delay of the presynaptic spike, the larger is the decrease in the mean weight.}
\end{figure}

The STDP curve shown in the Main Text %\Cref{fig:STDP-prediction} 
arises from a series of simulations with varying spike-timings. \Cref{si:fig:STDP-variable-dynamics} \textbf{(A)} and \textbf{(B)} show Synaptic Filter variables for the cases $t_{\rm pre} - t_{\rm post} = \pm 10$ms, linked to the positive and negative STDP lobe respectively. The strength of potentiation in \textbf{(A)} is related to the amount of presynaptic activation present when the postsynaptic spike occurs. The reason is that the product $y_t x^\epsilon_t$ is proportional to the update of the mean in \Cref{eq:si:dot-mu}. The strength of depression in \textbf{(B)} depends on amplitude of the mean of the bias when the presynaptic spike occurs. The mean bias increases the expected firing rate $\gamma_t$, which modulates the update via the term $-\gamma_t x^\epsilon_t$. Thus the timescale of the negative lobe is directly associated with the prior timescale of the bias $\tau_{\rm ou,bias}$, which controls how quickly the bias returns to its equilibrium value.
From this mechanistic perspective on the STDP protocol, it becomes clear that the negative lobe dependence on the presence of the bias. However, this result do not qualitatively depend on including the dynamics of the covariance matrix.

\begin{figure}[h!]
\begin{minipage}{\myFigureWidth\linewidth}
\begin{tabular}{ll}
{\bf (A)} & {\bf (B)} \\
\begin{minipage}{0.45\textwidth}
\includegraphics[width=\textwidth,trim={0cm 0cm 0cm 0cm},clip]{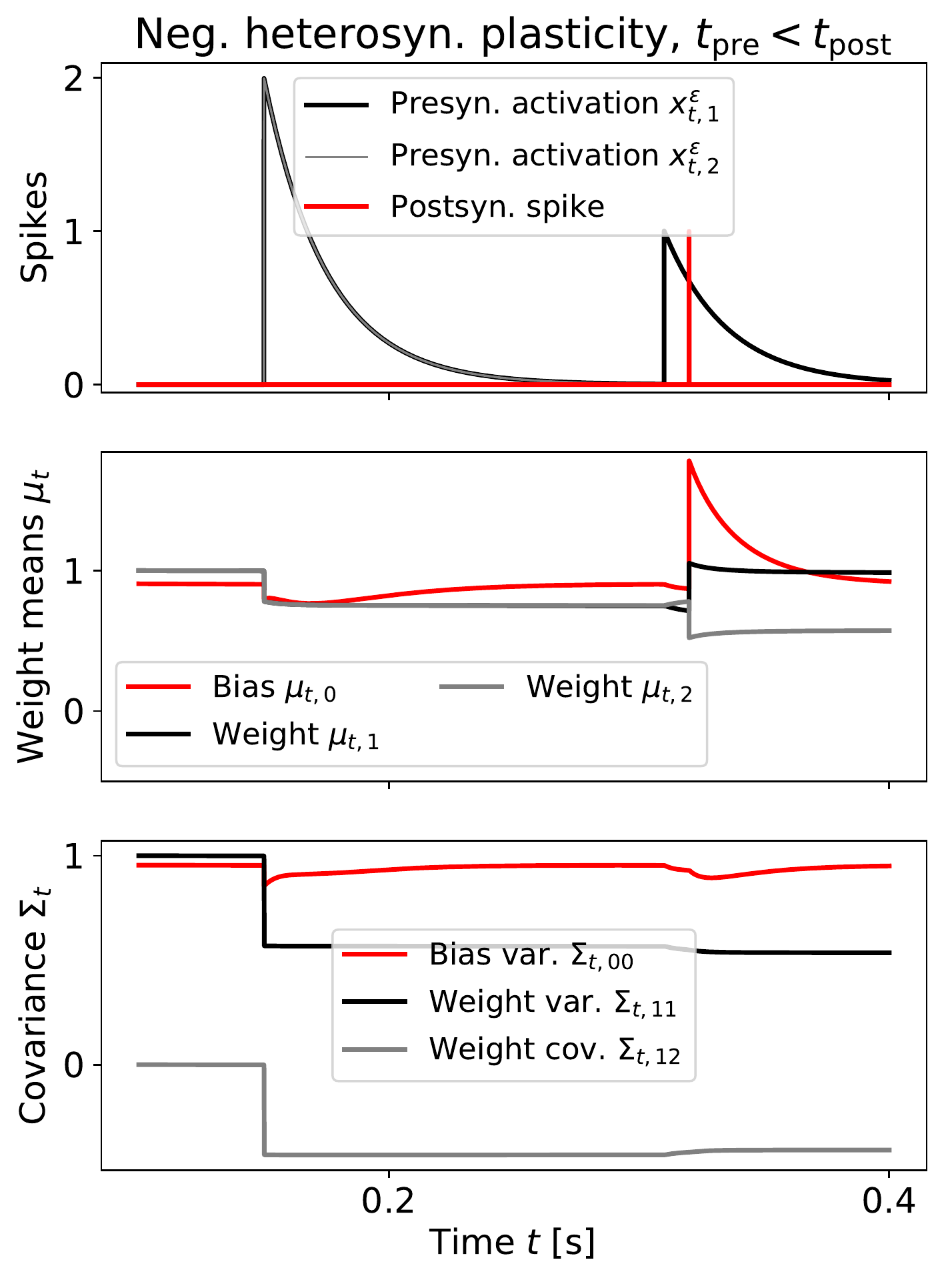}
\end{minipage} &
\begin{minipage}{0.45\textwidth}
\includegraphics[width=\textwidth]{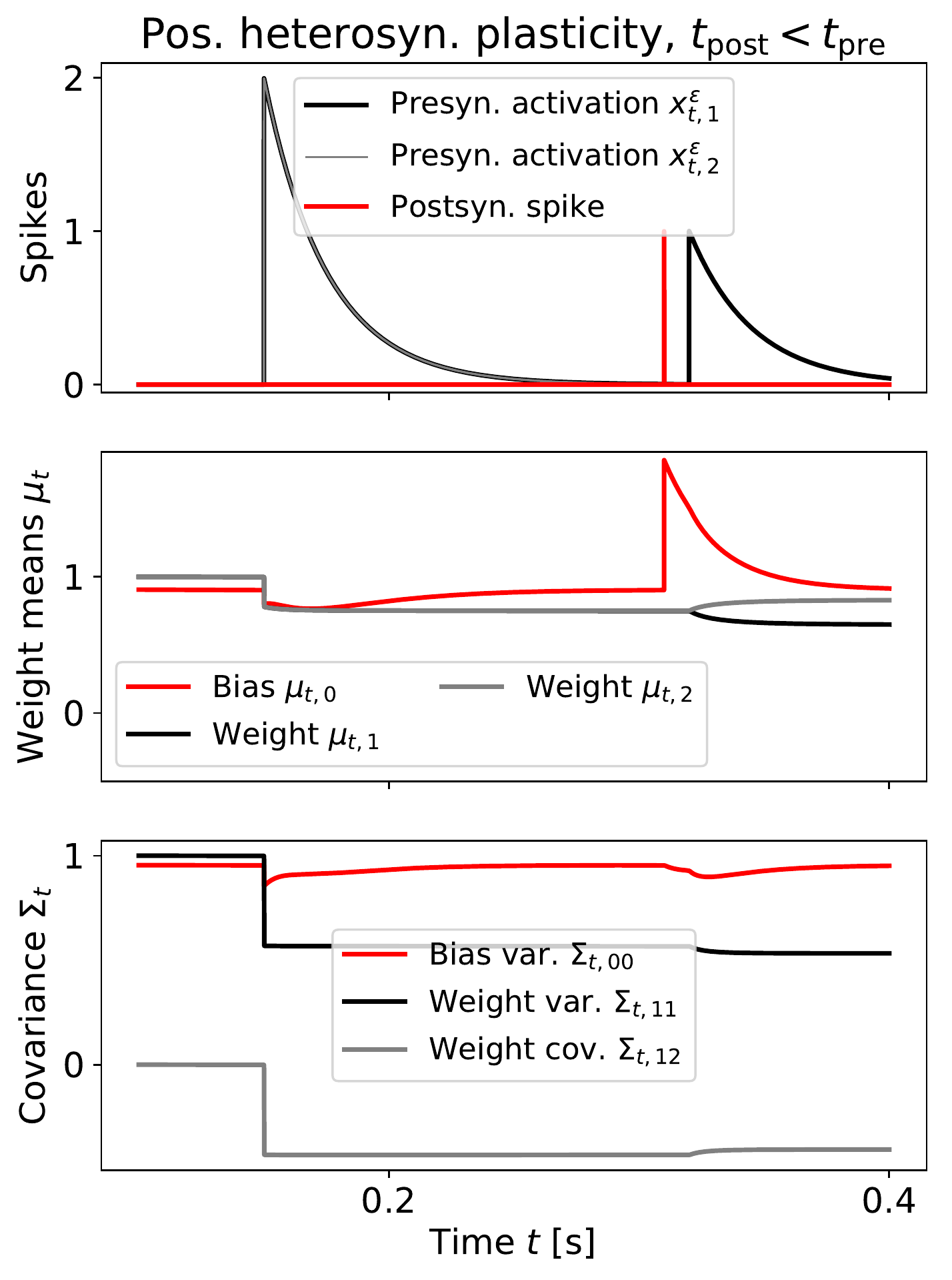}
\end{minipage}
\end{tabular}
\end{minipage}
\caption{ \label{si:fig:HETERO-variable-dynamics} The dynamics of the variables of the Synaptic Filter during the heterosynaptic protocol with preconditioning. \textbf{(A, top)} shows the presynaptic activation (black, gray) of both synapses and the postsynaptic spikes (red) for a pre-before-post protocol. The initial waiting time of 100ms has been removed. \textbf{(A, middle)} shows the mean weights (black, gray) and the bias (red). The bias returns to its equilibrium value after the preconditioning protocol. At $t=300$ms, the pre-post protocol is applied to the first weight, leading to potentiation and depression in the first and second mean weight respectively. \textbf{(A, bottom)} shows elements of the covariance matrix. The preconditioning protocol reduces the variance values. In particular, the covariance between the weights drops below zero. A non-zero value of the negative covariance causes the heterosynaptic plasticity.
\textbf{(B)} shows a post-before-pre protocol. In contrast to the previous case, the arrival of a presynaptic spike at $t=300$ms causes homosynaptic depression (black) and heterosynaptic potentiation (gray), as shown in \textbf{(B, middle)}.}
\end{figure}

The simulations of the heterosynaptic plasticity protocol %(shown in \Cref{fig:hetero-prediction}) 
are similar to the STDP protocol; however, they include a preconditioning protocol and an additional synaptic weight, whose change in strength we label as heterosynaptic plasticity. \Cref{si:fig:HETERO-variable-dynamics} \textbf{(A)} and \textbf{(B)} show Synaptic Filter variables for the cases $t_{\rm pre} - t_{\rm post} = \pm 10$ms. The preconditioning protocol consists of two presynaptic spikes at both inputs with minimal delay. This correlated input causes negative weight correlations via \Cref{eq:si:dot-sigma2}. When the pre-before-post, shown in \textbf{(A)}, and post-before-pre, shown in \textbf{(B)}, protocols are applied, the negative weight correlation between the first and the second weight leads to opposing directions of plasticity. Mathematically, this is manifested in the prefactor $\Sigma_t x^\epsilon_t$ in the updates, \Cref{eq:si:dot-mu}. When weight correlations are present, the covariance matrix converts presynaptic activation in the first weight into a non-zero coefficient in the second weight, i.e., the matrix prefactor mixes presynaptic activation between inputs. Thus, the results for heterosynaptic plasticity depend on the presence of weight correlations.

% figure STDP

%
% figure HETERO

%
%% Sec 6
\subsection{Negativity of weight correlations}
In the Materials and Methods in the Main Text, we showed that the weight correlations, i.e., the off-diagonal elements of the covariance matrix $\Sigma$, are always negative in two dimensions. In the following, we show that for constant input $x_t^{\epsilon} \equiv x_0 = {\rm const}$ the off-diagonal elements are negative for any $d$.

To show this, we represent the covariance matrix in the orthogonal, normalised basis: $\hat{x}_i^\top \hat{x}_j = \delta_{ij}$ for all $i,j \in (0, \dots, d-1)$. The basis is chosen such that its first basis vector is parallel to the input vector: $\hat{x}_{0}^\top x_0 = || x_0 ||$. In this basis, the covariance matrix is:
\begin{align}
    \Sigma = \sum_{i=0}^{d-1} a_i \hat{x}_{i} \hat{x}_{i}^{\top},
    \label{si:eq:decomposition}
\end{align}
where $a_i \geq 0$ because the covariance is positive semi-definite. Using this representation in the covariance update \Cref{eq:si:dot-sigma2} and projecting the dynamics onto the basis yields update equations for the coefficients:
\begin{align}
    \label{si:eq:coef-decay}
    \dot{a}_i = \hat{x}_i^\top \dot{\Sigma} \hat{x}_i = - \delta_{i0} \gamma a_i^2 ||x_0||^2 - 2 \tau_{\rm ou}^{-1} (a_i - \sigma^2_{\rm ou}),
\end{align}
where we used $\Sigma_{\rm ou} = \mathbb{1} \sigma_{\rm ou}^2$ to obtain the second term. 

The initial condition of the covariance matrix is $\Sigma = \mathbb{1}\sigma_{\rm ou}^2$. Thus, the initial condition of the coefficients is $a_{i} = \sigma^2_{\rm ou}$. It follows from \Cref{si:eq:coef-decay} that the updates of the coefficients are non-positive $\dot{a}_i \leq 0$. The values of $a_i$ decrease until a fixed point is reached. This implies, because the elements of the first basis vector are non-negative, $\hat{x}_{0,k} \geq 0$ for all $k \in (0, \dots, d-1)$, that all elements of the covariance matrix, including the off-diagonals, decrease as well until the fixed points of the coefficients $a_i$ are reached.

Empirically, we find that weight correlations are non-positive in the case of time-dependent inputs as well. The following analysis of a change from one static input to another supports this observation because it concludes that the first coefficient still has the fastest reduction rate after the change. 

The time-dependent input vector is $x_t^\epsilon$. Changes in the input vector correspond to a transformation of the basis vectors $\{ \hat{x}_i \}$, used in \Cref{si:eq:decomposition}. Under such transformations, the first basis vector $\hat{x}_t^\epsilon$ remains in the first quadrant because its entries are always non-negative. Without loss of generality, we assume that such a transformation is characterised by a rotation $\phi \in [0,\pi/2]$ in the plane of the first two vectors of the basis and under the constraint that the first basis vector remains in the first quadrant. Thus, the first and second coefficient change as follows: 
\begin{align}
    \label{si:eq:a0dash}
    a_0' = \cos(\phi) a_0 + \sin(\phi) a_1 \\
    \label{si:eq:a1dash}
    a_1' = -\sin(\phi) a_0 + \cos(\phi) a_1.
\end{align}
From the analysis of the static case and the presence of an additional decay term in \Cref{si:eq:coef-decay} for the first coefficient, we conclude that $a_0 < a_1$ before the change of the input vector occurs. Thus \Cref{si:eq:a0dash} implies that $a_0' > a_0$, which leads to a faster decay of the first coefficient after the change of input. \Cref{si:eq:a1dash} implies that $a_1' < a_1$, resulting in a slower recovery to the equilibrium value $\sigma^2_{\rm ou}$ than without the change in input. However, the effect of changing the basis vectors is neglected here.

Intuitively, the transformation, which corresponds to changes in the input, shifts the direction of maximal reduction of the covariance away from the direction in which the covariance has been maximally reduced under the previous input. Thus, the coefficient $a_0$ remains the one with the highest rate of reduction compared to the other coefficients. 

Turning this argument into a proof in the case of time-varying inputs would require the investigation of all edge cases of the coefficient dynamics and is beyond the scope of this work.

% Bibliography
\bibliography{pnas-sample}

\end{document}